\documentclass[10pt]{iopart}

\newcommand{\hc}{\hat{c}^{\phantom{\dagger}}}
\newcommand{\hcd}{\hat{c}^{\dagger}}

\newcommand{\eq}{\begin{equation}}
\newcommand{\eqx}{\end{equation}}
\newcommand{\eqn}{\begin{eqnarray}}
\newcommand{\eqnx}{\end{eqnarray}}

\newcommand{\Gammam}{\Omega}

\newcommand{\veco}{{\bf 0}}
\newcommand{\vecr}{{\bf r}}
\newcommand{\vecB}{{\bf B}}
\newcommand{\vecU}{{\bf U}}
\newcommand{\vecc}{{\bf c}}
\newcommand{\vecd}{{\bf d}}
\newcommand{\vecI}{{\bf I}}

\newcommand{\uu}{{u}}
\newcommand{\vv}{{v}}
\newcommand{\ze}{{\zeta}}
\newcommand{\et}{{\eta}}

\newcommand{\uua}{{u}}
\newcommand{\vva}{{v}}
\newcommand{\zea}{{\zeta}}
\newcommand{\xia}{{\xi}}

\newcommand{\ag}{{a}}
\newcommand{\xx}{{x}}
\newcommand{\yy}{{y}}
\newcommand{\zz}{{z}}
\newcommand{\al}{{\alpha}}
\newcommand{\be}{{\beta}}
\newcommand{\ga}{{\gamma}}

\newcommand{\xs}{{x'}}
\newcommand{\ys}{{y'}}
\newcommand{\as}{{\alpha}'}
\renewcommand{\bs}{{\beta}'}

\newcommand{\aga}{{a}}
\newcommand{\xxa}{{x}}
\newcommand{\yya}{{y}}
\newcommand{\zza}{{z}}
\newcommand{\ala}{{\alpha}}
\newcommand{\bea}{{\beta}}
\newcommand{\gaa}{{\gamma}}

\newcommand{\agz}{{a}}
\newcommand{\xxz}{{x'}}
\newcommand{\yyz}{{y'}}
\newcommand{\zzz}{{z}}
\newcommand{\alz}{{\alpha'}}
\newcommand{\bez}{{\beta'}}
\newcommand{\gaz}{{\gamma}}

\newcommand{\sqaa}{{\frac{1}{2}}}
\newcommand{\sqab}{{\frac{3}{2}}}
\newcommand{\sqac}{{\sqrt{3}}}
\newcommand{\sqad}{{\frac{1}{\sqrt{3}}}}
\newcommand{\sqae}{{\frac{\sqrt{3}}{2}}}
\newcommand{\sqaf}{{\frac{1}{3}}}
\newcommand{\sqag}{{\frac{2}{3}}}
\newcommand{\sqah}{{\frac{\sqrt{3}}{4}}}
\newcommand{\sqai}{{}}
\newcommand{\sqaj}{{\frac{4}{3}}}
\newcommand{\sqak}{{2}}

\newcommand{\sqan}{{\frac{1}{\sqrt{12}}}}
\newcommand{\sqao}{{\frac{2}{\sqrt{3}}}}
\newcommand{\sqap}{{\frac{3}{4}}}
\newcommand{\sqaq}{{\frac{1}{4}}}
\newcommand{\sqar}{{\frac{9}{8}}}
\newcommand{\sqas}{{\frac{1}{6}}}
\newcommand{\sqat}{{\frac{5}{6}}}
\newcommand{\sqau}{{\frac{4}{9}}}
\newcommand{\sqav}{{\frac{1}{192}}}
\newcommand{\sqaw}{{\frac{4096}{6250}}}

\newcommand{\sqay}{{\frac{8}{9}}}
\newcommand{\sqaz}{{\frac{7}{9}}}

\newcommand{\sqba}{{\frac{16}{9}}}
\newcommand{\sqbb}{{\frac{5}{3}}}
\newcommand{\sqbc}{{\frac{1}{18}}}
\newcommand{\sqbd}{{\frac{2}{18}}}
\newcommand{\sqbe}{{\frac{5}{4}}}
\newcommand{\sqbf}{{\frac{45}{64}}}
\newcommand{\sqbg}{{\frac{13}{64}}}
\newcommand{\sqbh}{{\frac{17}{18}}}
\newcommand{\sqbi}{{\frac{7}{18}}}
\newcommand{\sqbj}{{\frac{161}{192}}}
\newcommand{\sqbk}{{\frac{65}{192}}}
\newcommand{\sqbl}{{\frac{371}{576}}}
\newcommand{\sqbm}{{\frac{659}{576}}}
\newcommand{\sqbn}{{\frac{31}{48}}}

\newcommand{\sqbo}{{\frac{1}{3}\sqrt{\frac{5}{3}}}}
\newcommand{\sqbp}{{\frac{2}{3}\sqrt{\frac{5}{3}}}}
\newcommand{\sqbq}{{\frac{1}{9}\sqrt{\frac{5}{3}}}}
\newcommand{\sqbr}{{\frac{4}{27}\sqrt{5}}}
\newcommand{\sqbs}{{\frac{1}{6}\sqrt{\frac{5}{3}}}}
\newcommand{\sqbt}{{\frac{1}{2}\sqrt{\frac{5}{3}}}}
\newcommand{\sqbu}{{\sqrt{\frac{5}{3}}}}
\newcommand{\sqbv}{{\frac{5}{4}\sqrt{\frac{5}{3}}}}
\newcommand{\sqbw}{{\frac{13}{64}\sqrt{\frac{5}{3}}}}
\newcommand{\sqbx}{{\frac{61}{192}\sqrt{\frac{5}{3}}}}
\newcommand{\sqby}{{\frac{77}{48}\sqrt{\frac{5}{3}}}}
\newcommand{\sqbz}{{\frac{1}{8}}}
\newcommand{\sqca}{{\frac{3}{8}}}
\newcommand{\sqcb}{{\frac{\sqrt{3}}{8}}}
\newcommand{\sqcc}{{3\frac{\sqrt{3}}{8}}}
\newcommand{\sqcd}{{3\frac{\sqrt{3}}{4}}}
\newcommand{\sqce}{{\frac{3}{16}}}
\newcommand{\sqcf}{{\frac{27}{16}}}
\newcommand{\sqcg}{{\frac{9}{16}\sqrt{3}}}
\newcommand{\sqch}{{\frac{1}{16}\sqrt{3}}}
\newcommand{\sqci}{{\frac{9}{8}\sqrt{3}}}
\newcommand{\sqcj}{{\frac{9}{4}}}
\newcommand{\sqck}{{\frac{3}{2}\sqrt{3}}}
\newcommand{\sqcl}{{\frac{7}{8}}}
\newcommand{\sqcm}{{\frac{1}{10}}}
\newcommand{\sqcn}{{\frac{1}{20}}}
\newcommand{\sqco}{{\frac{3}{5}}}
\newcommand{\sqcp}{{\frac{2}{15}}}
\newcommand{\sqcq}{{\frac{4}{5}}}
\newcommand{\sqcr}{{\frac{2}{9}}}
\newcommand{\sqcs}{{\frac{1}{60}}}
\newcommand{\sqct}{{\frac{1}{36}}}
\newcommand{\sqcu}{{\frac{1}{12}}}
\newcommand{\sqcv}{{\frac{1}{15}}}
\newcommand{\sqcw}{{\frac{1}{30}}}
\newcommand{\sqcx}{{\frac{4}{15}}}
\newcommand{\sqcy}{{\frac{1}{2\sqrt{15}}}}
\newcommand{\sqcz}{{\frac{247}{1440}}}

\newcommand{\sqda}{{\frac{73}{180}}}
\newcommand{\sqdb}{{\frac{2}{\sqrt{15}}}}
\newcommand{\sqdc}{{\frac{1}{8\sqrt{15}}}}
\newcommand{\sqdd}{{\frac{1}{\sqrt{15}}}}
\newcommand{\sqde}{{\frac{4}{5\sqrt{15}}}}
\newcommand{\sqdf}{{\frac{1}{8}\sqrt{\frac{3}{5}}}}
\newcommand{\sqdg}{{\frac{1}{16}\sqrt{\frac{3}{5}}}}
\newcommand{\sqdh}{{\frac{2}{5\sqrt{15}}}}
\newcommand{\sqdi}{{\frac{1}{32}\sqrt{\frac{3}{5}}}}
\newcommand{\sqdj}{{ \frac{49}{160\sqrt{15}} }}
\newcommand{\sqdk}{{\frac{\sqrt{15}}{32}}}
\newcommand{\sqdl}{{\frac{1}{32}}}
\newcommand{\sqdm}{{\frac{31}{60}}}
\newcommand{\sqdn}{{\frac{8}{15}}}
\newcommand{\sqdo}{{\frac{3}{20}}}
\newcommand{\sqdp}{{\frac{7}{30}}}
\newcommand{\sqdq}{{\frac{3}{10}}}
\newcommand{\sqdr}{{\sqrt{\frac{3}{5}}}}
\newcommand{\sqds}{{\frac{9}{4}\sqrt{\frac{3}{5}}}}
\newcommand{\sqdt}{{\frac{3}{8}\sqrt{\frac{3}{5}}}}
\newcommand{\sqdu}{{2\sqrt{\frac{3}{5}}}}
\newcommand{\sqdv}{{\frac{1}{4\sqrt{15}}}}
\newcommand{\sqdw}{{\frac{1}{4}\sqrt{\frac{3}{5}}}}
\newcommand{\sqdx}{{\frac{1}{20\sqrt{15}}}}
\newcommand{\sqdy}{{\frac{3}{16}\sqrt{\frac{3}{5}}}}
\newcommand{\sqdz}{{\frac{1}{40\sqrt{15}}}}

\newcommand{\sqea}{{\frac{9}{8}\sqrt{\frac{3}{5}}}}
\newcommand{\sqeb}{{\frac{13}{8\sqrt{15}}}}
\newcommand{\sqec}{{\frac{1}{8\sqrt{15}}}}
\newcommand{\sqed}{\frac{13}{16\sqrt{15}}{}}
\newcommand{\sqee}{{\frac{1}{16\sqrt{15}}}}

\newcommand{\sqeg}{{\frac{257}{80\sqrt{15}}}}
\newcommand{\sqeh}{{\frac{79}{160}}}
\newcommand{\sqei}{{\frac{7}{320}}}

\newcommand{\sqek}{{\frac{379}{1920}}}
\newcommand{\sqel}{{\frac{1}{3\sqrt{15}}}}
\newcommand{\sqem}{{\frac{1}{32\sqrt{15}}}}
\newcommand{\sqen}{{\frac{2}{3\sqrt{15}}}}
\newcommand{\sqeo}{{\frac{1}{12\sqrt{15}}}}
\newcommand{\sqep}{{\frac{\sqrt{15}}{360}}}

\newcommand{\sqet}{{\frac{221}{800}\sqrt{\frac{3}{5}}}}

\newcommand{\sqew}{{\frac{1}{16}}}

\newcommand{\sqfa}{{\frac{2}{10}}}
\newcommand{\sqfb}{{\frac{5}{8}}}

\newcommand{\sqfd}{{\frac{15}{16}}}

\newcommand{\sqfj}{{\frac{3}{4}\sqrt{\frac{3}{5}}}}

\newcommand{\sqfl}{{2\sqrt{\frac{3}{5}}}}

\newcommand{\sqfv}{{\frac{7}{16}\sqrt{\frac{3}{5}}}}

\newcommand{\sqfx}{{\frac{9}{5}}}

\newcommand{\sqgd}{{\frac{73}{64\sqrt{15}}}}
\newcommand{\sqge}{{\frac{9\sqrt{15}}{128}}}

\newcommand{\sqgg}{{\frac{73}{32\sqrt{15}}}}
\newcommand{\sqgh}{{\frac{71}{64\sqrt{15}}}}

 \newcommand{\sqgj}{{\frac{2}{5}                          }}
 \newcommand{\sqgk}{{\frac{25}{372}                       }}
 \newcommand{\sqgl}{{\frac{73}{372}                       }}
 \newcommand{\sqgm}{{\frac{175}{372}                      }}
 \newcommand{\sqgn}{{\frac{85}{186}                       }}
 \newcommand{\sqgo}{{\frac{5}{186}                        }}
 \newcommand{\sqgp}{{\frac{5}{31}                         }}
 \newcommand{\sqgq}{{\frac{4}{93}                         }}
 \newcommand{\sqgr}{{\frac{5}{124}                        }}
 \newcommand{\sqgs}{{\frac{20}{93}                        }}
 \newcommand{\sqgt}{{\frac{265}{279}                      }}
 \newcommand{\sqgu}{{\frac{25}{93}                        }}
 \newcommand{\sqgv}{{\frac{211}{744}                      }}
 \newcommand{\sqgw}{{\frac{175}{744}                      }}
 \newcommand{\sqgx}{{\frac{15}{62}                        }}
 \newcommand{\sqgy}{{\frac{85}{372}                       }}
 \newcommand{\sqgz}{{\frac{5}{372}                        }}
 \newcommand{\sqha}{{\frac{5}{62}                         }}
 \newcommand{\sqhb}{{\frac{2}{93}                         }}
 \newcommand{\sqhc}{{\frac{5}{248}                        }}
 \newcommand{\sqhd}{{\frac{10}{93}                        }}
 \newcommand{\sqhe}{{\frac{100}{279}                      }}
 \newcommand{\sqhf}{{\frac{25}{186}                       }}
 \newcommand{\sqhg}{{\frac{25}{124}                       }}
 \newcommand{\sqhh}{{\frac{115}{248}                      }}
 \newcommand{\sqhi}{{\frac{735}{248}                      }}
 \newcommand{\sqhj}{{\frac{189}{62}                       }}
 \newcommand{\sqhk}{{\frac{66}{31}                        }}
 \newcommand{\sqhl}{{\frac{21}{124}                       }}
 \newcommand{\sqhm}{{\frac{16}{31}                        }}
 \newcommand{\sqhn}{{\frac{42}{155}                       }}
 \newcommand{\sqho}{{\frac{63}{620}                       }}
 \newcommand{\sqhp}{{\frac{63}{248}                       }}
 \newcommand{\sqhq}{{\frac{177}{62}                       }}
 \newcommand{\sqhr}{{\frac{6}{31}                         }}
 \newcommand{\sqhs}{{\frac{5}{12}                         }}
 \newcommand{\sqht}{{\frac{8}{3}                          }}
 \newcommand{\sqhu}{{\frac{76}{155}                       }}
 \newcommand{\sqhv}{{\frac{57}{310}                       }}
 \newcommand{\sqhw}{{\frac{200}{279}                      }}
 \newcommand{\sqhx}{{\frac{1}{31}                         }}
 \newcommand{\sqhy}{{\frac{1}{558}                        }}
 \newcommand{\sqhz}{{\frac{1}{930}                        }}
 \newcommand{\sqia}{{\frac{1}{372}                        }}
 \newcommand{\sqib}{{\frac{101}{558}                      }}
 \newcommand{\sqic}{{\frac{146}{837}                      }}
 \newcommand{\sqid}{{\frac{15}{124}                       }}
 \newcommand{\sqie}{{\frac{69}{248}                       }}
 \newcommand{\sqif}{{\frac{441}{248}                      }}
 \newcommand{\sqig}{{\frac{567}{310}                      }}
 \newcommand{\sqih}{{\frac{21}{31}                        }}
 \newcommand{\sqii}{{\frac{9}{31}                         }}
 \newcommand{\sqij}{{\frac{396}{775}                      }}
 \newcommand{\sqik}{{\frac{69}{62}                        }}
 \newcommand{\sqil}{{\frac{49}{62}                        }}
 \newcommand{\sqim}{{\frac{15}{31}                        }}
 \newcommand{\sqin}{{\frac{81}{620}                       }}
 \newcommand{\sqio}{{\frac{11}{744}                       }}
 \newcommand{\sqip}{{\frac{148}{837}                      }}
 \newcommand{\sqiq}{{\frac{43}{992}                       }}
 \newcommand{\sqir}{{\frac{25}{496}                       }}
 \newcommand{\sqis}{{\frac{41}{620}                       }}
 \newcommand{\sqit}{{\frac{11}{496}                       }}
 \newcommand{\sqiu}{{\frac{29}{310}                       }}
 \newcommand{\sqiv}{{\frac{9}{248}                        }}
 \newcommand{\sqiw}{{\frac{116}{775}                      }}
 \newcommand{\sqix}{{\frac{43}{248}                       }}
 \newcommand{\sqiy}{{\frac{16}{93}                        }}
 \newcommand{\sqiz}{{\frac{2}{31}                         }}
 \newcommand{\sqja}{{\frac{25}{279}                       }}
 \newcommand{\sqjb}{{\frac{5}{24}                         }}
 \newcommand{\sqjc}{{\frac{42}{31}                        }}
 \newcommand{\sqjd}{{\frac{125}{558}                      }}
 \newcommand{\sqje}{{\frac{176}{465}                      }}
 \newcommand{\sqjf}{{\frac{239}{744}                      }}
 \newcommand{\sqjg}{{\frac{305}{558}                      }}
 \newcommand{\sqjh}{{\frac{10}{279}                       }}
 \newcommand{\sqji}{{\frac{84}{155}                       }}
 \newcommand{\sqjj}{{\frac{8}{93}                         }}
 \newcommand{\sqjk}{{\frac{61}{279}                       }}
 \newcommand{\sqjl}{{\frac{320}{837}                      }}
 \newcommand{\sqjm}{{\frac{40}{279}                       }}
 \newcommand{\sqjn}{{\frac{189}{775}                      }}
 \newcommand{\sqjo}{{\frac{706}{837}                      }}
 \newcommand{\sqjp}{{\frac{425}{837}                      }}
 \newcommand{\sqjq}{{\frac{425}{8928}                     }}
 \newcommand{\sqjr}{{\frac{101}{2232}                     }}
 \newcommand{\sqjs}{{\frac{35}{1116}                      }}
 \newcommand{\sqjt}{{\frac{59}{1116}                      }}
 \newcommand{\sqju}{{\frac{4}{1395}                       }}
 \newcommand{\sqjv}{{\frac{425}{2232}                     }}
 \newcommand{\sqjw}{{\frac{189}{1240}                     }}
 \newcommand{\sqjx}{{\frac{421}{8928}                     }}
 \newcommand{\sqjy}{{\frac{505}{4464}                     }}
 \newcommand{\sqjz}{{\frac{373}{4464}                     }}
 \newcommand{\sqka}{{\frac{251}{2790}                     }}
 \newcommand{\sqkb}{{\frac{251}{4650}                     }}
 \newcommand{\sqkc}{{\frac{391}{7440}                     }}
 \newcommand{\sqkd}{{\frac{421}{2232}                     }}
 \newcommand{\sqke}{{\frac{353}{1116}                     }}
 \newcommand{\sqkf}{{\frac{87}{1550}                      }}
 \newcommand{\sqkg}{{\frac{117}{2480}                     }}
 \newcommand{\sqkh}{{\frac{305}{2232}                     }}
 \newcommand{\sqki}{{\frac{925}{2232}                     }}
 \newcommand{\sqkj}{{\frac{691}{1116}                     }}
 \newcommand{\sqkk}{{\frac{491}{2790}                     }}
 \newcommand{\sqkl}{{\frac{205}{1674}                     }}
 \newcommand{\sqkm}{{\frac{61}{1116}                      }}
 \newcommand{\sqkn}{{\frac{743}{1116}                     }}
 \newcommand{\sqko}{{\frac{352}{1395}                     }}
 \newcommand{\sqkp}{{\frac{352}{2325}                     }}
 \newcommand{\sqkq}{{\frac{239}{1860}                     }}
 \newcommand{\sqkr}{{\frac{233}{1550}                     }}
 \newcommand{\sqks}{{\frac{572}{2325}                     }}
 \newcommand{\sqkt}{{\frac{473}{2790}                     }}
 \newcommand{\sqku}{{\frac{5\sqrt{\frac{5}{3}}}{186}      }}
 \newcommand{\sqkv}{{\frac{4\sqrt{\frac{3}{5}}}{31}       }}
 \newcommand{\sqkw}{{\frac{2\sqrt{\frac{5}{3}}}{31}       }}
 \newcommand{\sqkx}{{\frac{12\sqrt{\frac{3}{5}}}{155}     }}
 \newcommand{\sqky}{{\frac{10\sqrt{\frac{5}{3}}}{93}      }}
 \newcommand{\sqkz}{{\frac{8\sqrt{\frac{5}{3}}}{93}       }}
 \newcommand{\sqla}{{\frac{7\sqrt{\frac{5}{3}}}{372}      }}
 \newcommand{\sqlb}{{\frac{16}{93\sqrt{15}}               }}
 \newcommand{\sqlc}{{\frac{16}{155\sqrt{15}}              }}
 \newcommand{\sqld}{{\frac{8}{31\sqrt{15}}                }}
 \newcommand{\sqle}{{\frac{7\sqrt{\frac{5}{3}}}{93}       }}
 \newcommand{\sqlf}{{\frac{\sqrt{\frac{5}{3}}}{4}         }}
 \newcommand{\sqlg}{{\frac{19}{31\sqrt{15}}               }}
 \newcommand{\sqlh}{{\frac{4}{\sqrt{15}}                  }}
 \newcommand{\sqli}{{\frac{5\sqrt{15}}{248}               }}
 \newcommand{\sqlj}{{\frac{11\sqrt{\frac{5}{3}}}{93}      }}
 \newcommand{\sqlk}{{\frac{5\sqrt{\frac{5}{3}}}{31}       }}
 \newcommand{\sqll}{{\frac{3\sqrt{15}}{62}                }}
 \newcommand{\sqlm}{{\frac{2\sqrt{\frac{3}{5}}}{31}       }}
 \newcommand{\sqln}{{\frac{11\sqrt{\frac{5}{3}}}{31}      }}
 \newcommand{\sqlo}{{\frac{5\sqrt{15}}{62}                }}
 \newcommand{\sqlp}{{\frac{29\sqrt{\frac{5}{3}}}{2976}    }}
 \newcommand{\sqlq}{{\frac{\sqrt{\frac{5}{3}}}{8}         }}
 \newcommand{\sqlr}{{\frac{29}{248\sqrt{15}}              }}
 \newcommand{\sqls}{{\frac{7}{31\sqrt{15}}                }}
 \newcommand{\sqlt}{{\frac{77\sqrt{\frac{5}{3}}}{744}     }}
 \newcommand{\sqlu}{{\frac{211\sqrt{\frac{5}{3}}}{372}    }}
 \newcommand{\sqlv}{{\frac{32\sqrt{\frac{3}{5}}}{155}     }}
 \newcommand{\sqlw}{{\frac{83}{124\sqrt{15}}              }}
 \newcommand{\sqlx}{{\frac{77\sqrt{\frac{5}{3}}}{186}     }}
 \newcommand{\sqly}{{\frac{80\sqrt{\frac{5}{3}}}{279}     }}
 \newcommand{\sqlz}{{\frac{56\sqrt{\frac{5}{3}}}{93}      }}
 \newcommand{\sqma}{{\frac{32\sqrt{\frac{3}{5}}}{31}      }}
 \newcommand{\sqmb}{{\frac{637}{186\sqrt{15}}             }}
 \newcommand{\sqmc}{{\frac{32}{31\sqrt{15}}               }}
 \newcommand{\sqmd}{{\frac{128}{465\sqrt{15}}             }}
 \newcommand{\sqme}{{\frac{32\sqrt{\frac{5}{3}}}{93}      }}
 \newcommand{\sqmf}{{\frac{113\sqrt{\frac{5}{3}}}{744}    }}
 \newcommand{\sqmg}{{\frac{73\sqrt{\frac{3}{5}}}{62}      }}
 \newcommand{\sqmh}{{\frac{529}{744\sqrt{15}}             }}
 \newcommand{\sqmi}{{\frac{133}{124\sqrt{15}}             }}
 \newcommand{\sqmj}{{\frac{174\sqrt{\frac{3}{5}}}{155}    }}
 \newcommand{\sqmk}{{\frac{73}{248\sqrt{15}}              }}
 \newcommand{\sqml}{{\frac{863}{372\sqrt{15}}             }}
 \newcommand{\sqmm}{{\frac{113\sqrt{\frac{5}{3}}}{186}    }}
 \newcommand{\sqmn}{{\frac{123\sqrt{\frac{3}{5}}}{155}    }}
 \newcommand{\sqmo}{{\frac{143\sqrt{\frac{3}{5}}}{155}    }}
 \newcommand{\sqmp}{{\frac{19\sqrt{\frac{5}{3}}}{248}     }}
 \newcommand{\sqmq}{{\frac{107}{372\sqrt{15}}             }}
 \newcommand{\sqmr}{{\frac{214}{465\sqrt{15}}             }}
 \newcommand{\sqms}{{\frac{107}{620\sqrt{15}}             }}
 \newcommand{\sqmt}{{\frac{19\sqrt{\frac{5}{3}}}{62}      }}
 \newcommand{\sqmu}{{\frac{25\sqrt{15}}{124}              }}
 \newcommand{\sqmv}{{\frac{242}{93\sqrt{15}}              }}
 \newcommand{\sqmw}{{\frac{107}{62\sqrt{15}}              }}
 \newcommand{\sqmx}{{\frac{25\sqrt{\frac{5}{3}}}{496}     }}
 \newcommand{\sqmy}{{\frac{1279}{744\sqrt{15}}            }}
 \newcommand{\sqmz}{{\frac{29\sqrt{\frac{5}{3}}}{279}     }}
 \newcommand{\sqna}{{\frac{29\sqrt{\frac{5}{3}}}{744}     }}
 \newcommand{\sqnb}{{\frac{823}{620\sqrt{15}}             }}
 \newcommand{\sqnc}{{\frac{20\sqrt{\frac{5}{3}}}{31}      }}
 \newcommand{\sqnd}{{\frac{152}{155\sqrt{15}}             }}
 \newcommand{\sqne}{{\frac{19\sqrt{\frac{3}{5}}}{155}     }}
 \newcommand{\sqnf}{{\frac{56}{93\sqrt{15}}               }}
 \newcommand{\sqng}{{\frac{716}{465\sqrt{15}}             }}
 \newcommand{\sqnh}{{\frac{179}{310\sqrt{15}}             }}
 \newcommand{\sqni}{{\frac{91\sqrt{\frac{3}{5}}}{310}     }}
 \newcommand{\sqnj}{{\frac{131\sqrt{\frac{3}{5}}}{310}    }}
 \newcommand{\sqnk}{{\frac{863}{1488\sqrt{15}}            }}
 \newcommand{\sqnl}{{\frac{1279}{2976\sqrt{15}}           }}
 \newcommand{\sqnm}{{\frac{2297}{1488\sqrt{15}}           }}
 \newcommand{\sqnn}{{\frac{134}{279\sqrt{15}}             }}
 \newcommand{\sqno}{{\frac{1337}{1550}                    }}
 \newcommand{\sqnp}{{\frac{1004}{6975}                    }}
 \newcommand{\sqnq}{{\frac{2816}{6975}                    }}
 \newcommand{\sqnr}{{\frac{6827}{13950}                   }}
 \newcommand{\sqns}{{\frac{37\sqrt{15}}{124}              }}
 \newcommand{\sqnt}{{\frac{343}{62\sqrt{15}}              }}
 \newcommand{\sqnu}{{\frac{253\sqrt{\frac{5}{3}}}{248}    }}
 \newcommand{\sqnv}{{\frac{309\sqrt{\frac{3}{5}}}{155}    }}
 \newcommand{\sqnw}{{\frac{452\sqrt{\frac{5}{3}}}{279}    }}
 \newcommand{\sqnx}{{\frac{256\sqrt{\frac{5}{3}}}{279}    }}
 \newcommand{\sqny}{{\frac{1123\sqrt{\frac{5}{3}}}{1488}  }}
 \newcommand{\sqnz}{{\frac{1646}{465\sqrt{15}}            }}
 \newcommand{\sqoa}{{\frac{1144}{279\sqrt{15}}            }}
 \newcommand{\sqob}{{\frac{73}{150}                       }}
 \newcommand{\sqoc}{{\frac{7}{6}                          }}
 \newcommand{\sqod}{{\frac{16}{25}                        }}
 \newcommand{\sqoe}{{\frac{11}{30}                        }}
 \newcommand{\sqof}{{\frac{106}{45}                       }}
 \newcommand{\sqog}{{\frac{62}{25}                        }}
 \newcommand{\sqoh}{{\frac{171}{100}                      }}
 \newcommand{\sqoi}{{\frac{7}{2}                          }}
 \newcommand{\sqoj}{{\frac{48}{25}                        }}
 \newcommand{\sqok}{{\frac{12}{5}                         }}
 \newcommand{\sqol}{{\frac{6}{5}                          }}
 \newcommand{\sqom}{{\frac{106}{15}                       }}
 \newcommand{\sqon}{{\frac{186}{25}                       }}
 \newcommand{\sqoo}{{\frac{17}{10}                        }}
 \newcommand{\sqop}{{\frac{63}{20}                        }}
 \newcommand{\sqoq}{{\frac{13}{2}                         }}
 \newcommand{\sqor}{{\frac{63}{10}                        }}
 \newcommand{\sqos}{{\frac{31}{24}                        }}
 \newcommand{\sqot}{{\frac{1169}{300}                     }}
 \newcommand{\sqou}{{\frac{19}{30}                        }}
 \newcommand{\sqov}{{\frac{14}{3}                         }}
 \newcommand{\sqow}{{\frac{571}{100}                      }}
 \newcommand{\sqox}{{\frac{227}{60}                       }}
 \newcommand{\sqoy}{{\frac{21}{10}                        }}
 \newcommand{\sqoz}{{\frac{1433}{90}                      }}
 \newcommand{\sqpa}{{\frac{811}{50}                       }}
 \newcommand{\sqpb}{{\frac{25}{6}                         }}
 \newcommand{\sqpc}{{\frac{13}{96}                        }}
 \newcommand{\sqpd}{{\frac{181}{600}                      }}
 \newcommand{\sqpe}{{\frac{7}{12}                         }}
 \newcommand{\sqpf}{{\frac{53}{150}                       }}
 \newcommand{\sqpg}{{\frac{16}{15}                        }}
 \newcommand{\sqph}{{\frac{98}{75}                        }}
 \newcommand{\sqpi}{{\frac{13}{24}                        }}
 \newcommand{\sqpj}{{\frac{143}{2400}                     }}
 \newcommand{\sqpk}{{\frac{7}{40}                         }}
 \newcommand{\sqpl}{{\frac{199}{600}                      }}
 \newcommand{\sqpm}{{\frac{11}{80}                        }}
 \newcommand{\sqpn}{{\frac{11}{120}                       }}
 \newcommand{\sqpo}{{\frac{1}{40}                         }}
 \newcommand{\sqpp}{{\frac{7}{60}                         }}
 \newcommand{\sqpq}{{\frac{19}{300}                       }}
 \newcommand{\sqpr}{{\frac{51}{50}                        }}
 \newcommand{\sqps}{{\frac{189}{100}                      }}
 \newcommand{\sqpt}{{\frac{21}{20}                        }}
 \newcommand{\sqpu}{{\frac{9}{10}                         }}
 \newcommand{\sqpv}{{\frac{33}{10}                        }}
 \newcommand{\sqpw}{{\frac{189}{50}                       }}
 \newcommand{\sqpx}{{\frac{5}{96}                         }}
 \newcommand{\sqpy}{{\frac{7}{120}                        }}
 \newcommand{\sqpz}{{\frac{727}{2400}                     }}
 \newcommand{\sqqa}{{\frac{91}{600}                       }}
 \newcommand{\sqqb}{{\frac{163}{240}                      }}
 \newcommand{\sqqc}{{\frac{9}{40}                         }}
 \newcommand{\sqqd}{{\frac{191}{180}                      }}
 \newcommand{\sqqe}{{\frac{391}{300}                      }}
 \newcommand{\sqqf}{{\frac{149}{800}                      }}
 \newcommand{\sqqg}{{\frac{17}{40}                        }}
 \newcommand{\sqqh}{{\frac{17}{200}                       }}
 \newcommand{\sqqi}{{\frac{41}{80}                        }}
 \newcommand{\sqqj}{{\frac{77}{60}                        }}
 \newcommand{\sqqk}{{\frac{117}{100}                      }}
 \newcommand{\sqql}{{\frac{87}{160}                       }}
 \newcommand{\sqqm}{{\frac{51}{20}                        }}
 \newcommand{\sqqn}{{\frac{29}{60}                        }}
 \newcommand{\sqqo}{{\frac{17}{15}                        }}
 \newcommand{\sqqp}{{\frac{43}{18}                        }}
 \newcommand{\sqqq}{{\frac{53}{30}                        }}
 \newcommand{\sqqr}{{\frac{29}{150}                       }}
 \newcommand{\sqqs}{{\frac{34}{75}                        }}
 \newcommand{\sqqt}{{\frac{13}{45}                        }}
 \newcommand{\sqqu}{{\frac{53}{75}                        }}
 \newcommand{\sqqv}{{\frac{151}{40\sqrt{15}}              }}
 \newcommand{\sqqw}{{\frac{29}{5\sqrt{15}}                }}
 \newcommand{\sqqx}{{\frac{19}{4\sqrt{15}}                }}
 \newcommand{\sqqy}{{\frac{43}{3\sqrt{15}}                }}
 \newcommand{\sqqz}{{\frac{83}{5\sqrt{15}}                }}
 \newcommand{\sqra}{{\frac{13\sqrt{\frac{5}{3}}}{12}      }}
 \newcommand{\sqrb}{{\frac{7}{2\sqrt{15}}                 }}
 \newcommand{\sqrc}{{\frac{9\sqrt{\frac{3}{5}}}{20}       }}
 \newcommand{\sqrd}{{\frac{16}{5\sqrt{15}}                }}
 \newcommand{\sqre}{{\frac{32}{3\sqrt{15}}                }}
 \newcommand{\sqrf}{{\frac{32}{5\sqrt{15}}                }}
 \newcommand{\sqrg}{{\frac{11}{100\sqrt{15}}              }}
 \newcommand{\sqrh}{{\frac{\sqrt{\frac{3}{5}}}{5}         }}
 \newcommand{\sqri}{{\frac{32}{25\sqrt{15}}               }}
 \newcommand{\sqrj}{{\frac{4\sqrt{\frac{3}{5}}}{5}        }}
 \newcommand{\sqrk}{{\frac{3\sqrt{\frac{3}{5}}}{5}        }}
 \newcommand{\sqrl}{{\frac{1}{5\sqrt{15}}                 }}
 \newcommand{\sqrm}{{\frac{2\sqrt{\frac{3}{5}}}{5}        }}
 \newcommand{\sqrn}{{\frac{\sqrt{\frac{3}{5}}}{10}        }}
 \newcommand{\sqro}{{\frac{6\sqrt{\frac{3}{5}}}{5}        }}
 \newcommand{\sqrp}{{\frac{26}{25\sqrt{15}}               }}
 \newcommand{\sqrq}{{\frac{161}{80\sqrt{15}}              }}
 \newcommand{\sqrr}{{\frac{73}{20\sqrt{15}}               }}
 \newcommand{\sqrs}{{\frac{7\sqrt{\frac{3}{5}}}{8}        }}
 \newcommand{\sqru}{{\frac{\sqrt{\frac{5}{3}}}{12}        }}
 \newcommand{\sqrv}{{\frac{17}{6\sqrt{15}}                }}
 \newcommand{\sqrw}{{\frac{73}{10\sqrt{15}}               }}
 \newcommand{\sqrx}{{\frac{53}{40\sqrt{15}}               }}
 \newcommand{\sqry}{{\frac{23}{20\sqrt{15}}               }}
 \newcommand{\sqrz}{{\frac{19}{3\sqrt{15}}                }}
 \newcommand{\sqsa}{{\frac{8\sqrt{\frac{3}{5}}}{5}        }}
 \newcommand{\sqsb}{{\frac{7}{4\sqrt{15}}                 }}
 \newcommand{\sqsc}{{\frac{53}{32\sqrt{15}}               }}
 \newcommand{\sqsd}{{\frac{5\sqrt{\frac{5}{3}}}{8}        }}
 \newcommand{\sqse}{{\frac{\sqrt{\frac{5}{3}}}{24}        }}
 \newcommand{\sqsf}{{\frac{77}{12\sqrt{15}}               }}

\begin{document}

\title{Coulomb matrix elements in multi-orbital Hubbard models}
\author{J\"org B\"unemann$^{1}$ and Florian Gebhard$^2$}

\address{$^1$Institut f\"ur Physik, BTU Cottbus-Senftenberg, P.O.\ Box 101344, 03013 Cottbus, 
Germany}
\address{$^2$Fachbereich Physik, Philipps Universit\"at Marburg,
D-35032 Marburg, Germany}

\begin{abstract}
Coulomb matrix elements are needed in all studies in solid-state theory that 
are based on Hubbard-type multi-orbital models.  
Due to symmetries, the matrix elements are not independent. 
We determine a set of independent  Coulomb parameters
for a $d$-shell and a $f$-shell and all point groups 
with up to $16$ elements ($O_h$, $O$, $T_d$,  $T_h$, $D_{6h}$, and  $D_{4h}$). 
Furthermore, we express all other matrix elements as a function
of the independent Coulomb parameters.
Apart from the solution of the general point-group problem 
we investigate in detail the spherical approximation and  
first-order corrections to the spherical approximation. 
\end{abstract}

\submitto{\JPCM}

\ioptwocol


\section{Introduction}
\label{sec:intro}

One of the main weaknesses of state-of-the-art band-structure methods 
is their frequent failure to describe the electronic properties of systems
with partially filled $d$-shells or $f$-shells. The orbitals of such shells 
are well localised which often leads to substantial correlation 
effects. These effects are not captured by effective single-particle approaches 
such as the common methods based on Density-Functional Theory (DFT). Over 
the past 15 years it has therefore been a general trend to 
combine ab-initio methods with many-particle approaches that permit a
more sophisticated treatment of local correlations.

To this end, one usually separates the two-particle interactions into non-local 
and local terms.  While the former are treated in the standard 
(e.g., DFT) way, the latter are studied by means 
of many-particle methods such as Dynamical Mean Field 
Theory~\cite{georges1996} or the 
variational  Gutzwiller approach~\cite{buenemann1998,buenemann2014a}.
In general, the local Coulomb interaction (`Hubbard interaction')
has the form
\begin{eqnarray}
\label{hm}
\hat{H}_{\rm C}=\sum_i\sum_{{\scriptstyle b_1,b_2}\atop {\scriptstyle b_3,b_4}}
U^i_{b_1,b_2,b_3,b_4}\sum_{\sigma,\sigma'}
\hcd_{i,b_1,\sigma}\hcd_{i,b_2,\sigma'}\hc_{i,b_3,\sigma'}\hc_{i,b_4,\sigma},\nonumber \\[-9pt]
\end{eqnarray}
where $i$, $b_l$, and $\sigma, \sigma'$ denote the lattice site, (correlated) 
orbital indices, and spin indices, respectively. Depending on the number $n_{\rm o}$ of 
correlated orbitals per site, 
there can be up to $n^4_{\rm o}$ non-zero Coulomb-interaction parameters
$U^i_{b_1,b_2,b_3,b_4}$. In most cases, however, this number 
is much smaller due to the point-group symmetry at the lattice site $i$. 
Moreover, 
the  non-zero parameters are not independent, but can be expressed by
a sub-set of independent parameters. 

It is the purpose of this work to analyse in detail 
which Coulomb matrix elements vanish due to symmetry
and which of them are independent
for all high-symmetry point groups in the most relevant cases 
of a $d$-shell or an $f$-shell. 
Note that the lattice-site index $i$ enters 
the problem
only through its   point-group symmetry and will therefore be dropped in our
following considerations.

The special case of a $d$-shell in a cubic environment was
analysed in the textbook by Sugano, Tanabe, and 
Kamimura~\cite{sugano1970}. 
In contrast to our approach, however,
their method cannot be readily applied to
other systems. To reduce the number of independent parameters, one often
employs the `spherical approximation', in which the matrix elements are 
calculated with atomic wave functions, i.e., without a crystal 
field~ \cite{slater,condon,racah}. We shall derive the spherical approximation in 
a different way by using the same method that we develop for the full point-group
problem. This enables us to formulate also a systematic first-order 
correction to the spherical approximation.

In solids we often face the situation that not all of the $d$-orbitals or the $f$-orbitals
are partially filled. In such cases, only the sub-sets of partially filled orbitals must 
be included in the Hubbard interaction~(\ref{hm}). Note that the general 
results which 
we present in this work can be readily applied in all these cases as well.
One just needs to drop all those 
matrix elements which contain orbitals that are not taken into 
account in~(\ref{hm}). 

Our work is organised as follows. In section~\ref{axcv} we develop the general 
approach for the analysis of Coulomb matrix elements  
which is used throughout this work. The appropriate orbital basis for 
a $d$-shell and an $f$-shell are introduced in section~\ref{cfs}. In the following 
sections~\ref{hhh}, \ref{spha1}, and~\ref{axcvdd}, we 
analyse the Coulomb matrix elements 
for, (i) the full point group, (ii), the spherical approximation, 
and, (iii), a first-order correction to the spherical approximation, respectively. 
A summary, in section~\ref{summary}, closes
our presentation. Most of the explicit results are deferred to four appendices.

\section{General formalism}
\label{axcv}

Depending on the point-group symmetry at the site of a correlated atom 
in a lattice,
the $n_{\rm o}=5$ $d$-orbitals or $n_{\rm o}=7$  $f$-orbitals split up into orbitals 
$\varphi_{b}(\vecr)$ ($b=1,\ldots,n_{\rm o}$)
with a maximally three-fold degeneracy. 
Each orbital belongs to an  irreducible representation $\Gamma^p$ 
of the point group $G^{\rm point}$  at the transition-metal site.
The occurring representations $\Gamma^p$ are obtained from a reduction of 
the $j=2$ and 
 $j=3$ representations $\Gamma^j$ of the full rotational group $O(3)$,
\begin{equation}\label{lll}
\Gamma^j=\sum_pn_p^j\Gamma^p \;.
\end{equation}
We will introduce the coefficients $n_p^j$ for all relevant point groups in 
section~\ref{cfs}.

Without spin-orbit coupling,
the orbitals $\varphi_{b}(\vecr)$ can be represented by real wave functions. 
The $g$  point group operations, described by three-dimensional 
orthogonal matrices $\tilde{D}_l$ ($l=1,\ldots,g$), 
have isomorph unitary operators $\hat{T}_l$ in the Hilbert space, defined by
 \begin{equation}
\hat{T}_l\Psi(\vecr)=\Psi(\tilde{D}_l\vecr)\;.
\end{equation}
Note that, when applied to many-particle wave functions, 
the operator transforms all spatial coordinates simultaneously,
 \begin{equation}
\label{qwe}
\hat{T}_l\Psi(\vecr_1,\ldots,\vecr_N)=\Psi(\tilde{D}_l\vecr_1,\ldots,\tilde{D}_l\vecr_N)\;.
\end{equation}
The behaviour of orbital wave functions under point-group transformations 
is well defined, 
\begin{equation}
\label{sda}
\hat{T}_l\varphi_{b}(\vecr)=\sum_{b'}\Gamma^{p}_{b',b}(l)\varphi_{b}(\vecr)\;,
\end{equation}
where $\Gamma^{p}_{b',b}(l)$ denote the elements of the 
irreducible representation  matrices $\tilde{\Gamma}^{p}(l)$.
These matrices are documented
for all $32$ crystallographic point groups in the literature, see, e.g., 
Ref.~\cite{koster1963}.  In the following we drop the label $p$ in 
$\Gamma^{p}_{b',b}(l)$ because its information is already included in
the orbital indices $b,b'$.
 
To set up the local Hamiltonian~(\ref{hm}), 
we need to determine the $D_d=5^4=625$ (or $D_f=7^4=2401$) 
Coulomb matrix elements
\begin{eqnarray}\label{sdf}
&&  U_{b_1,b_2,b_3,b_4}\\\nonumber
&&=\int{\rm d}^3r\int{\rm d}^3r'\varphi_{b_1}(\vecr)
  \varphi_{b_2}(\vecr')f(\vecr,\vecr')\varphi_{b_3}(\vecr')
  \varphi_{b_4}(\vecr)\;,
\end{eqnarray}
where $f(\vecr,\vecr')$
 is the screened Coulomb interaction. 
 The exact form of this interaction is usually not known, however, 
  our analysis only requires that it 
 is invariant under all transformations of the respective point group.
When we insert $\hat{1}=\hat{T}^{\dagger}_l\hat{T}_l$ 
on the left-hand-side and on the right-hand-side 
of $f(\vecr,\vecr)$ in~(\ref{sdf}) and use
\begin{equation}\label{asd}
[\hat{T}_l,f(\vecr,\vecr')]_-=0\;,
\end{equation}
we find
\begin{equation}\label{jk}
U_{b_1,\ldots,b_4}=\sum_{\bar{b}_1,\ldots,\bar{b}_4}\Gamma_{\bar{b}_1,b_1}(l)\dots
 \Gamma_{\bar{b}_4,b_4}(l)U_{\bar{b}_1,\ldots,\bar{b}_4}\;.
\end{equation}
With the multiple index 
\begin{equation}\label{bdef}
\vecB=(b_1,\ldots,b_4)
\end{equation}
and the vectors 
$\vecU$ with components $U_{\vecB}$, equations~(\ref{jk}) assume the compact form
 \begin{equation}\label{jk2}
\vecU=\tilde{\Gammam}(l)\cdot \vecU\;\;\;(l=1,\ldots,g)\;.
\end{equation}
Here we introduced the product matrices $\tilde{\Gammam}(l)$
with the elements
\begin{eqnarray}\label{jk2b}
\Gammam_{\vecB,\bar{\vecB}}(l)
&=&\Gammam_{(\bar{b}_1,\ldots,\bar{b}_4),(b_1,\ldots,b_4)}(l)\\
&\equiv&\Gamma_{\bar{b}_1,b_1}(l)\dots
 \Gamma_{\bar{b}_4,b_4}(l)\;.
\end{eqnarray}
Equation~(\ref{jk2}) shows that we need to calculate the space of 
joint eigenvectors of all $g$ matrices  $\tilde{\Gammam}(l)$ with 
the same eigenvalue $\lambda=1$. Suppose we have found a $d$-dimensional
(orthogonal and normalised) basis $\vecc^{(k)}$ ($k=1,\ldots,d$) of this space  
and a basis $\vecd^{(k')}$ ($k'=1,\ldots,D-d$) of its orthogonal complement.
Then, since all matrices $\tilde{\Gammam}(l)$ are regular,  equation~(\ref{jk2})
leads to
\begin{eqnarray}\label{sdy}
I^{\rm k}&=&\sum_{\vecB}c^{(k)}_{\vecB}U_{\vecB}\;,\\
0&=&\sum_{\vecB}d^{(k')}_{\vecB}U_{\vecB}\;,\label{sdy2}
\end{eqnarray}
where we introduced $d$ independent Coulomb parameters  $I^{\rm k}$.
A simple inversion of these equations gives us all   Coulomb matrix elements 
$U_{\vecB}$ as a function of the independent parameters  $I^{\rm k}$. Since
any rotation of the vectors $\vecc^{(k)}$ is permitted there is a freedom 
in the choice of the parameters  $I^{\rm k}$. We shall find it most 
convenient to chose them as a set of $d$ independent matrix elements 
$U^{({\rm i})}_{\vecB}$, see section~\ref{hhh}.  

The number $d$ of independent matrix elements can be determined by the following 
group-theoretical considerations without a complete solution of equations~(\ref{jk2}).
The matrices   $\tilde{\Gammam}(l)$ define the product representation of four
representations  $\Gamma^{j}$, 
\begin{equation}\label{hbsdd}
\Gammam=\Gamma^{j}\bigotimes\Gamma^{j}\bigotimes
\Gamma^{j}\bigotimes\Gamma^{j}\;.
\end{equation}
With~(\ref{lll}) this equation reads
\begin{eqnarray}\label{scf0}
\Gammam&=&\sum_{p_1,p_2,p_3,p_4}n^j_{p_1}n^j_{p_2}n^j_{p_3}n^j_{p_4}
\Gamma^{p_1,p_2,p_3,p_4}\;, \\\label{scf}
\Gamma^{p_1,p_2,p_3,p_4}&\equiv& 
\Gamma^{p_1}\bigotimes\Gamma^{p_2}\bigotimes\Gamma^{p_3}
\bigotimes\Gamma^{p_4}\:.
\end{eqnarray}
The reduction of~(\ref{scf}) into irreducible components can be derived from 
the multiplication tables for the
point groups~\cite{koster1963}. For example, in cubic symmetry, there is 
one contribution in~(\ref{scf0})  with all $p_i$ belonging to an $E_g$ representation. 
Its reduction is given  as
 \begin{equation}\label{hbs2}
\Gamma^{E_{\rm g},E_{\rm g},E_{\rm g},E_{\rm g}}=
3 \Gamma^{A_{1{\rm g}}}+3\Gamma^{A_{2{\rm g}}}+5\Gamma^{E_{\rm g}}\;.
\end{equation}
In this way we can determine all coefficients in the reduction of $\Gammam$,
\begin{equation}
\Gammam=\sum_p n(p) \Gamma^p \;. \label{780}
\end{equation}
To find the joint eigenvectors to the eigenvalue $\lambda=1$ 
in (\ref{jk2}) is mathematically equivalent to the determination 
of the space that belongs to the totally 
symmetric representation $A_{1{\rm g}}$ on the right hand side of (\ref{780}). 
Its dimension $n(A_{1{\rm g}})$ is just the number $d$ of independent 
Coulomb matrix elements  $I^{\rm k}$. For example, we can conclude 
from (\ref{hbs2}) that in a pure $E_{\rm g}$ shell 
there would be three such parameters. Alternatively, we can 
calculate  $n(A_{1{\rm g}})$ with the general formula \cite{streitwolf1971}
\begin{equation}\label{axc}
n(p)=\frac{1}{g}\sum_{l}(\chi(l))^*\chi^p(l)
\end{equation}
for the coefficients in~(\ref{780})
where $\chi(l))$, $\chi^p(l)$ 
are the characters of $\Gammam$,  $\Gamma^p$, respectively. 
For $p=A_{1{\rm g}}$ (i.e., $\chi^p(l)=1$) this equation reads
\begin{equation}\label{sfg}
d=n(A_{1{\rm g}})=\frac{1}{g}\sum_{l}\sum_{\vecB}\Gammam_{\vecB,\vecB}(l)\;.
\end{equation}
 
Thus far, we only made use of the commutator relation~(\ref{asd}) and of the 
transformation behaviour  (\ref{sda}) of the orbitals. Hence, our 
analysis applies to rather general  matrix 
elements such as
 \begin{eqnarray}\label{sdf2}
&&  U'_{b_1,b_2,b_3,b_4} =\int{\rm d}^3r_1\int{\rm d}^3r_2
\int{\rm d}^3r_3\int{\rm d}^3r_4
\\\nonumber
&&\varphi^{p_1}_{b_1}(\vecr_1)
  \varphi^{p_2}_{b_2}(\vecr_2)
f(\vecr_1,\vecr_2,\vecr_3,\vecr_4)\varphi^{p_3}_{b_3}(\vecr_3)
  \varphi^{p_4}_{b_4}(\vecr_4)\;,
\end{eqnarray}
as long as $f(\vecr_1,\vecr_2,\vecr_3,\vecr_4)$ commutes with all
  $\hat{T}_{l}$. Our physical Coulomb matrix elements~(\ref{sdf}), however, 
 have the additional permutation symmetries
\begin{eqnarray}\label{yxc2ee}
 U_{b_1,b_2,b_3,b_4} &=&U_{b_1,b_3,b_2,b_4}\;,\\
 U_{b_1,b_2,b_3,b_4} &=&U_{b_4,b_2,b_3,b_1}\;,\\\label{yxc2ee2}
 U_{b_1,b_2,b_3,b_4} &=&U_{b_2,b_1,b_4,b_3}\;.
 \end{eqnarray}
These permutations define a group $G^{\rm perm}$ with the eight elements
\begin{eqnarray}\label{perm}
(1,2,3,4)&\to&(1,2,3,4)\equiv \hat{{\cal{P}}}_1\;, \\\nonumber
         &\to&(1,3,2,4)\equiv \hat{{\cal{P}}}_2\;, \\\nonumber
         &\to&(4,2,3,1)\equiv \hat{{\cal{P}}}_3\;, \\\nonumber
         &\to&(4,3,2,1)\equiv \hat{{\cal{P}}}_4\;, \\\nonumber
         &\to&(2,1,4,3)\equiv \hat{{\cal{P}}}_5\;, \\\nonumber
         &\to&(2,4,1,3)\equiv \hat{{\cal{P}}}_6\;, \\\nonumber
         &\to&(3,1,4,2)\equiv \hat{{\cal{P}}}_7\;, \\\nonumber
         &\to&(3,4,1,2)\equiv \hat{{\cal{P}}}_8\;. \nonumber
\end{eqnarray}

The matrices $\tilde{P}(i)$ with the elements
\begin{equation}\label{ozt}
\tilde{P}_{\vecB,\vecB'}(i)\equiv  \langle \vecB | \hat{{\cal{P}}}_i  | \vecB' \rangle
\end{equation}
form a $D$-dimensional representation of $G^{\rm perm}$. The permutation symmetry
of our Coulomb matrix elements can then be cast into the same form as in~(\ref{jk2}),
 \begin{equation}\label{izt}
\vecU=\tilde{P}(i)\cdot \vecU \quad (i=1,\ldots,8)\;.
\end{equation}
Therefore, we need to find the space of joint eigen\-vectors $\vecc^{(k)}$  to the 
eigenvalue $\lambda=1$, not only of the matrices $\tilde{\Gammam}(l)$ 
but also of $\tilde{P}(i)$. The dimension  of this space is
smaller than that without the additional permutation symmetries. 
This reduces the number $d$ of independent Coulomb parameters $I^{\rm k}$. It can 
also be determined by group-theoretical arguments, i.e., 
without an explicit solution of equations~(\ref{jk2}), (\ref{izt}), 
as we explain in~\ref{gg1}.

\section{Crystal-field splitting}
\label{cfs}

In this work we study $d$-orbitals and $f$-orbitals in environments that are 
described by crystallographic point groups with up to $16$ elements. These 
groups are $O_h$, $O$, $T_d$,  $T_h$, $D_{6h}$, and  $D_{4h}$. It turns out
that the Coulomb integrals for the groups   $O_h$, $O$ and $T_d$ are the 
same. Hence, we only need to study four different cases.

As a starting 
point for our further considerations, we introduce the proper orbital basis
states in all point-group environments. The situation is simplest for a
$d$-shell since here we can set up irreducible spaces for all 
our point groups with the same basis,
\begin{eqnarray}\label{varphi}
\varphi_{v}(r,\varphi,\theta)&=&\frac{1}{\sqrt{2}}R_{v}(r)
\left[Y_{2,2}(\varphi,\theta)+ Y_{2,-2}(\varphi,\theta) \right]\nonumber \\
&&\sim(x^2-y^2)\;,\\
\varphi_{u}(r,\varphi,\theta)&=&R_{u}(r)Y_{2,0}(\varphi,\theta)\sim (3z^2-r^2)\;.
\\
\varphi_{\zeta}(r,\varphi,\theta)&=&-\frac{\rm i}{\sqrt{2}}R_{\zeta}(r)
\left[Y_{2,2}(\varphi,\theta)- Y_{2,-2}(\varphi,\theta) \right]\nonumber \\
&&\sim xy\;,\label{varzeta}\\
\varphi_{\eta}(r,\varphi,\theta)&=&-\frac{1}{\sqrt{2}}R_{\eta}(r)
\left[Y_{2,1}(\varphi,\theta)- Y_{2,-1}(\varphi,\theta) \right]\nonumber \\
&&\sim xz\;,\\
\varphi_{\xi}(r,\varphi,\theta)&=&\frac{\rm i}{\sqrt{2}}R_{\xi}(r)
\left[Y_{2,1}(\varphi,\theta)+ Y_{2,-1}(\varphi,\theta) \right]\nonumber \\
&&\sim yz\;.
\label{aa2}
\end{eqnarray}
Here, we introduced the `spherical harmonic' functions 
$Y_{l,m}(\varphi,\theta)$ ($m=-l,\ldots,l$)~\cite{messiah1963}, and the
unspecified radial wave functions $R_{b}(r)$.
 Note that, although the basis states are the same, the  irreducible spaces
 (i.e., also the orbital degeneracies) depend on the specific point group. They
 are shown in table \ref{tableone}.
\begin{table}[ht]
\centering
\begin{tabular}{l|c|c}
  $G^{\rm point}$ &  $ \Gamma^p$  &$\{ \varphi_b\}$  \\  \hline \hline
$[O_h, O, T_d,  T_h]$&$[E_{\rm g},E, E, E_{\rm g}]$   &$\varphi_{u}$, $\varphi_{v}$  \\
&$[T_{2{\rm g}},T_2 ,T_2,T_{\rm g}]$   &$\varphi_{\zeta}$, $\varphi_{\eta}$, $\varphi_{\xi}$  \\
  \hline  
$D_{6h}$&$A_{1{\rm g}}$&   $\varphi_{u}$\\  
&$E_{2{\rm g}}$&   $\varphi_{v}$, $\varphi_{\zeta}$\\  
&$E_{1{\rm g}}$& $\varphi_{\eta}$, $\varphi_{\xi}$\\
  \hline  
$D_{4h}$&$A_{1{\rm g}}$&   $\varphi_{u}$\\ 
&$B_{1{\rm g}}$&   $\varphi_{v}$\\ \  
&$B_{2{\rm g}}$&   $\varphi_{\zeta}$\\ 
&  $E_{{\rm g}}$& $\varphi_{\eta}$, $\varphi_{\xi}$
\end{tabular}
\caption{Irreducible representations $\Gamma^p$ and corresponding basis 
 states $ \varphi_b$ for $d$-orbitals in environments 
that belong to the crystallographic point groups 
$O_h, O, T_d,  T_h,D_{6h},D_{4h}$. 
\label{tableone}}
\end{table}

For the treatment of an $f$-shell we introduce 
two different sets of basis states, namely the `axial basis'
\begin{eqnarray}\label{bb1}
\varphi_{a}(r,\varphi,\theta))&=&\frac{\rm i}{\sqrt{2}}R_{a}(r)
\left[Y_{3,-2}(\varphi,\theta)- Y_{3,2}(\varphi,\theta) \right] \nonumber \\
&&\sim xyz\;,\\
\varphi_{x'}(r,\varphi,\theta))&=&\frac{1}{\sqrt{2}}R_{x'}
\left[Y_{3,-1}(\varphi,\theta)- Y_{3,1}(\varphi,\theta) \right]  \nonumber \\
&&\sim x(5z^2-r^2)\;,\\
\varphi_{y'}(r,\varphi,\theta)&=&\frac{\rm i}{\sqrt{2}}R_{y'}(r)
\left[Y_{3,-1}(\varphi,\theta)+ Y_{3,1}(\varphi,\theta) \right]  \nonumber \\
&&\sim y(5z^2-r^2)\;,\\
\varphi_{z}(r,\varphi,\theta)&=&R_{z}(r)Y_{3,0}(\varphi,\theta)
\sim z(5z^2-3r^2)\;,\\
\varphi_{\alpha'}(r,\varphi,\theta)&=&\frac{1}{\sqrt{2}}R_{\alpha'}(r)
\left[Y_{3,-3}(\varphi,\theta)- Y_{3,3}(\varphi,\theta) \right] \nonumber \\
&&\sim x(x^2-3y^2)\;,\\
\varphi_{\beta'}(r,\varphi,\theta)&=&\frac{\rm i}{\sqrt{2}}R_{\beta'}(r)
\left[Y_{3,-3}(\varphi,\theta)+ Y_{3,3}(\varphi,\theta) \right] \nonumber \\
&&\sim y(3x^2-y^2)\;,\\
\varphi_{\gamma}(r,\varphi,\theta)&=&\frac{1}{\sqrt{2}}R_{\gamma}(r)
\left[Y_{3,-2}(\varphi,\theta)+ Y_{3,2}(\varphi,\theta) \right] \nonumber \\
&&\sim z(x^2-y^2)\;,
\end{eqnarray}
and the `cubic basis'
\begin{eqnarray}
\varphi_{a}(r,\varphi,\theta))&\sim& xyz\;,
\\
\varphi_{x}(r,\varphi,\theta))&\sim&x(5x^2-r^2)\;,
\\
\varphi_{y}(r,\varphi,\theta)&\sim& y(5y^2-r^2)\;,
\\
\varphi_{z}(r,\varphi,\theta)&\sim& z(5z^2-r^2)\;,
\\
\varphi_{\alpha}(r,\varphi,\theta)&\sim& x(y^2-z^2)\;,
\\
\varphi_{\beta}(r,\varphi,\theta)&\sim&y(x^2-z^2)\;,
\\\label{bb2}
\varphi_{\gamma}(r,\varphi,\theta)&\sim&z(x^2-y^2)\;.
\end{eqnarray}
With these basis sets of states we can set up irreducible spaces for all our 
point groups. The results are summarised in table~\ref{tabletwo}.
\begin{table}[ht]
\centering
\begin{tabular}{l|c|c}
 $G^{\rm point}$  &  $ \Gamma^p$    &  $\{ \varphi_b\}$   \\  \hline\hline
$[O_h, O, T_d]$  &$[A_{2{\rm u}}, A_2,A_1]$& $\varphi_{a}$  \\
&$[T_{1{\rm u}}, T_{1}, T_2]$&$\varphi_{x}$ , $\varphi_{y}$, $\varphi_{z}$  \\
&$[T_{2{\rm u}}, T_{2},T_1]$& $\varphi_{\alpha}$, $\varphi_{\beta}$, $\varphi_{\gamma}$  \\
   \hline
$T_h$&$A_{{\rm u}}$& $\varphi_{a}$  \\
&$T_{{\rm u}}$&$\varphi_{x}$, $\varphi_{y}$, $\varphi_{z}$  \\
&$T_{{\rm u}}$& $\varphi_{\alpha}$, $\varphi_{\beta}$, $\varphi_{\gamma}$  \\
   \hline  
$D_{6h}$&$A_{2{\rm u}}$&   $\varphi_{z}$\\ 
&$B_{1{\rm u}}$&   $\varphi_{\alpha'}$\\  
&$B_{2{\rm u}}$& $\varphi_{\beta'}$ \\  
&$E_{1{\rm u}}$&   $\varphi_{x'}$, $\varphi_{y'}$\\   
&$E_{2{\rm u}}$& $\varphi_{a}$, $\varphi_{\gamma}$\\
 \hline  
$D_{4h}$&$A_{2{\rm u}}$&   $\varphi_{z}$\\ 
&$B_{1{\rm u}}$&   $\varphi_{a}$\\  
&$B_{2{\rm u}}$& $\varphi_{\gamma}$ \\  
&$E_{{\rm u}}$&   $\varphi_{x'}$, $\varphi_{y'}$\\   
&$E_{{\rm u}}$& $\varphi_{\alpha'}$, $\varphi_{\beta'}$\\
\end{tabular}
\caption{Irreducible representations $\Gamma^p$ and corresponding basis 
 states $ \varphi_b$ for $f$-orbitals in environments 
that belong to the crystallographic point groups 
$O_h, O, T_d,  T_h,D_{6h},D_{4h}$. 
\label{tabletwo}}
\end{table}

\section{Coulomb matrix elements: full point group environment}\label{hhh}
First, we need to find the $d$-dimensional  basis of joint 
eigenvectors $\vecc^{(k)}$ of the matrices $\tilde{\Gammam}(l)$
and $\tilde{P}(i)$ and a basis $\vecd^{(k')}$ 
of their orthogonal complement. This linear algebra problem is  solved
by standard algorithms provided by {\sc Lapack}. 

Second, we have 
 to solve equations~(\ref{sdy}), (\ref{sdy2}). The vectors
   $\vecc^{(k)}$ that are provided by our numerical algorithm 
 are somewhat arbitrary because any rotation among these vectors 
 is permitted. 
 Hence, the independent parameters $I_k$ will usually be rather complicated
 linear combinations of Coulomb parameters $U_{\vecB}$. 
 We therefore prefer to look for a set of $d$ independent matrix elements
 $U^{({\rm i})}_{\vecB}$ (serving as parameters $I_k$) and $D-d$ dependent parameters 
 $U^{({\rm d})}_{\vecB}$. When we introduce the corresponding vectors 
 $\vecU^{({\rm i}),({\rm d})}$ (and $\vecI$ for $I_k$), 
we can write the inversion of  
equations~(\ref{sdy}), (\ref{sdy2}) as
\begin{equation}\label{ztr}
\left(
\begin{array}{c}
\vecU^{({\rm i})}\\
\vecU^{({\rm d})}
\end{array}
\right)
=
\left(
\begin{array}{cc}
\tilde{C}^{({\rm i,i})} & \tilde{C}^{({\rm i,d})}   \\
  \tilde{C}^{({\rm d,i})}   & \tilde{C}^{({\rm d,d})}   \\
\end{array}
\right)
\left(
\begin{array}{c}
\vecI\\
\veco
\end{array}
\right)\;,
\end{equation}
where the matrix in~(\ref{ztr}) has the form
\begin{eqnarray}\nonumber
\biggl(
\begin{array}{@{}cc@{}}
\tilde{C}^{({\rm i,i})} & \tilde{C}^{({\rm i,d})}   \\
  \tilde{C}^{({\rm d,i})}   & \tilde{C}^{({\rm d,d})}   \\
\end{array}
\biggr)=
\Bigl(
\vecc^{(1)},\ldots,\vecc^{(d)},\vecd^{(1)},\ldots,\vecd^{(D-d)}
\Bigr)\;.
\end{eqnarray}
Note that we can write equations~(\ref{sdy}), (\ref{sdy2}) in the 
form~(\ref{ztr}) because we are still free to chose the order of the indices in
$\vecB$. We now demand that equation~(\ref{ztr}) has a unique solution for 
  $\vecU^{({\rm d})}$ and $\vecI$ as a function of  $\vecU^{({\rm i})}$. 
 This is the case when the matrix $\tilde{C}^{({\rm i,i})}$ is regular, i.e., 
\begin{equation}
 |\tilde{C}^{({\rm i,i})}|\neq 0\;.
\end{equation}
With this condition, we can systematically set up our list of independent 
parameters $U^{({\rm i})}_{\vecB}$ and determine the dependent parameters through
\begin{equation}
\vecU^{({\rm d})}= \tilde{C}^{({\rm d,i})} \big( \tilde{C}^{({\rm i,i})}\big)^{-1}\
\vecU^{({\rm i})}\;.
\end{equation}

With our formalism, we calculated the independent
parameters $U^{({\rm i})}_{\vecB}$ and their relationship with 
all  dependent parameters $U^{({\rm d})}_{\vecB}$ for a $d$-shell and an $f$-shell
and for all the point groups introduced in section~\ref{cfs}. As an example,
we show the results for  $d$-orbitals in a cubic environment in 
this section. The corresponding results for all other groups and/or $f$-orbitals
are presented in~\ref{hhh2}.

For a more convenient reading, we introduce the following notations
for the Coulomb parameters
\begin{eqnarray}\label{cc1}
U(b)&\equiv& U_{b,b,b,b}\;,\\
U(b,b')&\equiv&U_{b,b',b',b}\;,\\
J(b,b')&\equiv&U_{b,b',b,b'}\;,\\
T(\bar{b};b,b')&\equiv&U_{b,\bar{b},\bar{b},b'}\;,\\
A(\bar{b};b,b')&\equiv&U_{b,\bar{b},b',\bar{b}}\;,\\\label{cc2}
S(b_1,b_2;b_3,b_4)&\equiv&U_{b_1,b_2;b_3,b_4}\;,
\end{eqnarray}
which we use throughout this work.
It is implicitly understood that all indices are mutually different in multi-orbital
Coulomb parameters as, e.g. in eq.~(\ref{cc2}).

Equation~(\ref{ztr}) still leaves a lot of freedom in our choice of 
the independent parameters. We prefer to have as many independent parameters
as possible that have an intuitive physical meaning. Hence we prioritise them
along the order set by equations~(\ref{cc1})-(\ref{cc2}). This means that we 
first try to maximise the number of independent parameters of the 
form $U(b)$, then of the form $U(b,b')$, and so forth. In the remaining 
ambiguity with respect to the orbitals we prioritise orbitals 
along the order in equations~(\ref{varphi})-(\ref{aa2}), (\ref{bb1})-(\ref{bb2}). 
For example, if we had to choose between
$U(v)$  and $U(u)$  we would work with  $U(v)$ as an independent parameter.

For the five $d$-orbitals in a cubic environment, i.e., for the point groups
$O_h$,  $O$, and $T_d$, we find $d=10$ independent 
parameters, in agreement with reference~\cite{sugano1970}. 
These can, for example, be chosen as 
\begin{eqnarray}
 &&
 U(\vv)             , 
 U(\ze)             , 
 U(\vv,\uu)         , 
 U(\vv,\ze)         , 
 U(\uu,\ze)         ,\nonumber 
 \\
 &&
 U(\ze,\et)         , 
 J(\vv,\ze)         , 
 J(\vv,\et)         , 
 J(\ze,\et)         , 
 S(\vv,\ze;\xi,\et) \; .
 \end{eqnarray}
The dependent parameters are 
\begin{eqnarray}\label{dd1}
 U(\uu)            &=&
 \sqai U(\vv)            
 \;,\\
 U(\et)            &=&
 \sqai U(\ze)            
 \;,\\
 U(\xi)            &=&
 \sqai U(\ze)            
 \;,\\
 U(\vv,\et)        &=&
 \sqaq U(\vv,\ze)        
 +\sqap U(\uu,\ze)        
 \;,\\
 U(\vv,\xi)        &=&
 \sqaq U(\vv,\ze)        
 +\sqap U(\uu,\ze)        
 \;,\\
 U(\uu,\et)        &=&
 \sqap U(\vv,\ze)        
 +\sqaq U(\uu,\ze)        
 \;,\\
 U(\uu,\xi)        &=&
 \sqap U(\vv,\ze)        
 +\sqaq U(\uu,\ze)        
 \;,\\
 U(\ze,\xi)        &=&
 \sqai U(\ze,\et)        
 \;,\\
 U(\et,\xi)        &=&
 \sqai U(\ze,\et)        
 \;,\\
 J(\vv,\uu)        &=&
 \sqaa U(\vv)            
 -\sqaa U(\vv,\uu)        
 \;,\\
 J(\uu,\ze)        &=&
 -\sqaf J(\vv,\ze)        
 +\sqaj J(\vv,\et)        
 \;,\\
 J(\vv,\xi)        &=&
 \sqai J(\vv,\et)        
 \;,\\
 J(\uu,\et)        &=&
 \sqag J(\vv,\ze)        
 +\sqaf J(\vv,\et)        
 \;,\\
 J(\uu,\xi)        &=&
 \sqag J(\vv,\ze)        
 +\sqaf J(\vv,\et)        
 \;,\\
 J(\ze,\xi)        &=&
 \sqai J(\ze,\et)        
 \;,\\
 J(\et,\xi)        &=&
 \sqai J(\ze,\et)        
 \;,\\
 T(\et;\vv,\uu)    &=&
 -\sqah U(\vv,\ze)        
 +\sqah U(\uu,\ze)        
 \;,\\
 T(\xi;\vv,\uu)    &=&
 \sqah U(\vv,\ze)        
 -\sqah U(\uu,\ze)        
 \;,\\
 A(\et;\vv,\uu)    &=&
 -\sqad J(\vv,\ze)        
 +\sqad J(\vv,\et)        
 \;,\\
 A(\xi;\vv,\uu)    &=&
 \sqad J(\vv,\ze)        
 -\sqad J(\vv,\et)        
 \;,\\
 S(\vv,\ze;\et,\xi)&=&
 -\sqai S(\vv,\ze;\xi,\et)
 \;,\\
 S(\uu,\ze;\et,\xi)&=&
 \sqad S(\vv,\ze;\xi,\et)
 \;,\\
 S(\uu,\ze;\xi,\et)&=&
 \sqad S(\vv,\ze;\xi,\et)
 \;,\\\label{dd2}
 S(\uu,\et;\xi,\ze)&=&
 -\sqao S(\vv,\ze;\xi,\et)
 \;.
 \end{eqnarray}
In this list, as in all corresponding lists in this work, we specify only
one of the (up to eight) Coulomb parameters that differ just by a 
permutation of the form~(\ref{perm}). Moreover, all parameters
that are not listed vanish due to symmetry.

Note that the coefficients in all lists of dependent parameters come out of our numerical 
algorithm in digital form. We wrote a separate code
that reliably identifies the analytical form of these digits, e.g.,
0.5773502691896258 is identified as $1/\sqrt{3}$.  
This program also generates the \LaTeX\ code of all formulae.
Therefore, we are confident that they are free of misprints.

\section{Spherical approximation}
\label{spha1}

The number of independent parameters determined in section~\ref{hhh} 
and~\ref{hhh2} varies between 
 $10$ ($d$-orbitals in a $O_h$ environment)   and $65$ ($f$-orbitals 
in a $D_{4h}$ environment). When a full $d$-shell or $f$-shell was considered
such numbers are too large if we aimed to determine them 
from meaningful fits to experiments. The localised nature of these orbitals, 
however, allows us to formulate sensible approximations that reduce the 
number of independent parameters significantly. The simplest one is the
`spherical approximation' which makes two assumptions.
\begin{itemize}
\item[(i)] The radial wave functions in equations~(\ref{varphi})-(\ref{aa2})
and in equations~(\ref{bb1})-(\ref{bb2}) 
are assumed to be orbital-independent. 
This means that the $n_o=5$ $d$-orbitals or $n_o=7$ $f$-orbitals form a representation 
space of the total angular momentum with $j=2$ and $j=3$, respectively. 
\item[ii)] The two-particle interaction 
in (\ref{sdf})  is assumed to be invariant under
all orthogonal transformations, i.e., 
\begin{equation}
 f(\vecr,\vecr')=f(|\vecr-\vecr'|).
\end{equation}
\end{itemize} 
As a generalisation of  
equations~(\ref{jk2}) and (\ref{jk2b}), these two assumptions lead to
 \begin{eqnarray}\label{jk2c}
\vecU&=&\tilde{\Gammam}(\tilde{D})\cdot \vecU\;,\\
\Gammam_{\vecB,\bar{\vecB}}(\tilde{D})
&=&\Gamma_{\bar{b}_1,b_1}(\tilde{D})\dots
 \Gamma_{\bar{b}_4,b_4}(\tilde{D})\;.
\end{eqnarray}
where $\tilde{D}$ can be any real, orthogonal matrix. When the matrices $\tilde{D}$
are chosen randomly, already {\sl two\/} of this infinite number of equations
contain all the information, and remain to be evaluated. 
Combined with the permutation 
equations~(\ref{izt}),  we can use the method of section~\ref{hhh} 
to determine a set of independent Coulomb integrals as well as 
their relationship with the dependent parameters. 
  
For five $d$-orbitals, we obtain three independent parameters, which we may chose as 
\begin{eqnarray}\label{345}
 &&
 U(\vv)             , 
 U(\vv,\uu)         , 
 U(\vv,\ze)         \; . 
 \end{eqnarray}
The dependent parameters are then given by
 \begin{eqnarray}\label{ss1}
 U(\uu)            &=&
 \sqai U(\vv)            
 \;, \\   
 U(\ze)            &=&
 \sqai U(\vv)            
 \;, \\   
 U(\et)            &=&
 \sqai U(\vv)            
 \;, \\   
 U(\uu,\ze)        &=&
 \sqai U(\vv,\uu)        
 \;, \\   
 U(\vv,\et)        &=&
 \sqap U(\vv,\uu)        
 +\sqaq U(\vv,\ze)        
 \;, \\   
 U(\vv,\xi)        &=&
 \sqap U(\vv,\uu)        
 +\sqaq U(\vv,\ze)        
 \;, \\   
 U(\uu,\et)        &=&
 \sqaq U(\vv,\uu)        
 +\sqap U(\vv,\ze)        
 \;, \\   
 U(\ze,\et)        &=&
 \sqap U(\vv,\uu)        
 +\sqaq U(\vv,\ze)        
 \;, \\   
 U(\et,\xi)        &=&
 \sqap U(\vv,\uu)        
 +\sqaq U(\vv,\ze)        
 \;, \\   
 J(\vv,\uu)        &=&
 \sqaa U(\vv)            
 -\sqaa U(\vv,\uu)        
 \;, \\   
 J(\vv,\ze)        &=&
 \sqaa U(\vv)            
 -\sqaa U(\vv,\ze)        
 \;, \\   
 J(\vv,\et)        &=&
 \sqaa U(\vv)            
 -\sqca U(\vv,\uu)        
 -\sqbz U(\vv,\ze)        
 \;, \\   
 J(\uu,\ze)        &=&
 \sqaa U(\vv)            
 -\sqaa U(\vv,\uu)        
 \;, \\   
 J(\vv,\xi)        &=&
 \sqaa U(\vv)            
 -\sqca U(\vv,\uu)        
 -\sqbz U(\vv,\ze)        
 \;, \\   
 J(\uu,\et)        &=&
 \sqaa U(\vv)            
 -\sqbz U(\vv,\uu)        
 -\sqca U(\vv,\ze)        
 \;, \\   
 J(\ze,\et)        &=&
 \sqaa U(\vv)            
 -\sqca U(\vv,\uu)        
 -\sqbz U(\vv,\ze)        
 \;, \\   
 J(\et,\xi)        &=&
 \sqaa U(\vv)            
 -\sqca U(\vv,\uu)        
 -\sqbz U(\vv,\ze)        
 \;, \\   
 T(\et;\vv,\uu)    &=&
 \sqah U(\vv,\uu)        
 -\sqah U(\vv,\ze)        
 \;, \\   
 A(\et;\vv,\uu)    &=&
 -\sqcb U(\vv,\uu)        
 +\sqcb U(\vv,\ze)        
 \;, \\   
 S(\vv,\ze;\xi,\et)&=&
 -\sqca U(\vv,\uu)        
 +\sqca U(\vv,\ze)        
 \;, \\   
 S(\vv,\ze;\et,\xi)&=&
 \sqca U(\vv,\uu)        
 -\sqca U(\vv,\ze)        
 \;, \\   
 S(\uu,\et;\xi,\ze)&=&
 \sqah U(\vv,\uu)        
 -\sqah U(\vv,\ze)        
 \;, \\   \label{ss2}
 S(\uu,\ze;\et,\xi)&=&
 -\sqcb U(\vv,\uu)        
 +\sqcb U(\vv,\ze)        
 \;.  
 \end{eqnarray}
This list only contains all {\sl finite\/} dependent parameters that 
we specified in  section~\ref{hhh} and~\ref{hhh2}. 
All other Coulomb parameters  can be calculated 
as a function of~(\ref{345}) using the results given in 
these two sections. For convenience, we provide a list 
for all (non-zero) Coulomb parameters in the 
 supplementary material.

The corresponding results for $f$-orbitals are given in~\ref{hhh3}.

\section{First order corrections to the spherical approximation}
\label{axcvdd}

In cases where the spherical approximation is not accurate enough, 
 as reported, e.g., in Refs.~\cite{PhysRevB.90.165105,PhysRevLett.116.106402}, it
is desirable to have a method that systemically  improves it. For its 
derivation, we assume 
that the radial wave functions in~(\ref{varphi})-(\ref{aa2})
and~(\ref{bb1})-(\ref{bb2}) 
differ only slight from each other, i.e.,
\begin{equation}
R_b(r)\approx R(r) +\delta R_b(r)\;.
\end{equation}
Then, we can linearise the Coulomb matrix elements with respect to 
the small perturbations $\delta R_b(r)$, 
\begin{equation}
U_{b_1,b_2,b_3,b_4}\approx U^{\rm SA}_{b_1,b_2,b_3,b_4}+\delta U_{b_1,b_2,b_3,b_4} \; ,
\end{equation}
where $U^{\rm SA}_{b_1,b_2,b_3,b_4}$ is the corresponding
result from the spherical approximation and 
\begin{eqnarray}\label{asdf}
\delta U_{b_1,b_2,b_3,b_4}&\equiv& \bar{U}_{b'_1,b_2,b_3,b_4}+
\bar{U}_{b1,b'_2,b_3,b_4}\nonumber\\
&&+\bar{U}_{b_1,b_2,b'_3,b_4}
+\bar{U}_{b_1,b_2,b_3,b'_4}\;.
\end{eqnarray}
The matrix elements $\bar{U}_{\cdots}$ are defined as 
in~(\ref{sdf}) with orbitals $\varphi_{b_i}$ and $\varphi_{b'_i}$ that 
have radial wave function $R(r)$ and $\delta R_{b_i}(r)$, respectively. 

Our first aim is now to identify independent and dependent 
parameters  $\bar{U}_{\cdots}$ and their relationships. To this end, we enlarge 
our orbital basis (from $n_o$ to $n'_o$)  by introducing some 
auxiliary wave functions. These form 
a $j=2$ or $j=3$ representation spaces for each set of states
$\varphi_{b'_i}$ that belong to the same representation of the point group. 
For example, in the case of $d$-orbitals in a cubic environment, we 
end up with $n'_o=3n_o=15$ orbital wave functions. These are 
\begin{itemize}
\item[(i)] Five $d$-orbitals with a radial wave function $R(r)$.
\item[(ii)] Five $d$-orbitals with the same radial wave function 
$\delta R_{u}(r)=\delta R_{v}(r)$,  including three auxiliary $t_{2{\rm g}}$ orbitals.
\item[(iii)] Five $d$-orbitals with the same radial wave function 
$\delta R_{\zeta}(r)=\delta R_{\eta}(r)=\delta R_{\xi}(r)$,  including two
aux\-ili\-ary $e_{{\rm g}}$ orbitals.
\end{itemize}
We introduce the familiar multiple indices 
$\bar{\vecB}$ as in eq.~(\ref{bdef}), 
however, with the indices $b_i$ ranging from unity to $n'_o$. Each of the 
$n'_o/n_o$ orbital subspaces transforms like a $j=2$ or $j=3$
representation. Hence, we obtain the equation~(\ref{jk2c})
as in the spherical approximation but now for the parameters 
$\bar{U}_{\bar{\vecB}}$. Together with the corresponding equation of the 
form~(\ref{izt}) we can determine independent and dependent parameters 
$\bar{U}^{(\rm i)}_{\bar{\vecB}}$  and $\bar{U}^{(\rm d)}_{\bar{\vecB}}$ 
with the same method as in  section~\ref{hhh}.

In principle, the problem is solved with this approach because, with the 
parameters $\bar{U}^{\rm (i,d)}_{\bar{\vecB}}$ at hand, we are able to calculate 
 the variations~(\ref{asdf}) as a function of 
 $\bar{U}^{\rm (i)}_{\bar{\vecB}}$, 
\begin{equation}\label{axd}
\delta U_{\vecB}=\delta U_{\vecB}(\{  \bar{U}^{\rm (i)}_{\bar{\vecB}} \}) \; .
\end{equation}
However, this formulation leaves room for improvement for two reasons.
\begin{itemize}
\item[(i)] The naive selection of some independent parameters 
$\bar{U}^{\rm (i)}_{\vecB}$ will also lead to variations of parameters 
$U_{\vecB}$ which have already been chosen as independent parameters 
in the spherical approximation, see equations~(\ref{345}) and~(\ref{345b}).
Therefore, the spherical approximation and its first order corrections 
would be mixed up and could not be easily distinguished.
\item[(ii)] The parameters  $\bar{U}^{\rm (i)}_{\bar{\vecB}}$ have no intuitive 
physical meaning because only in their sum~(\ref{asdf}) they define 
the variation of a Coulomb matrix element.
\end{itemize}
Both problems can be readily addressed when we use the linear 
equations~(\ref{axd})
to express the $d'$ independent parameters $\bar{U}^{\rm (i)}_{\bar{\vecB}}$
by some set of $d'$ independent variations  $\delta U^{\rm (i)}_{\vecB}$. 
The latter set should contain the independent parameters 
from the spherical approximation so that we overcome the problem~(i).

As an example, we consider the familiar case of $d$-orbitals in 
a cubic environment. Here, we obtain as independent variations 
$\delta U^{\rm (i)}_{\vecB}$,
\begin{eqnarray}
 &&
 \delta U(\vva)                , 
 \delta U(\vva,\uua)           , 
\delta U(\vva,\zea)           , 
\nonumber
\\\label{jks}
 &&
\delta U(\zea)                , 
\delta U(\uua,\zea)           , 
\delta U(\zea,\eta)           \; .
 \end{eqnarray}
The first line contains the independent parameters from the 
spherical approximation. Since their relation to all 
other parameters  is already fully covered in 
equations~(\ref{dd1})-(\ref{dd2}), we only need to document 
the first-order changes introduced by the parameters in~(\ref{jks}).
In cubic symmetry these are 
 \begin{eqnarray}
 \delta J(\vva,\zea)          &=&
 \sqaq \delta U(\zea)\;,               
 \\
 \delta J(\vva,\eta)          &=&
 \sqaq \delta U(\zea)               
 -\sqca \delta U(\uua,\zea) \;,         
 \\
 \delta J(\zea,\eta)          &=&
 \sqaa \delta U(\zea)               
 -\sqaa \delta U(\zea,\eta)\;,          
 \\
 \delta S(\vva,\zea;\xia,\eta)&=&
 -\sqcf \delta U(\uua,\zea)          
 +\sqap \delta U(\zea,\eta) \;.         
 \end{eqnarray}
Note that, unlike the spherical approximation, the first-order correction
depends on the point group that is considered. We present the results
for all other point groups and/or $f$-orbitals in \ref{hhh4}. There
we drop the variations of the parameters~(\ref{345}) and~(\ref{345b})
because they are  
already covered by the corresponding formulae in section~\ref{spha1} 
and \ref{hhh3}.
 
Again, we provide the results for the independent parameters only that 
we specified in  section~\ref{hhh} and~\ref{hhh2}. 
All other Coulomb parameter variations  can be  calculated 
using the results given in these two sections. 
For convenience we provide a full list
of all (non-zero) Coulomb parameter variations in the supplementary material. 

\section{Summary}
\label{summary}

In this work, we presented a comprehensive study of symmetries
among Coulomb matrix elements of $d$-orbitals and $f$-orbitals in 
crystallographic point group environments. Such matrix elements are 
needed in all theoretical investigations that
are based on Hubbard-type multi-orbital models. For all considered point groups 
($O_h$, $O$, $T_d$,  $T_h$, $D_{6h}$, and  $D_{4h}$) we determined an irreducible 
sub-set of independent Coulomb matrix elements and their relationship with all 
other matrix elements. Besides this evaluation of the full point-group problem, we also 
present results for the spherical approximation and a first-order correction to it. 

Although our results are rather general in their inclusion 
of all orbitals of the  $d$-shell and $f$-shell, 
they can be readily applied to situations where
only a sub-set of orbitals needs to be taken into account in a theoretical 
study on a specific material.      

\ack
J.B.\ thanks Andreas Krebs for helpful discussions on linear algebra. 
F.G.\ thanks L.\ Veis for his suggestions at early stages of this work.
\appendix

\section{Number of independent Coulomb interaction parameters}\label{gg1}
The number $d$ of independent Coulomb interaction parameters is one of 
 the results which one obtains from the explicit solution 
 of equations~(\ref{jk2}), (\ref{izt}). Here we explain how $d$ can be 
 determined without that solution.

In equation~(\ref{ozt}) we introduced 
the eight ($D$-dimensional) representation
matrices $\tilde{P}(i)$ of  $G^{\rm perm}$. Note that these matrices are real and
symmetric because we have $\hat{{\cal{P}}}_i^{-1}=\hat{{\cal{P}}}_i$ for 
all our permutations. 

The permutation group dissects the basis states $|\vecB\rangle$
into disjoint sets $S^{(z)}$ of states that are connected by at least
one permutation, i.e.,
\begin{equation}
|\vecB\rangle,|\vecB'\rangle \in S^{(z)} \;\; \Leftrightarrow \;\; 
 \exists \hat{{\cal{P}}}_i \;\;{\rm with}  
\;\;\langle \vecB| \hat{{\cal{P}}}_i |\vecB'\rangle\neq 0\;\;.
\end{equation}
The number of elements  in $S^{(z)}$ is denoted as  $N_z$ ($=1,\ldots,8$).
For example, 
\begin{eqnarray}
N_z=1:S^{(z)}&=&\{(b_1, b_1,b_1,b_1)\}\; ;\\
N_z=4:S^{(z)}&=&\{(b_1, b_2,b_2,b_2),(b_2, b_1,b_2,b_2),\\\nonumber
&&(b_2, b_2,b_1,b_2),(b_2, b_2,b_2,b_1)\}\; ; \\
N_z=4:S^{(z)}&=&\{(b_1, b_1,b_2,b_2),(b_1, b_2,b_1,b_2),\\\nonumber
&&(b_2, b_1,b_2,b_1),(b_2, b_2,b_1,b_1)\}\;.
\end{eqnarray}
With a representative $|\vecB^{(z)}\rangle$ of each set $S^{(z)}$, we define the 
  states
\begin{eqnarray}
|\bar{\vecB}^{(z)}\rangle &\equiv& \frac{\sqrt{N_z}}{8}\sum_i \hat{{\cal{P}}}_i
|\vecB^{(z)}\rangle\\
&=& \frac{\sqrt{N_z}}{8}\sum_i \sum_{\vecB}P_{\vecB,\vecB^{(z)}}(i)|\vecB\rangle
\end{eqnarray}
that form an orthogonal and normalised basis for the 
($d_{\rm p}$-dimensional) 
space of all states  that obey equations~(\ref{izt}). With this basis
we may define the following  $d_{\rm p}$-dimensional representation of $G^{\rm point}$,
\begin{equation}
\Gammam_{z,z'}(l)\equiv  \langle \bar{\vecB}^{(z)} |\hat{T}_l |\bar{\vecB}^{(z')}\rangle\;.
\end{equation}
Using equation~(\ref{axc}) we then  obtain
 \begin{equation}
d=\frac{1}{g}\sum_{l}\sum_{z}\Gammam_{z,z}(l)\;.
\end{equation}
This equation can be further evaluated,
\begin{eqnarray}
d&=&\frac{1}{g}\sum_{z,l}
\langle \bar{\vecB}^{(z)} |\hat{T}_l |\bar{\vecB}^{(z)}\rangle\\\nonumber
&=&\frac{1}{64g}\sum_{z,l} N_z\sum_{i,j}\sum_{\vecB,\vecB'}
P_{\vecB,\vecB^{(z)}}(i)P_{\vecB',\vecB^{(z)}}(j)\Gammam_{\vecB,\vecB'}(l)\;.
\end{eqnarray}
Now we use the fact that, by choosing a different  
representative $|\vecB^{(z)}\rangle$, the state $|\bar{\vecB}^{(z)}\rangle$ 
remains unchanged. Hence, we find that
\begin{eqnarray}
&&\sum_{z} N_z\sum_{i,j}\sum_{\vecB,\vecB'}
P_{\vecB,\vecB^{(z)}}(i)P_{\vecB',\vecB^{(z)}}(j)\\
&&=\sum_{z} \frac{N_z }{N_z} \sum_{\vecB''\in S^{(z)}}
\sum_{i,j}\sum_{\vecB,\vecB'}
P_{\vecB,\vecB''}(i)P_{\vecB',\vecB''}(j)\;.
\end{eqnarray}
Since
\begin{equation}
\sum_{z} \sum_{\vecB''\in S^{(z)}}= \sum_{\vecB''}
\end{equation}
we finally obtain
\begin{eqnarray}
d&=&\frac{1}{64g}\sum_{l}\sum_{i,j}\sum_{\vecB,\vecB'}
P_{\vecB,\vecB'}(i \circ j)\Gammam_{\vecB,\vecB'}(l)\\
&=& \frac{1}{8g}\sum_{l}\sum_{i}\sum_{\vecB,\vecB'}
P_{\vecB,\vecB'}(i)\Gammam_{\vecB,\vecB'}(l)\\
&=&\frac{1}{8g}\sum_{l,i}\sum_{\vecB}  [\tilde{P}(i)\cdot \tilde{\Gammam}(l)]_{\vecB,\vecB} \:.
\end{eqnarray}
This equation is obviously a generalisation of~(\ref{sfg}) in the presence 
 of the additional permutation symmetry. 


\section{Coulomb matrix elements: full point group environment}\label{hhh2}
\subsection{$d$-orbitals, group $T_h$}
Independent parameters:
\begin{eqnarray}
 &&
 U(\vv)             , 
 U(\ze)             , 
 U(\vv,\uu)         , 
 U(\vv,\ze)         , 
 U(\uu,\ze)         , 
 \\\nonumber
 &&
 U(\vv,\et)         , 
 U(\ze,\et)         , 
 J(\vv,\ze)         , 
 J(\vv,\et)         , 
 J(\uu,\ze)         , 
 \\\nonumber
 &&
 J(\ze,\et)         , 
 S(\vv,\ze;\xi,\et) , 
 S(\vv,\et;\xi,\ze) , 
 \end{eqnarray}
Dependent parameters:
 \begin{eqnarray}
 U(\uu)            =
 \sqai U(\vv)            
 \:,\\\nonumber
 U(\et)            =
 \sqai U(\ze)            
 \:,\\\nonumber
 U(\xi)            =
 \sqai U(\ze)            
 \:,\\\nonumber
 U(\vv,\xi)        =
 \sqaa U(\vv,\ze)        
 +\sqab U(\uu,\ze)        
 -\sqai U(\vv,\et)        
 \:,\\\nonumber
 U(\uu,\et)        =
 \sqai U(\vv,\ze)        
 +\sqai U(\uu,\ze)        
 -\sqai U(\vv,\et)        
 \:,\\\nonumber
 U(\uu,\xi)        =
 \sqaa U(\vv,\ze)        
 -\sqaa U(\uu,\ze)        
 +\sqai U(\vv,\et)        
 \:,\\\nonumber
 U(\ze,\xi)        =
 \sqai U(\ze,\et)        
 \:,\\\nonumber
 U(\et,\xi)        =
 \sqai U(\ze,\et)        
 \:,\\\nonumber
 J(\vv,\uu)        =
 \sqaa U(\vv)            
 -\sqaa U(\vv,\uu)        
 \:,\\\nonumber
 J(\vv,\xi)        =
 \sqaa J(\vv,\ze)        
 -\sqai J(\vv,\et)        
 +\sqab J(\uu,\ze)        
 \:,\\\nonumber
 J(\uu,\et)        =
 \sqai J(\vv,\ze)        
 -\sqai J(\vv,\et)        
 +\sqai J(\uu,\ze)        
 \:,\\\nonumber
 J(\uu,\xi)        =
 \sqaa J(\vv,\ze)        
 +\sqai J(\vv,\et)        
 -\sqaa J(\uu,\ze)        
 \:,\\\nonumber
 J(\ze,\xi)        =
 \sqai J(\ze,\et)        
 \:,\\\nonumber
 J(\et,\xi)        =
 \sqai J(\ze,\et)        
 \:,\\\nonumber
 T(\ze;\vv,\uu)    =
 -\sqan U(\vv,\ze)        
 -\sqae U(\uu,\ze)        
 +\sqao U(\vv,\et)        
 \:,\\\nonumber
 T(\et;\vv,\uu)    =
 -\sqan U(\vv,\ze)        
 +\sqae U(\uu,\ze)        
 -\sqad U(\vv,\et)        
 \:,\\\nonumber
 T(\xi;\vv,\uu)    =
 \sqad U(\vv,\ze)        
 -\sqad U(\vv,\et)        
 \:,\\\nonumber
 A(\ze;\vv,\uu)    =
 -\sqan J(\vv,\ze)        
 +\sqao J(\vv,\et)        
 -\sqae J(\uu,\ze)        
 \:,\\\nonumber
 A(\et;\vv,\uu)    =
 -\sqan J(\vv,\ze)        
 -\sqad J(\vv,\et)        
 +\sqae J(\uu,\ze)        
 \:,\\\nonumber
 A(\xi;\vv,\uu)    =
 \sqad J(\vv,\ze)        
 -\sqad J(\vv,\et)        
 \:,\\\nonumber
 S(\vv,\ze;\et,\xi)=
 -\sqai S(\vv,\ze;\xi,\et)
 -\sqai S(\vv,\et;\xi,\ze)
 \:,\\\nonumber
 S(\uu,\ze;\et,\xi)=
 \sqad S(\vv,\ze;\xi,\et)
 -\sqad S(\vv,\et;\xi,\ze)
 \:,\\\nonumber
 S(\uu,\ze;\xi,\et)=
 \sqad S(\vv,\ze;\xi,\et)
 +\sqao S(\vv,\et;\xi,\ze)
 \:,\\\nonumber
 S(\uu,\et;\xi,\ze)=
 -\sqao S(\vv,\ze;\xi,\et)
 -\sqad S(\vv,\et;\xi,\ze)
 \:.
 \end{eqnarray}


\subsection{$d$-orbitals, group $D_{6h}$}
Independent parameters:
 \begin{eqnarray}
 &&
 U(\vv)             , 
 U(\uu)             , 
 U(\et)             , 
 U(\vv,\uu)         , 
 U(\vv,\ze)         , 
 \\  \nonumber
 &&
 U(\vv,\et)         , 
 U(\vv,\xi)         , 
 U(\uu,\et)         , 
 U(\et,\xi)         , 
 J(\vv,\uu)         , 
 \\  \nonumber
 &&
 J(\vv,\et)         , 
 J(\vv,\xi)         , 
 J(\uu,\et)         , 
 T(\ze;\vv,\uu)     , 
 T(\et;\vv,\uu)     , 
 \\  \nonumber
 &&
 A(\et;\vv,\uu)     , 
 S(\vv,\ze;\et,\xi) \; . 
 \end{eqnarray}
Dependent parameters:
 \begin{eqnarray}
 U(\ze)            =
 \sqai U(\vv)            
 \;, \\ \nonumber
 U(\xi)            =
 \sqai U(\et)            
 \;, \\ \nonumber
 U(\uu,\ze)        =
 \sqai U(\vv,\uu)        
 \;, \\ \nonumber
 U(\uu,\xi)        =
 \sqai U(\uu,\et)        
 \;, \\ \nonumber
 U(\ze,\et)        =
 \sqai U(\vv,\xi)        
 \;, \\ \nonumber
 U(\ze,\xi)        =
 \sqai U(\vv,\et)        
 \;, \\ \nonumber
 J(\vv,\ze)        =
 \sqaa U(\vv)            
 -\sqaa U(\vv,\ze)        
 \;, \\ \nonumber
 J(\uu,\ze)        =
 \sqai J(\vv,\uu)        
 \;, \\ \nonumber
 J(\uu,\xi)        =
 \sqai J(\uu,\et)        
 \;, \\ \nonumber
 J(\ze,\et)        =
 \sqai J(\vv,\xi)        
 \;, \\ \nonumber
 J(\ze,\xi)        =
 \sqai J(\vv,\et)        
 \;, \\ \nonumber
 J(\et,\xi)        =
 \sqaa U(\et)            
 -\sqaa U(\et,\xi)        
 \;, \\ \nonumber
 T(\xi;\vv,\uu)    =
 -\sqai T(\et;\vv,\uu)    
 \;, \\ \nonumber
 A(\vv;\vv,\uu)    =
 -\sqai T(\ze;\vv,\uu)    
 \;, \\ \nonumber
 A(\ze;\vv,\uu)    =
 \sqai T(\ze;\vv,\uu)    
 \;, \\ \nonumber
 A(\xi;\vv,\uu)    =
 -\sqai A(\et;\vv,\uu)    
 \;, \\ \nonumber
 S(\vv,\ze;\xi,\et)=
 -\sqai J(\vv,\et)        
 +\sqai J(\vv,\xi)        
 -\sqai S(\vv,\ze;\et,\xi)
 \;, \\ \nonumber
 S(\vv,\et;\xi,\ze)=
 -\sqaa U(\vv,\et)        
 +\sqaa U(\vv,\xi)        
 \;, \\ \nonumber
 S(\uu,\ze;\et,\xi)=
 \sqai A(\et;\vv,\uu)    
 \;, \\ \nonumber
 S(\uu,\ze;\xi,\et)=
 \sqai A(\et;\vv,\uu)    
 \;, \\ \nonumber
 S(\uu,\et;\xi,\ze)=
 \sqai T(\et;\vv,\uu)    
 \;.
 \end{eqnarray}


\subsection{$d$-orbitals, group $D_{4h}$}
Independent parameters:
 \begin{eqnarray}
 &&
 U(\vv)             , 
 U(\uu)             , 
 U(\ze)             , 
 U(\et)             , 
 U(\vv,\uu)         , 
 \\ \nonumber
 &&
 U(\vv,\ze)         , 
 U(\uu,\ze)         , 
 U(\vv,\et)         , 
 U(\uu,\et)         , 
 U(\ze,\et)         , 
 \\ \nonumber
 &&
 U(\et,\xi)         , 
 J(\vv,\uu)         , 
 J(\vv,\ze)         , 
 J(\vv,\et)         , 
 J(\uu,\ze)         , 
 \\ \nonumber
 &&
 J(\uu,\et)         , 
 J(\ze,\et)         , 
 J(\et,\xi)         , 
 T(\et;\vv,\uu)     , 
 A(\et;\vv,\uu)     , 
 \\ \nonumber
 &&
 S(\vv,\ze;\xi,\et) , 
 S(\uu,\ze;\et,\xi) , 
 S(\uu,\et;\xi,\ze) \; . 
 \end{eqnarray}
Dependent parameters:
 \begin{eqnarray}
 U(\xi)            =
 \sqai U(\et)            
 \;,  \\  \nonumber
 U(\vv,\xi)        =
 \sqai U(\vv,\et)        
 \;,  \\  \nonumber
 U(\uu,\xi)        =
 \sqai U(\uu,\et)        
 \;,  \\  \nonumber
 U(\ze,\xi)        =
 \sqai U(\ze,\et)        
 \;,  \\  \nonumber
 J(\vv,\xi)        =
 \sqai J(\vv,\et)        
 \;,  \\  \nonumber
 J(\uu,\xi)        =
 \sqai J(\uu,\et)        
 \;,  \\  \nonumber
 J(\ze,\xi)        =
 \sqai J(\ze,\et)        
 \;,  \\  \nonumber
 T(\xi;\vv,\uu)    =
 -\sqai T(\et;\vv,\uu)    
 \;,  \\  \nonumber
 A(\xi;\vv,\uu)    =
 -\sqai A(\et;\vv,\uu)    
 \;,  \\  \nonumber
 S(\vv,\ze;\et,\xi)=
 -\sqai S(\vv,\ze;\xi,\et)
 \;,  \\  \nonumber
 S(\uu,\ze;\xi,\et)=
 \sqai S(\uu,\ze;\et,\xi)
 \;.
 \end{eqnarray}


\subsection{$f$-orbitals, groups $O_{h}$, $O$, $T_{d}$}
Independent parameters:
 \begin{eqnarray}
 &&
 U(\ag)             , 
 U(\xx)             , 
 U(\al)             , 
 U(\ag,\xx)         , 
 U(\xx,\yy)         , 
 \\ \nonumber
 &&
 U(\ag,\al)         , 
 U(\xx,\al)         , 
 U(\xx,\be)         , 
 U(\al,\be)         , 
 J(\ag,\xx)         , 
 \\ \nonumber
 &&
 J(\xx,\yy)         , 
 J(\ag,\al)         , 
 J(\xx,\al)         , 
 J(\yy,\al)         , 
 J(\al,\be)         , 
 \\ \nonumber
 &&
 T(\yy;\xx,\al)     , 
 T(\be;\xx,\al)     , 
 A(\xx;\xx,\be)     , 
 A(\be;\xx,\al)     , 
 \\ \nonumber
 &&
 S(\ag,\xx;\zz,\yy) , 
 S(\ag,\xx;\be,\zz) , 
 S(\ag,\xx;\ga,\be) , 
 S(\ag,\al;\be,\zz) , 
 \\ \nonumber
 &&
 S(\xx,\al;\be,\yy) , 
S(\xx,\yy;\al,\be) , 
 S(\xx,\yy;\be,\al) \; . 
 \end{eqnarray}
Dependent parameters:
 \begin{eqnarray}
 U(\yy)            =
 \sqai U(\xx)            
 \; ,  \\ \nonumber 
 U(\zz)            =
 \sqai U(\xx)            
 \; ,  \\ \nonumber 
 U(\be)            =
 \sqai U(\al)            
 \; ,  \\ \nonumber 
 U(\ga)            =
 \sqai U(\al)            
 \; ,  \\ \nonumber 
 U(\ag,\yy)        =
 \sqai U(\ag,\xx)        
 \; ,  \\ \nonumber 
 U(\ag,\zz)        =
 \sqai U(\ag,\xx)        
 \; ,  \\ \nonumber 
 U(\xx,\zz)        =
 \sqai U(\xx,\yy)        
 \; ,  \\ \nonumber 
 U(\ag,\be)        =
 \sqai U(\ag,\al)        
 \; ,  \\ \nonumber 
 U(\yy,\zz)        =
 \sqai U(\xx,\yy)        
 \; ,  \\ \nonumber 
 U(\ag,\ga)        =
 \sqai U(\ag,\al)        
 \; ,  \\ \nonumber 
 U(\yy,\al)        =
 \sqai U(\xx,\be)        
 \; ,  \\ \nonumber 
 U(\zz,\al)        =
 \sqai U(\xx,\be)        
 \; ,  \\ \nonumber 
 U(\xx,\ga)        =
 \sqai U(\xx,\be)        
 \; ,  \\ \nonumber 
 U(\yy,\be)        =
 \sqai U(\xx,\al)        
 \; ,  \\ \nonumber 
 U(\zz,\be)        =
 \sqai U(\xx,\be)        
 \; ,  \\ \nonumber 
 U(\yy,\ga)        =
 \sqai U(\xx,\be)        
 \; ,  \\ \nonumber 
 U(\zz,\ga)        =
 \sqai U(\xx,\al)        
 \; ,  \\ \nonumber 
 U(\al,\ga)        =
 \sqai U(\al,\be)        
 \; ,  \\ \nonumber 
 U(\be,\ga)        =
 \sqai U(\al,\be)        
 \; ,  \\ \nonumber 
 J(\ag,\yy)        =
 \sqai J(\ag,\xx)        
 \; ,  \\ \nonumber 
 J(\ag,\zz)        =
 \sqai J(\ag,\xx)        
 \; ,  \\ \nonumber 
 J(\xx,\zz)        =
 \sqai J(\xx,\yy)        
 \; ,  \\ \nonumber 
 J(\ag,\be)        =
 \sqai J(\ag,\al)        
 \; ,  \\ \nonumber 
 J(\yy,\zz)        =
 \sqai J(\xx,\yy)        
 \; ,  \\ \nonumber 
 J(\xx,\be)        =
 \sqai J(\yy,\al)        
 \; ,  \\ \nonumber 
 J(\ag,\ga)        =
 \sqai J(\ag,\al)        
 \; ,  \\ \nonumber 
 J(\yy,\be)        =
 \sqai J(\xx,\al)        
 \; ,  \\ \nonumber 
 J(\xx,\ga)        =
 \sqai J(\yy,\al)        
 \; ,  \\ \nonumber 
 J(\zz,\al)        =
 \sqai J(\yy,\al)        
 \; ,  \\ \nonumber 
 J(\zz,\be)        =
 \sqai J(\yy,\al)        
 \; ,  \\ \nonumber 
 J(\yy,\ga)        =
 \sqai J(\yy,\al)        
 \; ,  \\ \nonumber 
 J(\zz,\ga)        =
 \sqai J(\xx,\al)        
 \; ,  \\ \nonumber 
 J(\al,\ga)        =
 \sqai J(\al,\be)        
 \; ,  \\ \nonumber 
 J(\be,\ga)        =
 \sqai J(\al,\be)        
 \; ,  \\ \nonumber 
 T(\xx;\yy,\be)    =
 \sqai T(\yy;\xx,\al)    
 \; ,  \\ \nonumber 
 T(\xx;\zz,\ga)    =
 \sqai T(\yy;\xx,\al)    
 \; ,  \\ \nonumber 
 T(\zz;\xx,\al)    =
 -\sqai T(\yy;\xx,\al)    
 \; ,  \\ \nonumber 
 T(\zz;\yy,\be)    =
 -\sqai T(\yy;\xx,\al)    
 \; ,  \\ \nonumber 
 T(\yy;\zz,\ga)    =
 -\sqai T(\yy;\xx,\al)    
 \; ,  \\ \nonumber 
 T(\al;\yy,\be)    =
 \sqai T(\be;\xx,\al)    
 \; ,  \\ \nonumber 
 T(\ga;\xx,\al)    =
 -\sqai T(\be;\xx,\al)    
 \; ,  \\ \nonumber 
 T(\al;\zz,\ga)    =
 \sqai T(\be;\xx,\al)    
 \; ,  \\ \nonumber 
 T(\be;\zz,\ga)    =
 -\sqai T(\be;\xx,\al)    
 \; ,  \\ \nonumber 
 T(\ga;\yy,\be)    =
 -\sqai T(\be;\xx,\al)    
 \; ,  \\ \nonumber 
 A(\yy;\xx,\al)    =
 \sqai A(\xx;\xx,\be)    
 \; ,  \\ \nonumber 
 A(\xx;\xx,\ga)    =
 \sqai A(\xx;\xx,\be)    
 \; ,  \\ \nonumber 
 A(\zz;\xx,\al)    =
 -\sqai A(\xx;\xx,\be)    
 \; ,  \\ \nonumber 
 A(\yy;\yy,\ga)    =
 -\sqai A(\xx;\xx,\be)    
 \; ,  \\ \nonumber 
 A(\zz;\yy,\be)    =
 -\sqai A(\xx;\xx,\be)    
 \; ,  \\ \nonumber 
 A(\al;\yy,\be)    =
 \sqai A(\be;\xx,\al)    
 \; ,  \\ \nonumber 
 A(\al;\zz,\ga)    =
 \sqai A(\be;\xx,\al)    
 \; ,  \\ \nonumber 
 A(\ga;\xx,\al)    =
 -\sqai A(\be;\xx,\al)    
 \; ,  \\ \nonumber 
 A(\be;\zz,\ga)    =
 -\sqai A(\be;\xx,\al)    
 \; ,  \\ \nonumber 
 A(\ga;\yy,\be)    =
 -\sqai A(\be;\xx,\al)    
 \; ,  \\ \nonumber 
 S(\ag,\yy;\zz,\xx)=
 \sqai S(\ag,\xx;\zz,\yy)
 \; ,  \\ \nonumber 
 S(\ag,\xx;\yy,\zz)=
 \sqai S(\ag,\xx;\zz,\yy)
 \; ,  \\ \nonumber 
 S(\ag,\zz;\al,\yy)=
 -\sqai S(\ag,\xx;\be,\zz)
 \; ,  \\ \nonumber 
 S(\ag,\zz;\be,\xx)=
 -\sqai S(\ag,\xx;\be,\zz)
 \; ,  \\ \nonumber 
 S(\ag,\xx;\ga,\yy)=
 \sqai S(\ag,\xx;\be,\zz)
 \; ,  \\ \nonumber 
 S(\ag,\yy;\al,\zz)=
 \sqai S(\ag,\xx;\be,\zz)
 \; ,  \\ \nonumber 
 S(\ag,\yy;\ga,\xx)=
 -\sqai S(\ag,\xx;\be,\zz)
 \; ,  \\ \nonumber 
 S(\ag,\zz;\al,\be)=
 \sqai S(\ag,\xx;\ga,\be)
 \; ,  \\ \nonumber 
 S(\ag,\al;\ga,\yy)=
 -\sqai S(\ag,\al;\be,\zz)
 \; ,  \\ \nonumber 
 S(\ag,\yy;\ga,\al)=
 -\sqai S(\ag,\xx;\ga,\be)
 \; ,  \\ \nonumber 
 S(\ag,\zz;\be,\al)=
 \sqai S(\ag,\xx;\ga,\be)
 \; ,  \\ \nonumber 
 S(\ag,\xx;\be,\ga)=
 \sqai S(\ag,\xx;\ga,\be)
 \; ,  \\ \nonumber 
 S(\ag,\yy;\al,\ga)=
 -\sqai S(\ag,\xx;\ga,\be)
 \; ,  \\ \nonumber 
 S(\ag,\be;\ga,\xx)=
 \sqai S(\ag,\al;\be,\zz)
 \; ,  \\ \nonumber 
 S(\xx,\zz;\al,\ga)=
 -\sqai S(\xx,\yy;\al,\be)
 \; ,  \\ \nonumber 
 S(\xx,\zz;\ga,\al)=
 -\sqai S(\xx,\yy;\be,\al)
 \; ,  \\ \nonumber 
 S(\xx,\al;\ga,\zz)=
 -\sqai S(\xx,\al;\be,\yy)
 \; ,  \\ \nonumber 
 S(\yy,\be;\ga,\zz)=
 \sqai S(\xx,\al;\be,\yy)
 \; ,  \\ \nonumber 
 S(\yy,\zz;\be,\ga)=
 \sqai S(\xx,\yy;\al,\be)
 \; ,  \\ \nonumber 
 S(\yy,\zz;\ga,\be)=
 \sqai S(\xx,\yy;\be,\al)
 \;.  \\ \nonumber 
 \end{eqnarray}


\subsection{$f$-orbitals, group $T_{h}$}
Independent parameters:
 \begin{eqnarray}
 &&
 U(\ag)             , 
 U(\xx)             , 
 U(\al)             , 
 U(\ag,\xx)         , 
 U(\xx,\yy)         , 
 \\ \nonumber
 &&
 U(\ag,\al)         , 
 U(\xx,\al)         , 
 U(\xx,\be)         , 
 U(\yy,\al)         , 
 U(\al,\be)         , 
 \\ \nonumber
 &&
 J(\ag,\xx)         , 
 J(\xx,\yy)         , 
 J(\ag,\al)         , 
 J(\xx,\al)         , 
 J(\yy,\al)         , 
 \\ \nonumber
 &&
 J(\xx,\be)         , 
 J(\al,\be)         , 
 T(\ag;\xx,\al)     , 
 T(\yy;\xx,\al)     , 
 T(\xx;\yy,\be)     , 
 \\ \nonumber
 &&
 T(\al;\yy,\be)     , 
 T(\be;\xx,\al)     , 
 A(\ag;\ag,\al)     , 
 A(\xx;\xx,\al)     , 
 A(\yy;\xx,\al)     , 
 \\ \nonumber
 &&
 A(\xx;\xx,\be)     , 
 A(\al;\xx,\al)     , 
 A(\be;\xx,\al)     , 
 A(\al;\yy,\be)     , 
 \\ \nonumber
 &&
 S(\ag,\xx;\zz,\yy) , 
 S(\ag,\yy;\ga,\xx) , 
 S(\ag,\xx;\ga,\yy) , 
 S(\ag,\xx;\zz,\be) , 
 \\ \nonumber
 &&
 S(\xx,\yy;\be,\al) , 
 S(\ag,\al;\ga,\yy) , 
 S(\ag,\xx;\ga,\be) , 
 S(\ag,\yy;\ga,\al) , 
\\ \nonumber
 &&
 S(\xx,\al;\be,\yy) , 
 S(\xx,\yy;\al,\be) , 
 S(\ag,\be;\ga,\al) \; . 
 \end{eqnarray}
Dependent parameters:
 \begin{eqnarray}
 U(\yy)            =
 \sqai U(\xx)            
 \;, \\  \nonumber
 U(\zz)            =
 \sqai U(\xx)            
 \;, \\  \nonumber
 U(\be)            =
 \sqai U(\al)            
 \;, \\  \nonumber
 U(\ga)            =
 \sqai U(\al)            
 \;, \\  \nonumber
 U(\ag,\yy)        =
 \sqai U(\ag,\xx)        
 \;, \\  \nonumber
 U(\ag,\zz)        =
 \sqai U(\ag,\xx)        
 \;, \\  \nonumber
 U(\xx,\zz)        =
 \sqai U(\xx,\yy)        
 \;, \\  \nonumber
 U(\ag,\be)        =
 \sqai U(\ag,\al)        
 \;, \\  \nonumber
 U(\yy,\zz)        =
 \sqai U(\xx,\yy)        
 \;, \\  \nonumber
 U(\ag,\ga)        =
 \sqai U(\ag,\al)        
 \;, \\  \nonumber
 U(\zz,\al)        =
 \sqai U(\xx,\be)        
 \;, \\  \nonumber
 U(\xx,\ga)        =
 \sqai U(\yy,\al)        
 \;, \\  \nonumber
 U(\yy,\be)        =
 \sqai U(\xx,\al)        
 \;, \\  \nonumber
 U(\zz,\be)        =
 \sqai U(\yy,\al)        
 \;, \\  \nonumber
 U(\yy,\ga)        =
 \sqai U(\xx,\be)        
 \;, \\  \nonumber
 U(\zz,\ga)        =
 \sqai U(\xx,\al)        
 \;, \\  \nonumber
 U(\al,\ga)        =
 \sqai U(\al,\be)        
 \;, \\  \nonumber
 U(\be,\ga)        =
 \sqai U(\al,\be)        
 \;, \\  \nonumber
 J(\ag,\yy)        =
 \sqai J(\ag,\xx)        
 \;, \\  \nonumber
 J(\ag,\zz)        =
 \sqai J(\ag,\xx)        
 \;, \\  \nonumber
 J(\xx,\zz)        =
 \sqai J(\xx,\yy)        
 \;, \\  \nonumber
 J(\ag,\be)        =
 \sqai J(\ag,\al)        
 \;, \\  \nonumber
 J(\yy,\zz)        =
 \sqai J(\xx,\yy)        
 \;, \\  \nonumber
 J(\ag,\ga)        =
 \sqai J(\ag,\al)        
 \;, \\  \nonumber
 J(\yy,\be)        =
 \sqai J(\xx,\al)        
 \;, \\  \nonumber
 J(\xx,\ga)        =
 \sqai J(\yy,\al)        
 \;, \\  \nonumber
 J(\zz,\al)        =
 \sqai J(\xx,\be)        
 \;, \\  \nonumber
 J(\zz,\be)        =
 \sqai J(\yy,\al)        
 \;, \\  \nonumber
 J(\yy,\ga)        =
 \sqai J(\xx,\be)        
 \;, \\  \nonumber
 J(\zz,\ga)        =
 \sqai J(\xx,\al)        
 \;, \\  \nonumber
 J(\al,\ga)        =
 \sqai J(\al,\be)        
 \;, \\  \nonumber
 J(\be,\ga)        =
 \sqai J(\al,\be)        
 \;, \\  \nonumber
 T(\ag;\yy,\be)    =
 \sqai T(\ag;\xx,\al)    
 \;, \\  \nonumber
 T(\ag;\zz,\ga)    =
 \sqai T(\ag;\xx,\al)    
 \;, \\  \nonumber
 T(\zz;\xx,\al)    =
 \sqai T(\xx;\yy,\be)    
 \;, \\  \nonumber
 T(\xx;\zz,\ga)    =
 \sqai T(\yy;\xx,\al)    
 \;, \\  \nonumber
 T(\zz;\yy,\be)    =
 \sqai T(\yy;\xx,\al)    
 \;, \\  \nonumber
 T(\yy;\zz,\ga)    =
 \sqai T(\xx;\yy,\be)    
 \;, \\  \nonumber
 T(\al;\zz,\ga)    =
 \sqai T(\be;\xx,\al)    
 \;, \\  \nonumber
 T(\ga;\xx,\al)    =
 \sqai T(\al;\yy,\be)    
 \;, \\  \nonumber
 T(\ga;\yy,\be)    =
 \sqai T(\be;\xx,\al)    
 \;, \\  \nonumber
 T(\be;\zz,\ga)    =
 \sqai T(\al;\yy,\be)    
 \;, \\  \nonumber
 A(\ag;\ag,\be)    =
 \sqai A(\ag;\ag,\al)    
 \;, \\  \nonumber
 A(\ag;\ag,\ga)    =
 \sqai A(\ag;\ag,\al)    
 \;, \\  \nonumber
 A(\xx;\xx,\ga)    =
 \sqai A(\yy;\xx,\al)    
 \;, \\  \nonumber
 A(\yy;\yy,\be)    =
 \sqai A(\xx;\xx,\al)    
 \;, \\  \nonumber
 A(\zz;\xx,\al)    =
 \sqai A(\xx;\xx,\be)    
 \;, \\  \nonumber
 A(\zz;\yy,\be)    =
 \sqai A(\yy;\xx,\al)    
 \;, \\  \nonumber
 A(\yy;\yy,\ga)    =
 \sqai A(\xx;\xx,\be)    
 \;, \\  \nonumber
 A(\zz;\zz,\ga)    =
 \sqai A(\xx;\xx,\al)    
 \;, \\  \nonumber
 A(\al;\zz,\ga)    =
 \sqai A(\be;\xx,\al)    
 \;, \\  \nonumber
 A(\ga;\xx,\al)    =
 \sqai A(\al;\yy,\be)    
 \;, \\  \nonumber
 A(\be;\yy,\be)    =
 \sqai A(\al;\xx,\al)    
 \;, \\  \nonumber
 A(\be;\zz,\ga)    =
 \sqai A(\al;\yy,\be)    
 \;, \\  \nonumber
 A(\ga;\yy,\be)    =
 \sqai A(\be;\xx,\al)    
 \;, \\  \nonumber
 A(\ga;\zz,\ga)    =
 \sqai A(\al;\xx,\al)    
 \;, \\  \nonumber
 S(\ag,\yy;\zz,\xx)=
 \sqai S(\ag,\xx;\zz,\yy)
 \;, \\  \nonumber
 S(\ag,\xx;\yy,\zz)=
 \sqai S(\ag,\xx;\zz,\yy)
 \;, \\  \nonumber
 S(\ag,\xx;\be,\zz)=
 \sqai S(\ag,\yy;\ga,\xx)
 \;, \\  \nonumber
 S(\ag,\yy;\zz,\al)=
 \sqai S(\ag,\xx;\zz,\be)
 \;, \\  \nonumber
 S(\ag,\zz;\al,\yy)=
 \sqai S(\ag,\yy;\ga,\xx)
 \;, \\  \nonumber
 S(\ag,\zz;\be,\xx)=
 \sqai S(\ag,\xx;\ga,\yy)
 \;, \\  \nonumber
 S(\ag,\yy;\al,\zz)=
 \sqai S(\ag,\xx;\ga,\yy)
 \;, \\  \nonumber
 S(\ag,\xx;\yy,\ga)=
 \sqai S(\ag,\xx;\zz,\be)
 \;, \\  \nonumber
 S(\ag,\zz;\al,\be)=
 \sqai S(\ag,\yy;\ga,\al)
 \;, \\  \nonumber
 S(\ag,\yy;\al,\ga)=
 \sqai S(\ag,\xx;\ga,\be)
 \;, \\  \nonumber
 S(\ag,\zz;\be,\al)=
 \sqai S(\ag,\xx;\ga,\be)
 \;, \\  \nonumber
 S(\ag,\be;\ga,\xx)=
 \sqai S(\ag,\al;\ga,\yy)
 \;, \\  \nonumber
 S(\ag,\xx;\be,\ga)=
 \sqai S(\ag,\yy;\ga,\al)
 \;, \\  \nonumber
 S(\ag,\al;\be,\zz)=
 \sqai S(\ag,\al;\ga,\yy)
 \;, \\  \nonumber
 S(\xx,\al;\ga,\zz)=
 \sqai S(\xx,\al;\be,\yy)
 \;, \\  \nonumber
 S(\xx,\zz;\ga,\al)=
 \sqai S(\xx,\yy;\be,\al)
 \;, \\  \nonumber
 S(\xx,\zz;\al,\ga)=
 \sqai S(\xx,\yy;\al,\be)
 \;, \\  \nonumber
 S(\ag,\al;\be,\ga)=
 \sqai S(\ag,\be;\ga,\al)
 \;, \\  \nonumber
 S(\ag,\al;\ga,\be)=
 \sqai S(\ag,\be;\ga,\al)
 \;, \\  \nonumber
 S(\yy,\zz;\ga,\be)=
 \sqai S(\xx,\yy;\be,\al)
 \;, \\  \nonumber
 S(\yy,\zz;\be,\ga)=
 \sqai S(\xx,\yy;\al,\be)
 \;, \\  \nonumber
 S(\yy,\be;\ga,\zz)=
 \sqai S(\xx,\al;\be,\yy)
 \;.
 \end{eqnarray}


\subsection{$f$-orbitals, group $D_{6h}$}
Independent parameters:
 \begin{eqnarray}
 &&
 U(\ag)             , 
 U(\xs)             , 
 U(\zz)             , 
 U(\as)             , 
 U(\bs)             , 
 \\ \nonumber
 &&
 U(\ag,\xs)         , 
 U(\ag,\ys)         , 
 U(\ag,\zz)         , 
 U(\xs,\ys)         , 
 U(\xs,\zz)         , 
 \\ \nonumber
 &&
 U(\ag,\as)         , 
 U(\xs,\as)         , 
 U(\ag,\bs)         , 
 U(\ag,\ga)         , 
 U(\xs,\bs)         , 
 \\ \nonumber
 &&
 U(\zz,\as)         , 
 U(\zz,\bs)         , 
 U(\as,\bs)         , 
 J(\ag,\xs)         , 
 J(\ag,\ys)         , 
 \\ \nonumber
 &&
 J(\ag,\zz)         , 
 J(\xs,\zz)         , 
 J(\ag,\as)         , 
 J(\ag,\bs)         , 
 J(\xs,\as)         , 
 \\ \nonumber
 &&
 J(\xs,\bs)         , 
 J(\zz,\as)         , 
 J(\zz,\bs)         , 
 J(\as,\bs)         , 
 T(\ag;\xs,\as)     , 
 \\ \nonumber
 &&
 T(\ag;\ys,\bs)     , 
 T(\ag;\zz,\ga)     , 
 T(\ys;\xs,\as)     , 
 T(\xs;\ys,\bs)     , 
 \\ \nonumber
 &&
 T(\xs;\zz,\ga)     , 
 A(\ag;\ag,\as)     , 
 A(\ag;\ag,\bs)     , 
 A(\xs;\xs,\ga)     , 
 \\ \nonumber
 &&
 S(\ag,\xs;\ga,\ys) , 
 S(\ag,\xs;\zz,\bs) , 
 S(\ag,\ys;\zz,\as) , 
 \\ \nonumber
 &&
 S(\ag,\zz;\as,\ys) , 
 S(\ag,\zz;\bs,\xs) , 
 S(\ag,\ys;\as,\zz) , 
 \\ \nonumber
 && 
 S(\ag,\xs;\bs,\zz) , 
 S(\xs,\ys;\bs,\as) ,
 S(\ag,\as;\ga,\bs) \; . 
 \end{eqnarray}
Dependent parameters:
  \begin{eqnarray}
 U(\ys)            =
 \sqai U(\xs)            
 \;, \\ \nonumber
 U(\ga)            =
 \sqai U(\ag)            
 \;, \\ \nonumber
 U(\ys,\zz)        =
 \sqai U(\xs,\zz)        
 \;, \\ \nonumber
 U(\ys,\as)        =
 \sqai U(\xs,\as)        
 \;, \\ \nonumber
 U(\xs,\ga)        =
 \sqai U(\ag,\ys)        
 \;, \\ \nonumber
 U(\ys,\bs)        =
 \sqai U(\xs,\bs)        
 \;, \\ \nonumber
 U(\ys,\ga)        =
 \sqai U(\ag,\xs)        
 \;, \\ \nonumber
 U(\zz,\ga)        =
 \sqai U(\ag,\zz)        
 \;, \\ \nonumber
 U(\as,\ga)        =
 \sqai U(\ag,\as)        
 \;, \\ \nonumber
 U(\bs,\ga)        =
 \sqai U(\ag,\bs)        
 \;, \\ \nonumber
 J(\xs,\ys)        =
 \sqaa U(\xs)            
 -\sqaa U(\xs,\ys)        
 \;, \\ \nonumber
 J(\ys,\zz)        =
 \sqai J(\xs,\zz)        
 \;, \\ \nonumber
 J(\ys,\as)        =
 \sqai J(\xs,\as)        
 \;, \\ \nonumber
 J(\ag,\ga)        =
 \sqaa U(\ag)            
 -\sqaa U(\ag,\ga)        
 \;, \\ \nonumber
 J(\ys,\bs)        =
 \sqai J(\xs,\bs)        
 \;, \\ \nonumber
 J(\xs,\ga)        =
 \sqai J(\ag,\ys)        
 \;, \\ \nonumber
 J(\ys,\ga)        =
 \sqai J(\ag,\xs)        
 \;, \\ \nonumber
 J(\zz,\ga)        =
 \sqai J(\ag,\zz)        
 \;, \\ \nonumber
 J(\as,\ga)        =
 \sqai J(\ag,\as)        
 \;, \\ \nonumber
 J(\bs,\ga)        =
 \sqai J(\ag,\bs)        
 \;, \\ \nonumber
 T(\ys;\zz,\ga)    =
 -\sqai T(\xs;\zz,\ga)    
 \;, \\ \nonumber
 T(\ga;\xs,\as)    =
 -\sqai T(\ag;\xs,\as)    
 \;, \\ \nonumber
 T(\ga;\ys,\bs)    =
 -\sqai T(\ag;\ys,\bs)    
 \;, \\ \nonumber
 A(\xs;\xs,\as)    =
 -\sqai T(\ys;\xs,\as)    
 \;, \\ \nonumber
 A(\ag;\ag,\ga)    =
 \sqai T(\ag;\zz,\ga)    
 \;, \\ \nonumber
 A(\xs;\xs,\bs)    =
 \sqai T(\xs;\ys,\bs)    
 \;, \\ \nonumber
 A(\ys;\xs,\as)    =
 \sqai T(\ys;\xs,\as)    
 \;, \\ \nonumber
 A(\ys;\ys,\bs)    =
 -\sqai T(\xs;\ys,\bs)    
 \;, \\ \nonumber
 A(\ys;\ys,\ga)    =
 -\sqai A(\xs;\xs,\ga)    
 \;, \\ \nonumber
 A(\ga;\xs,\as)    =
 -\sqai A(\ag;\ag,\as)    
 \;, \\ \nonumber
 A(\ga;\ys,\bs)    =
 -\sqai A(\ag;\ag,\bs)    
 \;, \\ \nonumber
 A(\ga;\zz,\ga)    =
 -\sqai T(\ag;\zz,\ga)    
 \;, \\ \nonumber
 S(\ag,\ys;\zz,\xs)=
 \sqai A(\xs;\xs,\ga)    
 \;, \\ \nonumber
 S(\ag,\xs;\ys,\zz)=
 \sqai T(\xs;\zz,\ga)    
 \;, \\ \nonumber
 S(\ag,\xs;\zz,\ys)=
 \sqai A(\xs;\xs,\ga)    
 \;, \\ \nonumber
 S(\ag,\xs;\ys,\ga)=
 \sqaa U(\ag,\xs)        
 -\sqaa U(\ag,\ys)        
 \;, \\ \nonumber
 S(\ag,\ys;\ga,\xs)=
 \sqai J(\ag,\xs)        
 -\sqai J(\ag,\ys)        
 -\sqai S(\ag,\xs;\ga,\ys)
 \;, \\ \nonumber
 S(\ag,\xs;\ga,\bs)=
 \sqai A(\ag;\ag,\bs)    
 \;, \\ \nonumber
 S(\ag,\ys;\ga,\as)=
 -\sqai A(\ag;\ag,\as)    
 \;, \\ \nonumber
 S(\ag,\as;\ga,\ys)=
 -\sqai A(\ag;\ag,\as)    
 \;, \\ \nonumber
 S(\ag,\bs;\ga,\xs)=
 \sqai A(\ag;\ag,\bs)    
 \;, \\ \nonumber
 S(\ag,\ys;\as,\ga)=
 -\sqai T(\ag;\xs,\as)    
 \;, \\ \nonumber
 S(\ag,\xs;\bs,\ga)=
 \sqai T(\ag;\ys,\bs)    
 \;, \\ \nonumber
 S(\xs,\ys;\as,\bs)=
 -\sqai S(\xs,\ys;\bs,\as)
 \;, \\ \nonumber
 S(\xs,\zz;\ga,\as)=
 -\sqai S(\ag,\ys;\as,\zz)
 \;, \\ \nonumber
 S(\xs,\zz;\as,\ga)=
 -\sqai S(\ag,\zz;\as,\ys)
 \;, \\ \nonumber
 S(\xs,\as;\ga,\zz)=
 -\sqai S(\ag,\ys;\zz,\as)
 \;, \\ \nonumber
 S(\ag,\bs;\ga,\as)=
 -\sqai S(\ag,\as;\ga,\bs)
 \;, \\ \nonumber
 S(\ys,\zz;\ga,\bs)=
 \sqai S(\ag,\xs;\bs,\zz)
 \;, \\ \nonumber
 S(\ys,\zz;\bs,\ga)=
 \sqai S(\ag,\zz;\bs,\xs)
 \;, \\ \nonumber
 S(\ys,\bs;\ga,\zz)=
 \sqai S(\ag,\xs;\zz,\bs)
 \;, \\ \nonumber
 \end{eqnarray}


\subsection{$f$-orbitals, group $D_{4h}$}

Independent parameters:
 \begin{eqnarray}
 &&
 U(\ag)             , 
 U(\xs)             , 
 U(\zz)             , 
 U(\as)             , 
 U(\ga)             , 
 \\ \nonumber
 &&
 U(\ag,\xs)         , 
 U(\ag,\zz)         , 
 U(\xs,\ys)         , 
 U(\xs,\zz)         , 
 U(\ag,\as)         , 
 \\ \nonumber
 &&
 U(\xs,\as)         , 
 U(\ag,\ga)         , 
 U(\xs,\bs)         , 
 U(\zz,\as)         , 
 U(\xs,\ga)         , 
 \\ \nonumber
 &&
 U(\zz,\ga)         , 
 U(\as,\bs)         , 
 U(\as,\ga)         , 
 J(\ag,\xs)         , 
 J(\ag,\zz)         , 
 \\ \nonumber
 &&
 J(\xs,\ys)         , 
 J(\xs,\zz)         , 
 J(\ag,\as)         , 
 J(\xs,\as)         , 
 J(\ys,\as)         , 
 \\ \nonumber
 &&
 J(\ag,\ga)         , 
 J(\xs,\ga)         , 
 J(\zz,\as)         , 
 J(\as,\bs)         , 
 J(\zz,\ga)         , 
 \\ \nonumber
 &&
 J(\as,\ga)         , 
 T(\ag;\xs,\as)     , 
 T(\xs;\ys,\bs)     , 
 T(\zz;\xs,\as)     , 
 \\ \nonumber
 &&
 T(\xs;\zz,\ga)     , 
 T(\as;\ys,\bs)     , 
 T(\as;\zz,\ga)     , 
 T(\ga;\xs,\as)     , 
 \\ \nonumber
 &&
 A(\ag;\ag,\as)     , 
 A(\xs;\xs,\as)     , 
 A(\ys;\xs,\as)     , 
 A(\zz;\xs,\as)     , 
 \\ \nonumber
 &&
 A(\xs;\xs,\ga)     , 
 A(\as;\xs,\as)     , 
 A(\bs;\xs,\as)     , 
 A(\ga;\xs,\as)     , 
 \\ \nonumber
 &&
 A(\as;\zz,\ga)     , 
 S(\ag,\xs;\zz,\ys) , 
 S(\ag,\xs;\ys,\zz) , 
 S(\ag,\xs;\ga,\ys) , 
 \\ \nonumber
 &&
 S(\ag,\xs;\zz,\bs) , 
 S(\ag,\zz;\as,\ys) , 
 S(\ag,\ys;\as,\zz) , 
 \\ \nonumber
 &&
 S(\ag,\bs;\ga,\xs) , 
 S(\xs,\as;\bs,\ys) , 
 S(\xs,\ys;\as,\bs) , 
 \\ \nonumber
 &&
 S(\ag,\zz;\bs,\as) , 
 S(\xs,\ys;\bs,\as) , 
 S(\ag,\as;\bs,\zz) , 
\\ \nonumber
 &&
 S(\ag,\ys;\as,\ga) , 
 S(\ag,\xs;\ga,\bs) , 
 S(\xs,\zz;\as,\ga) , 
\\ \nonumber
 &&
 S(\xs,\as;\ga,\zz) , 
 S(\xs,\zz;\ga,\as) , 
 S(\ag,\as;\ga,\bs) \; . 
 \end{eqnarray}
Dependent parameters:
 \begin{eqnarray}
 U(\ys)            =
 \sqai U(\xs)            
 \;, \\   \nonumber
 U(\bs)            =
 \sqai U(\as)            
 \;, \\   \nonumber
 U(\ag,\ys)        =
 \sqai U(\ag,\xs)        
 \;, \\   \nonumber
 U(\ag,\bs)        =
 \sqai U(\ag,\as)        
 \;, \\   \nonumber
 U(\ys,\zz)        =
 \sqai U(\xs,\zz)        
 \;, \\   \nonumber
 U(\ys,\as)        =
 \sqai U(\xs,\bs)        
 \;, \\   \nonumber
 U(\ys,\bs)        =
 \sqai U(\xs,\as)        
 \;, \\   \nonumber
 U(\zz,\bs)        =
 \sqai U(\zz,\as)        
 \;, \\   \nonumber
 U(\ys,\ga)        =
 \sqai U(\xs,\ga)        
 \;, \\   \nonumber
 U(\bs,\ga)        =
 \sqai U(\as,\ga)        
 \;, \\   \nonumber
 J(\ag,\ys)        =
 \sqai J(\ag,\xs)        
 \;, \\   \nonumber
 J(\ag,\bs)        =
 \sqai J(\ag,\as)        
 \;, \\   \nonumber
 J(\ys,\zz)        =
 \sqai J(\xs,\zz)        
 \;, \\   \nonumber
 J(\xs,\bs)        =
 \sqai J(\ys,\as)        
 \;, \\   \nonumber
 J(\ys,\bs)        =
 \sqai J(\xs,\as)        
 \;, \\   \nonumber
 J(\zz,\bs)        =
 \sqai J(\zz,\as)        
 \;, \\   \nonumber
 J(\ys,\ga)        =
 \sqai J(\xs,\ga)        
 \;, \\   \nonumber
 J(\bs,\ga)        =
 \sqai J(\as,\ga)        
 \;, \\   \nonumber
 T(\ag;\ys,\bs)    =
 -\sqai T(\ag;\xs,\as)    
 \;, \\   \nonumber
 T(\ys;\xs,\as)    =
 -\sqai T(\xs;\ys,\bs)    
 \;, \\   \nonumber
 T(\zz;\ys,\bs)    =
 -\sqai T(\zz;\xs,\as)    
 \;, \\   \nonumber
 T(\ys;\zz,\ga)    =
 -\sqai T(\xs;\zz,\ga)    
 \;, \\   \nonumber
 T(\bs;\xs,\as)    =
 -\sqai T(\as;\ys,\bs)    
 \;, \\   \nonumber
 T(\bs;\zz,\ga)    =
 -\sqai T(\as;\zz,\ga)    
 \;, \\   \nonumber
 T(\ga;\ys,\bs)    =
 -\sqai T(\ga;\xs,\as)    
 \;, \\   \nonumber
 A(\ag;\ag,\bs)    =
 -\sqai A(\ag;\ag,\as)    
 \;, \\   \nonumber
 A(\xs;\xs,\bs)    =
 -\sqai A(\ys;\xs,\as)    
 \;, \\   \nonumber
 A(\ys;\ys,\bs)    =
 -\sqai A(\xs;\xs,\as)    
 \;, \\   \nonumber
 A(\ys;\ys,\ga)    =
 -\sqai A(\xs;\xs,\ga)    
 \;, \\   \nonumber
 A(\zz;\ys,\bs)    =
 -\sqai A(\zz;\xs,\as)    
 \;, \\   \nonumber
 A(\as;\ys,\bs)    =
 -\sqai A(\bs;\xs,\as)    
 \;, \\   \nonumber
 A(\bs;\ys,\bs)    =
 -\sqai A(\as;\xs,\as)    
 \;, \\   \nonumber
 A(\bs;\zz,\ga)    =
 -\sqai A(\as;\zz,\ga)    
 \;, \\   \nonumber
 A(\ga;\ys,\bs)    =
 -\sqai A(\ga;\xs,\as)    
 \;, \\   \nonumber
 S(\ag,\ys;\zz,\xs)=
 \sqai S(\ag,\xs;\zz,\ys)
 \;, \\   \nonumber
 S(\ag,\zz;\bs,\xs)=
 -\sqai S(\ag,\zz;\as,\ys)
 \;, \\   \nonumber
 S(\ag,\ys;\zz,\as)=
 -\sqai S(\ag,\xs;\zz,\bs)
 \;, \\   \nonumber
 S(\ag,\ys;\ga,\xs)=
 -\sqai S(\ag,\xs;\ga,\ys)
 \;, \\   \nonumber
 S(\ag,\xs;\bs,\zz)=
 -\sqai S(\ag,\ys;\as,\zz)
 \;, \\   \nonumber
 S(\ag,\as;\ga,\ys)=
 \sqai S(\ag,\bs;\ga,\xs)
 \;, \\   \nonumber
 S(\ag,\xs;\bs,\ga)=
 \sqai S(\ag,\ys;\as,\ga)
 \;, \\   \nonumber
 S(\ag,\zz;\as,\bs)=
 \sqai S(\ag,\zz;\bs,\as)
 \;, \\   \nonumber
 S(\ag,\ys;\ga,\as)=
 \sqai S(\ag,\xs;\ga,\bs)
 \;, \\   \nonumber
 S(\ag,\bs;\ga,\as)=
 -\sqai S(\ag,\as;\ga,\bs)
 \;, \\   \nonumber
 S(\ys,\bs;\ga,\zz)=
 \sqai S(\xs,\as;\ga,\zz)
 \;, \\   \nonumber
 S(\ys,\zz;\ga,\bs)=
 \sqai S(\xs,\zz;\ga,\as)
 \;, \\   \nonumber
 S(\ys,\zz;\bs,\ga)=
 \sqai S(\xs,\zz;\as,\ga)
 \;. \\   \nonumber
 \end{eqnarray}


\section{Spherical approximation: $f$-orbitals}\label{hhh3}
\subsection{Cubic basis}
Independent parameters:
\begin{eqnarray}\label{345b}
 &&
 U(\ag)             , 
 U(\xx)             , 
 U(\ag,\xx)         , 
 U(\xx,\yy)         \; . 
 \end{eqnarray}
Dependent parameters:
  \begin{eqnarray}
 U(\al)            =
 \sqai U(\ag)            
 \;,  \\  \nonumber
 U(\ag,\al)        =
 -\sqay U(\ag)            
 +\sqay U(\xx)            
 -\sqaz U(\ag,\xx)        
  \\  \nonumber
 +\sqba U(\xx,\yy)        
 \;,  \\  \nonumber
 U(\xx,\al)        =
 \sqai U(\ag,\xx)        
 \;,  \\  \nonumber
 U(\xx,\be)        =
 -\sqat U(\ag)            
 +\sqat U(\xx)            
 -\sqag U(\ag,\xx)        
  \\  \nonumber
 +\sqbb U(\xx,\yy)        
 \;,  \\  \nonumber
 U(\yy,\al)        =
 -\sqat U(\ag)            
 +\sqat U(\xx)            
 -\sqag U(\ag,\xx)        
  \\  \nonumber
 +\sqbb U(\xx,\yy)        
 \;,  \\  \nonumber
 U(\al,\be)        =
 -\sqbc U(\ag)            
 +\sqbc U(\xx)            
 +\sqay U(\ag,\xx)        
  \\  \nonumber
 +\sqbd U(\xx,\yy)        
 \;,  \\  \nonumber
 J(\ag,\xx)        =
 -\sqap U(\ag)            
 +\sqbe U(\xx)            
 -\sqaa U(\ag,\xx)        
 \;,  \\  \nonumber
 J(\xx,\yy)        =
 \sqbf U(\ag)            
 -\sqbg U(\xx)            
 -\sqaa U(\xx,\yy)        
 \;,  \\  \nonumber
 J(\ag,\al)        =
 \sqbh U(\ag)            
 -\sqau U(\xx)            
 +\sqbi U(\ag,\xx)        
  \\  \nonumber
 -\sqay U(\xx,\yy)        
 \;,  \\  \nonumber
 J(\xx,\al)        =
 -\sqap U(\ag)            
 +\sqbe U(\xx)            
 -\sqaa U(\ag,\xx)        
 \;,  \\  \nonumber
 J(\yy,\al)        =
 \sqbj U(\ag)            
 -\sqbk U(\xx)            
 +\sqaf U(\ag,\xx)        
  \\  \nonumber
 -\sqat U(\xx,\yy)        
 \;,  \\  \nonumber
 J(\xx,\be)        =
 \sqbj U(\ag)            
 -\sqbk U(\xx)            
 +\sqaf U(\ag,\xx)        
  \\  \nonumber
 -\sqat U(\xx,\yy)        
 \;,  \\  \nonumber
 J(\al,\be)        =
 -\sqbl U(\ag)            
 +\sqbm U(\xx)            
 -\sqau U(\ag,\xx)        
  \\  \nonumber
 -\sqbc U(\xx,\yy)        
 \;,  \\  \nonumber
 T(\xx;\yy,\be)    =
 \sqbv U(\ag)            
 -\sqbv U(\xx)            
 +\sqbu U(\ag,\xx)        
  \\  \nonumber
 -\sqbu U(\xx,\yy)        
 \;,  \\  \nonumber
 T(\yy;\xx,\al)    =
 \sqbv U(\ag)            
 -\sqbv U(\xx)            
 +\sqbu U(\ag,\xx)        
  \\  \nonumber
 -\sqbu U(\xx,\yy)        
 \;,  \\  \nonumber
 T(\al;\yy,\be)    =
 \sqbs U(\ag)            
 -\sqbs U(\xx)            
 +\sqbo U(\ag,\xx)        
  \\  \nonumber
 -\sqbo U(\xx,\yy)        
 \;,  \\  \nonumber
 T(\be;\xx,\al)    =
 \sqbs U(\ag)            
 -\sqbs U(\xx)            
 +\sqbo U(\ag,\xx)        
  \\  \nonumber
 -\sqbo U(\xx,\yy)        
 \;,  \\  \nonumber
 A(\yy;\xx,\al)    =
 -\sqbw U(\ag)            
 +\sqbw U(\xx)        
  \\  \nonumber          
 -\sqbt U(\ag,\xx)  
 +\sqbt U(\xx,\yy)        
 \;,  \\  \nonumber
 A(\xx;\xx,\be)    =
 -\sqbw U(\ag)            
 +\sqbw U(\xx)           
  \\  \nonumber        
 -\sqbt U(\ag,\xx) 
 +\sqbt U(\xx,\yy)        
 \;,  \\  \nonumber
 A(\be;\xx,\al)    =
 -\sqbx U(\ag)            
 +\sqbx U(\xx)         
  \\  \nonumber          
 -\sqbs U(\ag,\xx) 
 +\sqbs U(\xx,\yy)        
 \;,  \\  \nonumber
 A(\al;\yy,\be)    =
 -\sqbx U(\ag)            
 +\sqbx U(\xx)        
  \\  \nonumber          
 -\sqbs U(\ag,\xx)  
 +\sqbs U(\xx,\yy)        
 \;,  \\  \nonumber
 S(\ag,\xx;\zz,\yy)=
 -\sqbq U(\ag)            
 +\sqbq U(\xx)            
 \;,  \\  \nonumber
 S(\ag,\xx;\ga,\yy)=
 \sqbn U(\ag)            
 -\sqbn U(\xx)            
 +\sqag U(\ag,\xx)        
  \\  \nonumber
 -\sqag U(\xx,\yy)        
 \;,  \\  \nonumber
 S(\ag,\yy;\ga,\xx)=
 -\sqbn U(\ag)            
 +\sqbn U(\xx)            
 -\sqag U(\ag,\xx)        
  \\  \nonumber
 +\sqag U(\xx,\yy)        
 \;,  \\  \nonumber
 S(\ag,\xx;\be,\zz)=
 \sqbn U(\ag)            
 -\sqbn U(\xx)            
 +\sqag U(\ag,\xx)        
  \\  \nonumber
 -\sqag U(\xx,\yy)        
 \;,  \\  \nonumber
 S(\xx,\al;\be,\yy)=
 \sqaw U(\ag)            
 -\sqaw U(\xx)            
 +\sqat U(\ag,\xx)        
  \\  \nonumber
 -\sqat U(\xx,\yy)        
 \;,  \\  \nonumber
 S(\xx,\yy;\al,\be)=
 \sqav U(\ag)            
 -\sqav U(\xx)            
 +\sqas U(\ag,\xx)        
  \\  \nonumber
 -\sqas U(\xx,\yy)        
 \;,  \\  \nonumber
 S(\xx,\yy;\be,\al)=
 -\sqar U(\ag)            
 +\sqar U(\xx)            
 -\sqai U(\ag,\xx)        
  \\  \nonumber
 +\sqai U(\xx,\yy)        
 \;,  \\  \nonumber
 S(\ag,\al;\be,\zz)=
 \sqby U(\ag)            
 -\sqby U(\xx)          
  \\  \nonumber        
 +\sqbr U(\ag,\xx)  
 -\sqbr U(\xx,\yy)        
 \;,  \\  \nonumber
 S(\ag,\al;\ga,\yy)=
 -\sqby U(\ag)            
 +\sqby U(\xx)           
  \\  \nonumber        
 -\sqbr U(\ag,\xx) 
 +\sqbr U(\xx,\yy)        
 \;,  \\  \nonumber
 S(\ag,\xx;\ga,\be)=
 -\sqbo U(\ag)            
 +\sqbo U(\xx)        
  \\  \nonumber           
 -\sqbp U(\ag,\xx) 
 +\sqbp U(\xx,\yy)        
 \;,  \\  \nonumber
 S(\ag,\yy;\ga,\al)=
 \sqbo U(\ag)            
 -\sqbo U(\xx)            
 +\sqbp U(\ag,\xx)        
  \\  \nonumber
 -\sqbp U(\xx,\yy)        
 \;.
 \end{eqnarray}

\subsection{Axial basis}
Independent parameters:
 \begin{eqnarray}
 &&
 U(\ag,\ag)         , 
 U(\xs,\xs)         , 
 U(\ag,\xs)         , 
 U(\ag,\zz)         , 
 \end{eqnarray}
Dependent parameters:
  \begin{eqnarray}
 U(\zz,\zz)        =
 \sqcv U(\ag,\ag)        
 +\sqdl U(\xs,\xs)        
\;, \\ \nonumber
 U(\as,\as)        =
 -\sqdm U(\ag,\ag)        
 +\sqdj U(\xs,\xs)        
\;, \\ \nonumber
 U(\bs,\bs)        =
 -\sqdm U(\ag,\ag)        
 +\sqdj U(\xs,\xs)        
\;, \\ \nonumber
 U(\ga,\ga)        =
 \sqai U(\ag,\ag)        
\;, \\ \nonumber
 U(\ag,\ys)        =
 \sqai U(\ag,\xs)        
\;, \\ \nonumber
 U(\xs,\ys)        =
 \sqcy U(\ag,\ag)        
 -\sqcy U(\xs,\xs)\\    \nonumber    
 +\sqcl U(\ag,\xs)        
 +\sqcc U(\ag,\zz)        
\;, \\ \nonumber
 U(\xs,\zz)        =
 -\sqdi U(\ag,\ag)        
 +\sqdi U(\xs,\xs)   \\    \nonumber      
 +\sqcq U(\ag,\xs)        
 -\sqcp U(\ag,\zz)        
\;, \\ \nonumber
 U(\ag,\as)        =
 \sqcc U(\ag,\xs)        
 +\sqcl U(\ag,\zz)        
\;, \\ \nonumber
 U(\xs,\as)        =
 \sqdk U(\ag,\ag)        
 -\sqdk U(\xs,\xs)        
 +\sqai U(\ag,\zz)        
\;, \\ \nonumber
 U(\ag,\bs)        =
 \sqcc U(\ag,\xs)        
 +\sqcl U(\ag,\zz)        
\;, \\ \nonumber
 U(\ag,\ga)        =
 \sqco U(\ag,\xs)        
 -\sqcc U(\ag,\zz)        
\;, \\ \nonumber
 U(\xs,\bs)        =
 \sqdk U(\ag,\ag)        
 -\sqdk U(\xs,\xs)        
 +\sqai U(\ag,\zz)        
\;, \\ \nonumber
 U(\zz,\as)        =
 \sqde U(\ag,\ag)        
 -\sqde U(\xs,\xs)        
 +\sqai U(\ag,\zz)        
\;, \\ \nonumber
 U(\xs,\ga)        =
 \sqai U(\ag,\xs)        
\;, \\ \nonumber
 U(\zz,\bs)        =
 \sqde U(\ag,\ag)        
 -\sqde U(\xs,\xs)        
 +\sqai U(\ag,\zz)        
\;, \\ \nonumber
 U(\zz,\ga)        =
 \sqai U(\ag,\zz)        
\;, \\ \nonumber
 U(\as,\bs)        =
 -\sqdj U(\ag,\ag)        
 +\sqdj U(\xs,\xs) \\    \nonumber        
 +\sqcn U(\ag,\xs)        
 -\sqcm U(\ag,\zz)        
\;, \\ \nonumber
 U(\as,\ga)        =
 \sqcc U(\ag,\xs)        
 +\sqcl U(\ag,\zz)        
\;, \\ \nonumber
 J(\ag,\xs)        =
 \sqcy U(\ag,\ag)        
 +\sqda U(\xs,\xs)        
 -\sqaa U(\ag,\xs)        
\;, \\ \nonumber
 J(\ag,\ys)        =
 \sqcy U(\ag,\ag)        
 +\sqda U(\xs,\xs)        
 -\sqaa U(\ag,\xs)        
\;, \\ \nonumber
 J(\ag,\zz)        =
\sqde U(\xs,\xs)   
 -\sqdf U(\ag,\ag)                
 -\sqaa U(\ag,\zz)        
\;, \\ \nonumber
 J(\xs,\ys)        =
 -\sqdn U(\ag,\ag)        
 +\sqdi U(\xs,\xs)\\    \nonumber         
 -\sqcf U(\ag,\xs)        
 -\sqce U(\ag,\zz)        
\;, \\ \nonumber
 J(\xs,\zz)        =
 \sqdh U(\ag,\ag)        
 +\sqcx U(\xs,\xs)        
 -\sqcj U(\ag,\xs)  \\    \nonumber       
 +\sqci U(\ag,\zz)        
\;, \\ \nonumber
 J(\ag,\as)        =
 -\sqcx U(\ag,\ag)        
 +\sqcw U(\xs,\xs)        
 -\sqce U(\ag,\xs) \\    \nonumber        
 -\sqcf U(\ag,\zz) 
\;, \\ \nonumber
 J(\ag,\bs)        =
 -\sqcx U(\ag,\ag)        
 +\sqcw U(\xs,\xs)        
 -\sqce U(\ag,\xs)     \\    \nonumber    
 -\sqcf U(\ag,\zz)        
\;, \\ \nonumber
 J(\xs,\as)        =
 \sqdg U(\ag,\ag)        
 +\sqcv U(\xs,\xs)        
 -\sqaa U(\ag,\zz)        
\;, \\ \nonumber
 J(\ys,\as)        =
 \sqdg U(\ag,\ag)        
 +\sqcv U(\xs,\xs)        
 -\sqaa U(\ag,\zz)        
\;, \\ \nonumber
 J(\xs,\bs)        =
 \sqdg U(\ag,\ag)        
 +\sqcv U(\xs,\xs)        
 -\sqaa U(\ag,\zz)        
\;, \\ \nonumber
 J(\ag,\ga)        =
 \sqaa U(\ag,\ag)        
 -\sqck U(\ag,\xs)        
 +\sqce U(\ag,\zz)        
\;, \\ \nonumber
 J(\xs,\ga)        =
 \sqcy U(\ag,\ag)        
 +\sqda U(\xs,\xs)        
 -\sqaa U(\ag,\xs)        
\;, \\ \nonumber
 J(\zz,\as)        =
 \sqaa U(\ag,\ag)        
 -\sqaa U(\ag,\zz)        
\;, \\ \nonumber
 J(\zz,\bs)        =
 \sqaa U(\ag,\ag)        
 -\sqaa U(\ag,\zz)        
\;, \\ \nonumber
 J(\as,\bs)        =
 \sqaa U(\ag,\ag)        
 -\sqch U(\ag,\xs)        
 +\sqcg U(\ag,\zz)        
\;, \\ \nonumber
 J(\zz,\ga)        =
\sqde U(\xs,\xs)   
 -\sqdf U(\ag,\ag)                
 -\sqaa U(\ag,\zz)        
\;, \\ \nonumber
 J(\as,\ga)        =
 -\sqcx U(\ag,\ag)        
 +\sqcw U(\xs,\xs)        
 -\sqce U(\ag,\xs)  \\    \nonumber       
 -\sqcf U(\ag,\zz)        
\;, \\ \nonumber
 T(\ag;\xs,\as)    =
 \sqcs U(\ag,\xs)        
 -\sqcs U(\ag,\zz)        
\;, \\ \nonumber
 T(\ag;\ys,\bs)    =
 -\sqcs U(\ag,\xs)        
 +\sqcs U(\ag,\zz)        
\;, \\ \nonumber
 T(\xs;\ys,\bs)    =
 \sqdd U(\ag,\ag)        
 -\sqdd U(\xs,\xs)        
\;, \\ \nonumber
 T(\ys;\xs,\as)    =
 -\sqdd U(\ag,\ag)        
 +\sqdd U(\xs,\xs)        
\;, \\ \nonumber
 T(\xs;\zz,\ga)    =
 \sqdc U(\ag,\ag)        
 -\sqdc U(\xs,\xs)  \\    \nonumber       
 -\sqct U(\ag,\xs)        
 +\sqct U(\ag,\zz)        
\;, \\ \nonumber
 T(\ga;\xs,\as)    =
 -\sqcs U(\ag,\xs)        
 +\sqcs U(\ag,\zz)        
\;, \\ \nonumber
 A(\ag;\ag,\as)    =
 \sqcz U(\ag,\ag)        
 -\sqcz U(\xs,\xs)\\    \nonumber         
 -\sqcr U(\ag,\xs)        
 +\sqcr U(\ag,\zz)        
\;, \\ \nonumber
 A(\ag;\ag,\bs)    =
 -\sqcz U(\ag,\ag)        
 +\sqcz U(\xs,\xs) \\    \nonumber        
 +\sqcr U(\ag,\xs)        
 -\sqcr U(\ag,\zz)        
\;, \\ \nonumber
 A(\xs;\xs,\as)    =
 \sqdd U(\ag,\ag)        
 -\sqdd U(\xs,\xs)        
\;, \\ \nonumber
 A(\ys;\xs,\as)    =
 -\sqdd U(\ag,\ag)        
 +\sqdd U(\xs,\xs)        
\;, \\ \nonumber
 A(\xs;\xs,\ga)    =
 -\sqdb U(\ag,\ag)        
 +\sqdb U(\xs,\xs)\\    \nonumber         
 +\sqcs U(\ag,\xs)        
 -\sqcs U(\ag,\zz)        
\;, \\ \nonumber
 A(\ga;\xs,\as)    =
 -\sqcz U(\ag,\ag)        
 +\sqcz U(\xs,\xs)\\    \nonumber         
 +\sqcr U(\ag,\xs)        
 -\sqcr U(\ag,\zz)  
\;, \\ \nonumber
 S(\ag,\xs;\zz,\ys)=
 -\sqdb U(\ag,\ag)        
 +\sqdb U(\xs,\xs)  \\    \nonumber       
 +\sqcs U(\ag,\xs)        
 -\sqcs U(\ag,\zz)        
\;, \\ \nonumber
 S(\ag,\xs;\ys,\zz)=
 \sqdc U(\ag,\ag)        
 -\sqdc U(\xs,\xs)  \\    \nonumber       
 -\sqct U(\ag,\xs)        
 +\sqct U(\ag,\zz)        
\;, \\ \nonumber
 S(\ag,\xs;\ga,\ys)=
 -\sqda U(\ag,\ag)        
 +\sqda U(\xs,\xs)\\    \nonumber         
 +\sqcd U(\ag,\xs)        
 -\sqcd U(\ag,\zz)        
\;, \\ \nonumber
 S(\ag,\xs;\zz,\bs)=
 -\sqcv U(\ag,\ag)        
 +\sqcv U(\xs,\xs)        
\;, \\ \nonumber
 S(\ag,\zz;\as,\ys)=
 \sqcc U(\ag,\xs)        
 -\sqcc U(\ag,\zz)        
\;, \\ \nonumber
 S(\ag,\zz;\bs,\xs)=
 -\sqcc U(\ag,\xs)        
 +\sqcc U(\ag,\zz)        
\;, \\ \nonumber
 S(\ag,\ys;\zz,\as)=
 \sqcv U(\ag,\ag)        
 -\sqcv U(\xs,\xs)        
\;, \\ \nonumber
 S(\ag,\ys;\as,\zz)=
 \sqcu U(\ag,\ag)        
 -\sqcu U(\xs,\xs)\\    \nonumber         
 -\sqcc U(\ag,\xs)        
 +\sqcc U(\ag,\zz)        
\;, \\ \nonumber
 S(\ag,\xs;\bs,\zz)=
 -\sqcu U(\ag,\ag)        
 +\sqcu U(\xs,\xs) \\    \nonumber        
 +\sqcc U(\ag,\xs)        
 -\sqcc U(\ag,\zz)        
\;, \\ \nonumber
 S(\ag,\bs;\ga,\xs)=
 -\sqcz U(\ag,\ag)        
 +\sqcz U(\xs,\xs)\\    \nonumber         
 +\sqcr U(\ag,\xs)        
 -\sqcr U(\ag,\zz)        
\;, \\ \nonumber
 S(\xs,\ys;\as,\bs)=
 \sqcv U(\ag,\ag)        
 -\sqcv U(\xs,\xs)\\    \nonumber         
 -\sqcc U(\ag,\xs)        
 +\sqcc U(\ag,\zz)        
\;, \\ \nonumber
 S(\xs,\ys;\bs,\as)=
 -\sqcv U(\ag,\ag)        
 +\sqcv U(\xs,\xs)\\    \nonumber         
 +\sqcc U(\ag,\xs)        
 -\sqcc U(\ag,\zz)        
\;, \\ \nonumber
 S(\ag,\ys;\as,\ga)=
 -\sqcs U(\ag,\xs)        
 +\sqcs U(\ag,\zz)        
\;, \\ \nonumber
 S(\ag,\xs;\ga,\bs)=
 -\sqcz U(\ag,\ag)        
 +\sqcz U(\xs,\xs)\\    \nonumber         
 +\sqcr U(\ag,\xs)        
 -\sqcr U(\ag,\zz)        
\;, \\ \nonumber
 S(\xs,\zz;\as,\ga)=
 -\sqcc U(\ag,\xs)        
 +\sqcc U(\ag,\zz)        
\;, \\ \nonumber
 S(\xs,\as;\ga,\zz)=
 -\sqcv U(\ag,\ag)        
 +\sqcv U(\xs,\xs)        
\;, \\ \nonumber
 S(\xs,\zz;\ga,\as)=
 -\sqcu U(\ag,\ag)        
 +\sqcu U(\xs,\xs)\\    \nonumber         
 +\sqcc U(\ag,\xs)        
 -\sqcc U(\ag,\zz)        
\;, \\ \nonumber
 S(\ag,\as;\ga,\bs)=
 -\sqcw U(\ag,\ag)        
 +\sqcw U(\xs,\xs) \\    \nonumber        
 +\sqab U(\ag,\xs)        
 -\sqab U(\ag,\zz)        \;.
 \end{eqnarray}


\section{First order correction to the spherical approximation}\label{hhh4}

\subsection{$d$-orbitals, group $T_{h}$}
Independent parameter variations:
\begin{eqnarray}
 &&
 \delta U(\zea)                , 
 \delta U(\uua,\zea)           , 
 \delta U(\zea,\eta)           \; . 
 \end{eqnarray}
Dependent parameter variations:
 \begin{eqnarray}
 \delta U(\vva,\eta)          =
 \sqap \delta U(\uua,\zea)          
 \;, \\ \nonumber
 \delta J(\vva,\zea)          =
 \sqaq \delta U(\zea)               
 \;, \\ \nonumber
 \delta J(\vva,\eta)          =
 \sqaq \delta U(\zea)               
 -\sqca \delta U(\uua,\zea)          
 \;, \\ \nonumber
 \delta J(\uua,\zea)          =
 \sqaq \delta U(\zea)               
 -\sqaa \delta U(\uua,\zea)          
 \;, \\ \nonumber
 \delta J(\zea,\eta)          =
 \sqaa \delta U(\zea)               
 -\sqaa \delta U(\zea,\eta)          
 \;, \\ \nonumber
 \delta S(\vva,\zea;\xia,\eta)=
 -\sqcf \delta U(\uua,\zea)          
 +\sqap \delta U(\zea,\eta)          
 \;.
 \end{eqnarray}

\subsection{$d$-orbitals, group $D_{6h}$}
Independent parameter variations:
\begin{eqnarray}
 &&
 \delta U(\uua)                , 
 \delta U(\eta)                , 
 \delta U(\uua,\zea)           , 
 \delta U(\eta,\xia)           , 
 \delta U(\vva,\eta)           , 
 \\ \nonumber
 &&
 \delta U(\uua,\eta)           \; . 
 \end{eqnarray}
Dependent parameter variations:
 \begin{eqnarray}
 \delta U(\vva,\xia)          =
 \sqai \delta U(\vva,\eta)          
 \;, \\ \nonumber
 \delta J(\vva,\uua)          =
 \sqaq \delta U(\uua)               
 \;, \\ \nonumber
 \delta J(\vva,\eta)          =
 \sqaq \delta U(\eta)               
 -\sqaa \delta U(\vva,\eta)          
 \;, \\ \nonumber
 \delta J(\vva,\xia)          =
 \sqaq \delta U(\eta)               
 -\sqaa \delta U(\vva,\eta)          
 \;, \\ \nonumber
 \delta J(\uua,\eta)          =
 \sqaq \delta U(\uua)               
 +\sqaq \delta U(\eta)               
 -\sqaa \delta U(\uua,\eta)          
 \;, \\ \nonumber
 \delta T(\eta;\vva,\uua)     =
 -\sqae \delta U(\uua,\zea)          
 +\sqac \delta U(\eta,\xia)           
 \\ \nonumber        
 -\sqck \delta U(\vva,\eta) 
 -\sqae \delta U(\uua,\eta)          
 \;, \\ \nonumber
 \delta A(\eta;\vva,\uua)     =
 \sqah \delta U(\uua,\zea)          
 -\sqae \delta U(\eta,\xia)        
 \\ \nonumber         
 +\sqcd \delta U(\vva,\eta)   
 +\sqah \delta U(\uua,\eta)          
 \;, \\ \nonumber
 \delta S(\vva,\zea;\eta,\xia)=
 -\sqap \delta U(\uua,\zea)          
 +\sqab \delta U(\eta,\xia)          
 -\sqcj \delta U(\vva,\eta)          
 \\ \nonumber
 -\sqap \delta U(\uua,\eta)          
 \;.
 \end{eqnarray}


\subsection{$d$-orbitals, group $D_{4h}$}
Independent parameter variations:
\begin{eqnarray}
 &&
 \delta U(\eta)                , 
 \delta U(\uua)                , 
 \delta U(\zea)                , 
 \delta U(\uua,\zea)           , 
 \delta U(\vva,\eta)           , 
 \\ \nonumber
 &&
 \delta U(\uua,\eta)           , 
 \delta U(\zea,\eta)           , 
 \delta U(\eta,\xia)           , 
 \delta T(\eta;\vva,\uua)      ; . 
 \end{eqnarray}
Dependent parameter variations:
 \begin{eqnarray}
 \delta J(\vva,\uua)          =
 \sqaq \delta U(\uua)               
 \;, \\ \nonumber
 \delta J(\vva,\zea)          =
 \sqaq \delta U(\zea)               
 \;, \\ \nonumber
 \delta J(\vva,\eta)          =
 \sqaq \delta U(\eta)               
 -\sqaa \delta U(\vva,\eta)          
 \;, \\ \nonumber
 \delta J(\uua,\zea)          =
 \sqaq \delta U(\uua)               
 +\sqaq \delta U(\zea)               
 -\sqaa \delta U(\uua,\zea)          
 \;, \\ \nonumber
 \delta J(\uua,\eta)          =
 \sqaq \delta U(\eta)               
 +\sqaq \delta U(\uua)               
 -\sqaa \delta U(\uua,\eta)          
 \;, \\ \nonumber
 \delta J(\zea,\eta)          =
 \sqaq \delta U(\eta)               
 +\sqaq \delta U(\zea)               
 -\sqaa \delta U(\zea,\eta)          
 \;, \\ \nonumber
 \delta J(\eta,\xia)          =
 \sqaa \delta U(\eta)               
 -\sqaa \delta U(\eta,\xia)          
 \;, \\  \nonumber
 \delta A(\eta;\vva,\uua)     =
 -\sqaa \delta T(\eta;\vva,\uua)     
 \;, \\ \nonumber
 \delta S(\vva,\zea;\xia,\eta)=
 -\sqca \delta U(\uua,\zea)          
 -\sqcl \delta U(\vva,\eta)          
 -\sqca \delta U(\uua,\eta)          
 \\  \nonumber
 +\sqaq \delta U(\zea,\eta)          
 +\sqaa \delta U(\eta,\xia)          
 -\sqcd \delta T(\eta;\vva,\uua)     
 \;, \\ \nonumber
 \delta S(\uua,\zea;\eta,\xia)=
 -\sqah \delta U(\uua,\zea)          
 -\sqah \delta U(\vva,\eta)          
 \\  \nonumber        
 +\sqah \delta U(\zea,\eta)  
 -\sqaa \delta T(\eta;\vva,\uua)     
 \;, \\ \nonumber
 \delta S(\uua,\eta;\xia,\zea)=
 \sqae \delta U(\uua,\zea)          
 +\sqae \delta U(\vva,\eta)            
 \\  \nonumber       
 -\sqae \delta U(\zea,\eta) 
 +\sqai \delta T(\eta;\vva,\uua)     
 \;. \\ \nonumber
 \end{eqnarray}


\subsection{$f$-orbitals, groups $O_h$, $O$, $T_d$}
Independent parameter variations:
\begin{eqnarray}
 &&
 \delta U(\ala)                , 
 \delta U(\aga,\ala)           , 
 \delta U(\xxa,\ala)           , 
 \delta U(\xxa,\bea)           , 
 \delta U(\ala,\bea)           , 
 \\ \nonumber
 &&
 \delta J(\aga,\xxa)           , 
 \delta J(\xxa,\yya)           , 
 \delta J(\ala,\bea)           \; .
 \end{eqnarray}
Dependent parameter variations:
 \begin{eqnarray}
 \delta J(\aga,\ala)          =
 \sqaq \delta U(\ala)               
 -\sqaa \delta U(\aga,\ala)          
  \;,\\ \nonumber
 \delta J(\xxa,\ala)          =
 -\sqcs \delta U(\ala)               
 -\sqaa \delta U(\xxa,\ala)          
 +\sqcq \delta U(\ala,\bea)          
 \\ \nonumber
 -\sqdn \delta J(\xxa,\yya)          
 +\sqdn \delta J(\ala,\bea)          
  \;,\\ \nonumber
 \delta J(\yya,\ala)          =
 \sqdp \delta U(\ala)               
 -\sqaa \delta U(\xxa,\bea)          
 +\sqcs \delta U(\ala,\bea)          
 \\ \nonumber
 +\sqdq \delta J(\xxa,\yya)          
 +\sqcw \delta J(\ala,\bea)          
  \;,\\ \nonumber
 \delta T(\yya;\xxa,\ala)     =
 \sqdh \delta U(\ala)               
 +\sqds \delta U(\xxa,\ala)           
 \\ \nonumber        
 -\sqdt \delta U(\xxa,\bea) 
 -\sqeg \delta U(\ala,\bea)          
 +\sqdu \delta J(\xxa,\yya)            
 \\ \nonumber       
 -\sqde \delta J(\ala,\bea)          
  \;,\\ \nonumber
 \delta T(\bea;\xxa,\ala)     =
 -\sqdv \delta U(\xxa,\ala)          
 -\sqeb \delta U(\xxa,\bea)          
 \\ \nonumber         
 +\sqbq \delta U(\ala,\bea) 
 -\sqen \delta J(\xxa,\yya)          
  \;,\\ \nonumber
 \delta A(\xxa;\xxa,\bea)     =
 \sqdz \delta U(\ala)               
 -\sqea \delta U(\xxa,\ala)           
 \\ \nonumber        
 +\sqdy \delta U(\xxa,\bea) 
 +\sqet \delta U(\ala,\bea)          
 -\sqdw \delta J(\xxa,\yya)           
 \\ \nonumber        
 -\sqdx \delta J(\ala,\bea)          
  \;,\\ \nonumber
 \delta A(\bea;\xxa,\ala)     =
 -\sqdf \delta U(\ala)               
 +\sqec \delta U(\xxa,\ala)          
 \\ \nonumber         
 +\sqed \delta U(\xxa,\bea) 
 -\sqdi \delta U(\ala,\bea)          
 -\sqeo \delta J(\xxa,\yya)          
 \\ \nonumber         
 +\sqdw \delta J(\ala,\bea)          
  \;,\\ \nonumber
 \delta S(\aga,\xxa;\zza,\yya)=
 \sqdf \delta J(\aga,\xxa)          
 -\sqde \delta J(\xxa,\yya)          
  \;,\\ \nonumber
 \delta S(\aga,\xxa;\bea,\zza)=
 \sqcw \delta U(\ala)               
 -\sqce \delta U(\aga,\ala)          
 +\sqeh \delta U(\xxa,\ala)          
 \\ \nonumber
 +\sqei \delta U(\xxa,\bea)          
 -\sqek \delta U(\ala,\bea)          
 -\sqbz \delta J(\aga,\xxa)          
 \\ \nonumber
 +\sqda \delta J(\xxa,\yya)          
 -\sqcv \delta J(\ala,\bea)          
  \;,\\ \nonumber
 \delta S(\aga,\xxa;\gaa,\bea)=
 \sqbq \delta U(\aga,\ala)          
 +\sqem \delta U(\xxa,\ala)          
 \\ \nonumber         
 +\sqgd \delta U(\xxa,\bea) 
 -\sqge \delta U(\ala,\bea)          
 +\sqeo \delta J(\xxa,\yya)          
  \;,\\ \nonumber
 \delta S(\aga,\ala;\bea,\zza)=
 \sqdd \delta U(\ala)               
 -\sqdc \delta U(\aga,\ala)         
 \\ \nonumber         
 -\sqee \delta U(\xxa,\ala)  
 -\sqgg \delta U(\xxa,\bea)          
 +\sqgh \delta U(\ala,\bea)         
 \\ \nonumber          
 -\sqdc \delta J(\aga,\xxa) 
 +\sqcy \delta J(\xxa,\yya)          
 -\sqdb \delta J(\ala,\bea)          
  \;,\\ \nonumber
 \delta S(\xxa,\yya;\ala,\bea)=
 -\sqcs \delta U(\ala)               
 +\sqcm \delta U(\xxa,\ala)          
 \\ \nonumber         
 -\sqcm \delta U(\xxa,\bea) 
 +\sqcs \delta U(\ala,\bea)          
 -\sqbc \delta J(\xxa,\yya)          
 \\ \nonumber         
 +\sqcw \delta J(\ala,\bea)          
  \;,\\ \nonumber
 \delta S(\xxa,\yya;\bea,\ala)=
 -\sqcp \delta U(\ala)               
 -\sqco \delta U(\xxa,\ala)          
 +\sqco \delta U(\xxa,\bea)          
 \\ \nonumber
 +\sqcp \delta U(\ala,\bea)          
 -\sqau \delta J(\xxa,\yya)          
 +\sqcq \delta J(\ala,\bea)          
  \;,\\ \nonumber
 \delta S(\xxa,\ala;\bea,\yya)=
 \sqcn \delta U(\ala)               
 +\sqaa \delta U(\xxa,\ala)          
 -\sqaa \delta U(\xxa,\bea)          
 \\ \nonumber
 -\sqcn \delta U(\ala,\bea)          
 +\sqas \delta J(\xxa,\yya)          
 -\sqcm \delta J(\ala,\bea)          
  \;.
 \end{eqnarray}


\subsection{$f$-orbitals, group $T_{h}$}
Independent parameter variations:
 \begin{eqnarray}
 &&
 \delta U(\ala)                , 
 \delta U(\aga,\ala)           , 
 \delta U(\xxa,\ala)           , 
 \delta U(\xxa,\bea)           , 
 \delta U(\ala,\bea)           , 
 \\ \nonumber
 &&
 \delta J(\aga,\xxa)           , 
 \delta J(\xxa,\yya)           , 
 \delta J(\ala,\bea)           \; . 
 \end{eqnarray}
Dependent parameter variations:
 \begin{eqnarray}
 \delta U(\yya,\ala)          =
 \sqai \delta U(\xxa,\bea)          
 \;, \\ \nonumber
 \delta J(\aga,\ala)          =
 \sqaq \delta U(\ala)               
 -\sqaa \delta U(\aga,\ala)          
 \;, \\ \nonumber
 \delta J(\xxa,\ala)          =
 -\sqcs \delta U(\ala)               
 -\sqaa \delta U(\xxa,\ala)          
 +\sqcq \delta U(\ala,\bea)          
 \\ \nonumber
 -\sqdn \delta J(\xxa,\yya)          
 +\sqdn \delta J(\ala,\bea)          
 \;, \\ \nonumber
 \delta J(\xxa,\bea)          =
 \sqdp \delta U(\ala)               
 -\sqaa \delta U(\xxa,\bea)          
 +\sqcs \delta U(\ala,\bea)          
 \\ \nonumber
 +\sqdq \delta J(\xxa,\yya)          
 +\sqcw \delta J(\ala,\bea)          
 \;, \\ \nonumber
 \delta J(\yya,\ala)          =
 \sqdp \delta U(\ala)               
 -\sqaa \delta U(\xxa,\bea)          
 +\sqcs \delta U(\ala,\bea)          
 \\ \nonumber
 +\sqdq \delta J(\xxa,\yya)          
 +\sqcw \delta J(\ala,\bea)          
 \;, \\ \nonumber
 \delta T(\yya;\xxa,\ala)     =
 \sqdh \delta U(\ala)               
 +\sqds \delta U(\xxa,\ala)          
 \\ \nonumber         
 -\sqdt \delta U(\xxa,\bea) 
 -\sqeg \delta U(\ala,\bea)          
 +\sqdu \delta J(\xxa,\yya)           
 \\ \nonumber        
 -\sqde \delta J(\ala,\bea)          
 \;, \\ \nonumber
 \delta T(\xxa;\yya,\bea)     =
 \sqdh \delta U(\ala)               
 +\sqds \delta U(\xxa,\ala)          
 \\ \nonumber        
 -\sqdt \delta U(\xxa,\bea)  
 -\sqeg \delta U(\ala,\bea)          
 +\sqdu \delta J(\xxa,\yya)         
 \\ \nonumber         
 -\sqde \delta J(\ala,\bea)          
 \;, \\ \nonumber
 \delta T(\bea;\xxa,\ala)     =
 -\sqdv \delta U(\xxa,\ala)          
 -\sqeb \delta U(\xxa,\bea)          
 \\ \nonumber         
 +\sqbq \delta U(\ala,\bea) 
 -\sqen \delta J(\xxa,\yya)          
 \;, \\ \nonumber
 \delta T(\ala;\yya,\bea)     =
 -\sqdv \delta U(\xxa,\ala)          
 -\sqeb \delta U(\xxa,\bea)          
 \\ \nonumber        
 +\sqbq \delta U(\ala,\bea)  
 -\sqen \delta J(\xxa,\yya)          
 \;, \\ \nonumber
 \delta A(\xxa;\xxa,\bea)     =
 \sqdz \delta U(\ala)               
 -\sqea \delta U(\xxa,\ala)         
 \\ \nonumber         
 +\sqdy \delta U(\xxa,\bea)  
 +\sqet \delta U(\ala,\bea)          
 -\sqdw \delta J(\xxa,\yya)         
 \\ \nonumber         
 -\sqdx \delta J(\ala,\bea)          
 \;, \\ \nonumber
 \delta A(\yya;\xxa,\ala)     =
 \sqdz \delta U(\ala)               
 -\sqea \delta U(\xxa,\ala)          
 \\ \nonumber         
 +\sqdy \delta U(\xxa,\bea) 
 +\sqet \delta U(\ala,\bea)          
 -\sqdw \delta J(\xxa,\yya)          
 \\ \nonumber         
 -\sqdx \delta J(\ala,\bea)          
 \;, \\ \nonumber
 \delta A(\bea;\xxa,\ala)     =
 -\sqdf \delta U(\ala)               
 +\sqec \delta U(\xxa,\ala)           
 \\ \nonumber        
 +\sqed \delta U(\xxa,\bea) 
 -\sqdi \delta U(\ala,\bea)          
 -\sqeo \delta J(\xxa,\yya)          
 \\ \nonumber         
 +\sqdw \delta J(\ala,\bea)          
 \;, \\ \nonumber
 \delta A(\ala;\yya,\bea)     =
 -\sqdf \delta U(\ala)               
 +\sqec \delta U(\xxa,\ala)        
 \\ \nonumber         
 +\sqed \delta U(\xxa,\bea)   
 -\sqdi \delta U(\ala,\bea)          
 -\sqeo \delta J(\xxa,\yya)         
 \\ \nonumber        
 +\sqdw \delta J(\ala,\bea)          
 \;, \\ \nonumber
 \delta S(\aga,\xxa;\zza,\yya)=
 \sqdf \delta J(\aga,\xxa)          
 -\sqde \delta J(\xxa,\yya)          
 \;, \\ \nonumber
 \delta S(\aga,\xxa;\gaa,\yya)=
 \sqcw \delta U(\ala)               
 -\sqce \delta U(\aga,\ala)          
 +\sqeh \delta U(\xxa,\ala)          
 \\ \nonumber
 +\sqei \delta U(\xxa,\bea)          
 -\sqek \delta U(\ala,\bea)          
 -\sqbz \delta J(\aga,\xxa)          
 \\ \nonumber
 +\sqda \delta J(\xxa,\yya)          
 -\sqcv \delta J(\ala,\bea)          
 \;, \\ \nonumber
 \delta S(\aga,\xxa;\gaa,\bea)=
 \sqbq \delta U(\aga,\ala)          
 +\sqem \delta U(\xxa,\ala)         
 \\ \nonumber         
 +\sqgd \delta U(\xxa,\bea)  
 -\sqge \delta U(\ala,\bea)          
 +\sqeo \delta J(\xxa,\yya)          
 \;, \\ \nonumber
 \delta S(\aga,\yya;\gaa,\xxa)=
 -\sqcw \delta U(\ala)               
 +\sqce \delta U(\aga,\ala)        
 \\ \nonumber          
 -\sqeh \delta U(\xxa,\ala)  
 -\sqei \delta U(\xxa,\bea)          
 +\sqek \delta U(\ala,\bea)           
 \\ \nonumber         
 +\sqbz \delta J(\aga,\xxa)
 -\sqda \delta J(\xxa,\yya)          
 +\sqcv \delta J(\ala,\bea)          
 \;, \\ \nonumber
 \delta S(\aga,\yya;\gaa,\ala)=
 -\sqbq \delta U(\aga,\ala)          
 -\sqem \delta U(\xxa,\ala)        
 \\ \nonumber         
 -\sqgd \delta U(\xxa,\bea)   
 +\sqge \delta U(\ala,\bea)          
 -\sqeo \delta J(\xxa,\yya)          
 \;, \\ \nonumber
 \delta S(\aga,\ala;\gaa,\yya)=
 -\sqdd \delta U(\ala)               
 +\sqdc \delta U(\aga,\ala)          
 \\ \nonumber         
 +\sqee \delta U(\xxa,\ala) 
 +\sqgg \delta U(\xxa,\bea)          
 -\sqgh \delta U(\ala,\bea)          
 \\ \nonumber         
 +\sqdc \delta J(\aga,\xxa) 
 -\sqcy \delta J(\xxa,\yya)          
 +\sqdb \delta J(\ala,\bea)          
 \;, \\ \nonumber
 \delta S(\xxa,\yya;\ala,\bea)=
 -\sqcs \delta U(\ala)               
 +\sqcm \delta U(\xxa,\ala)          
 \\ \nonumber         
 -\sqcm \delta U(\xxa,\bea) 
 +\sqcs \delta U(\ala,\bea)          
 -\sqbc \delta J(\xxa,\yya)          
 \\ \nonumber         
 +\sqcw \delta J(\ala,\bea)          
 \;, \\ \nonumber
 \delta S(\xxa,\yya;\bea,\ala)=
 -\sqcp \delta U(\ala)               
 -\sqco \delta U(\xxa,\ala)          
 +\sqco \delta U(\xxa,\bea)          
 \\ \nonumber
 +\sqcp \delta U(\ala,\bea)          
 -\sqau \delta J(\xxa,\yya)          
 +\sqcq \delta J(\ala,\bea)          
 \;, \\ \nonumber
 \delta S(\xxa,\ala;\bea,\yya)=
 \sqcn \delta U(\ala)               
 +\sqaa \delta U(\xxa,\ala)          
 -\sqaa \delta U(\xxa,\bea)          
 \\ \nonumber
 -\sqcn \delta U(\ala,\bea)          
 +\sqas \delta J(\xxa,\yya)          
 -\sqcm \delta J(\ala,\bea)          
 \;.
 \end{eqnarray}


\subsection{$f$-orbitals, group $D_{6h}$}
Independent parameter variations:
 \begin{eqnarray}
 &&
 \delta U(\zzz)                , 
 \delta U(\alz)                , 
 \delta U(\bez)                , 
 \delta U(\xxz,\yyz)           , 
 \delta U(\xxz,\zzz)           , 
 \\ \nonumber
 &&
 \delta U(\agz,\alz)           , 
 \delta U(\xxz,\alz)           , 
 \delta U(\agz,\bez)           , 
 \delta U(\agz,\gaz)           , 
 \delta U(\xxz,\bez)           , 
 \\ \nonumber
 &&
 \delta U(\zzz,\alz)           , 
 \delta U(\alz,\bez)           , 
 \delta J(\xxz,\alz)           , 
 \delta J(\agz,\xxz)           , 
 \delta J(\agz,\alz)           , 
 \\ \nonumber
 &&
 \delta J(\agz,\zzz)     \;       .
 \end{eqnarray}
Dependent parameter variations:
 \begin{eqnarray}
 \delta U(\zzz,\bez)          =
 \sqas \delta U(\zzz)               
 -\sqob \delta U(\alz)               
 -\sqdn \delta J(\xxz,\alz)          
 \\ \nonumber
 -\sqoc \delta U(\xxz,\yyz)          
 +\sqod \delta U(\agz,\alz)          
 -\sqoe \delta U(\xxz,\alz)          
 \\ \nonumber
 -\sqco \delta U(\agz,\bez)          
 -\sqcv \delta U(\agz,\gaz)          
 +\sqab \delta U(\xxz,\bez)          
 \\ \nonumber
 +\sqco \delta U(\zzz,\alz)          
 +\sqcm \delta U(\alz,\bez)          
 -\sqof \delta J(\agz,\xxz)          
 \\ \nonumber
 +\sqog \delta J(\agz,\alz)          
 -\sqag \delta J(\agz,\zzz)          
 \;, \\ \nonumber
 \delta J(\agz,\yyz)          =
 \sqai \delta J(\agz,\xxz)          
 \;, \\ \nonumber
 \delta J(\agz,\bez)          =
 -\sqaq \delta U(\alz)               
 +\sqaq \delta U(\bez)               
 +\sqaa \delta U(\agz,\alz)          
 \\ \nonumber
 -\sqaa \delta U(\agz,\bez)          
 +\sqai \delta J(\agz,\alz)          
 \;, \\ \nonumber
 \delta J(\xxz,\zzz)          =
 \sqaq \delta U(\zzz)               
 +\sqbz \delta U(\alz)               
 -\sqaa \delta U(\xxz,\zzz)          
 \\ \nonumber
 -\sqaq \delta U(\agz,\alz)          
 +\sqat \delta J(\agz,\xxz)          
 -\sqaa \delta J(\agz,\alz)          
 \;, \\ \nonumber
 \delta J(\xxz,\bez)          =
 \sqaa \delta U(\zzz)               
 -\sqoh \delta U(\alz)               
 +\sqaq \delta U(\bez)               
 \\ \nonumber
 -\sqco \delta J(\xxz,\alz)          
 -\sqoi \delta U(\xxz,\yyz)          
 +\sqoj \delta U(\agz,\alz)          
 \\ \nonumber
 +\sqok \delta U(\xxz,\alz)          
 -\sqfx \delta U(\agz,\bez)          
 -\sqfa \delta U(\agz,\gaz)          
 \\ \nonumber
 +\sqai \delta U(\xxz,\bez)          
 -\sqol \delta U(\zzz,\alz)          
 +\sqdq \delta U(\alz,\bez)          
 \\ \nonumber
 -\sqom \delta J(\agz,\xxz)          
 +\sqon \delta J(\agz,\alz)          
 -\sqak \delta J(\agz,\zzz)          
 \;, \\ \nonumber
 \delta J(\zzz,\alz)          =
 \sqfb \delta U(\zzz)               
 -\sqoo \delta U(\alz)               
 +\sqab \delta J(\xxz,\alz)          
 \\ \nonumber
 +\sqop \delta U(\agz,\alz)          
 +\sqap \delta U(\xxz,\alz)          
 -\sqaa \delta U(\zzz,\alz)          
 \\ \nonumber
 -\sqoq \delta J(\agz,\xxz)          
 +\sqor \delta J(\agz,\alz)          
 -\sqab \delta J(\agz,\zzz)          
 \;, \\ \nonumber
 \delta J(\zzz,\bez)          =
 \sqos \delta U(\zzz)               
 -\sqot \delta U(\alz)               
 +\sqaq \delta U(\bez)               
 \\ \nonumber
 -\sqou \delta J(\xxz,\alz)          
 -\sqov \delta U(\xxz,\yyz)          
 +\sqow \delta U(\agz,\alz)          
 \\ \nonumber
 +\sqox \delta U(\xxz,\alz)          
 -\sqok \delta U(\agz,\bez)          
 -\sqcq \delta U(\agz,\gaz)          
 \\ \nonumber
 +\sqab \delta U(\xxz,\bez)          
 -\sqoy \delta U(\zzz,\alz)          
 +\sqgj \delta U(\alz,\bez)          
 \\ \nonumber
 -\sqoz \delta J(\agz,\xxz)          
 +\sqpa \delta J(\agz,\alz)          
 -\sqpb \delta J(\agz,\zzz)          
 \;, \\ \nonumber
 \delta J(\alz,\bez)          =
 \sqaq \delta U(\alz)               
 +\sqaq \delta U(\bez)               
 -\sqaa \delta U(\alz,\bez)          
 \;, \\ \nonumber
 \delta T(\agz;\xxz,\alz)     =
 -\sqlf \delta U(\zzz)               
 +\sqqv \delta U(\alz)           
 \\ \nonumber             
 +\sqfl \delta J(\xxz,\alz) 
 +\sqbp \delta U(\xxz,\yyz)          
 -\sqqw \delta U(\agz,\alz)          
 \\ \nonumber         
 -\sqqx \delta U(\xxz,\alz) 
 +\sqbs \delta U(\agz,\gaz)          
 +\sqdr \delta U(\zzz,\alz)            
 \\ \nonumber        
 +\sqqy \delta J(\agz,\xxz)
 -\sqqz \delta J(\agz,\alz)          
 +\sqbu \delta J(\agz,\zzz)          
 \;, \\ \nonumber
 \delta T(\agz;\yyz,\bez)     =
 -\sqec \delta U(\alz)               
 +\sqbt \delta J(\xxz,\alz)          
 \\ \nonumber         
 +\sqra \delta U(\xxz,\yyz) 
 +\sqrb \delta U(\agz,\alz)          
 -\sqbt \delta U(\xxz,\alz)          
 \\ \nonumber         
 +\sqdb \delta U(\agz,\bez) 
 -\sqel \delta U(\agz,\gaz)          
 -\sqlf \delta U(\xxz,\bez)          
 \\ \nonumber         
 -\sqdw \delta U(\alz,\bez) 
 +\sqbp \delta J(\agz,\xxz)          
 -\sqdb \delta J(\agz,\alz)          
 \;, \\ \nonumber
 \delta T(\yyz;\xxz,\alz)     =
 -\sqrc \delta U(\alz)               
 -\sqdd \delta J(\xxz,\alz)           
 \\ \nonumber       
 +\sqrd \delta U(\agz,\alz)  
 -\sqcy \delta U(\xxz,\alz)          
 -\sqre \delta J(\agz,\xxz)          
 \\ \nonumber        
 +\sqrf \delta J(\agz,\alz)          
 \;, \\ \nonumber
 \delta T(\xxz;\yyz,\bez)     =
 \sqcy \delta U(\zzz)               
 -\sqrg \delta U(\alz)             
 \\ \nonumber         
 -\sqrh \delta J(\xxz,\alz)  
 -\sqrb \delta U(\xxz,\yyz)          
 -\sqri \delta U(\agz,\alz)           
 \\ \nonumber         
 +\sqrj \delta U(\xxz,\alz)
 -\sqrk \delta U(\agz,\bez)          
 -\sqrl \delta U(\agz,\gaz)         
 \\ \nonumber         
 +\sqfl \delta U(\xxz,\bez) 
 -\sqrm \delta U(\zzz,\alz)          
 +\sqrn \delta U(\alz,\bez)          
 \\ \nonumber          
 +\sqro \delta J(\agz,\xxz)
 +\sqrp \delta J(\agz,\alz)          
 -\sqdb \delta J(\agz,\zzz)          
 \;, \\ \nonumber
 \delta T(\xxz;\zzz,\gaz)     =
 \sqlf \delta U(\zzz)               
 -\sqrq \delta U(\alz)           
 \\ \nonumber             
 +\sqdw \delta J(\xxz,\alz) 
 +\sqbo \delta U(\xxz,\yyz)          
 -\sqlf \delta U(\xxz,\zzz)          
 \\ \nonumber         
 +\sqrr \delta U(\agz,\alz) 
 +\sqrs \delta U(\xxz,\alz)          
 +\sqru \delta U(\agz,\gaz)          
 \\ \nonumber         
 -\sqfj \delta U(\zzz,\alz) 
 -\sqrv \delta J(\agz,\xxz)          
 +\sqrw \delta J(\agz,\alz)          
 \\ \nonumber         
 -\sqbu \delta J(\agz,\zzz)          
 \;, \\ \nonumber
 \delta A(\agz;\agz,\alz)     =
 \sqlq \delta U(\zzz)               
 -\sqrx \delta U(\alz)                
 \\ \nonumber        
 +\sqcy \delta J(\xxz,\alz) 
 -\sqbo \delta U(\xxz,\yyz)          
 +\sqry \delta U(\agz,\alz)           
 \\ \nonumber        
 +\sqdr \delta U(\xxz,\alz) 
 -\sqru \delta U(\agz,\gaz)          
 -\sqfl \delta U(\zzz,\alz)          
 \\ \nonumber        
 -\sqrz \delta J(\agz,\xxz)  
 +\sqsa \delta J(\agz,\alz)          
 -\sqbt \delta J(\agz,\zzz)          
 \;, \\ \nonumber
 \delta A(\agz;\agz,\bez)     =
 -\sqlq \delta U(\zzz)               
 +\sqrx \delta U(\alz)                
 \\ \nonumber        
 -\sqcy \delta J(\xxz,\alz) 
 +\sqbo \delta U(\xxz,\yyz)          
 -\sqrj \delta U(\agz,\alz)         
 \\ \nonumber          
 -\sqsb \delta U(\xxz,\alz) 
 +\sqlf \delta U(\agz,\bez)          
 +\sqru \delta U(\agz,\gaz)          
 \\ \nonumber         
 -\sqlf \delta U(\xxz,\bez) 
 +\sqfl \delta U(\zzz,\alz)          
 +\sqrz \delta J(\agz,\xxz)          
 \\ \nonumber         
 -\sqsa \delta J(\agz,\alz) 
 +\sqbt \delta J(\agz,\zzz)          
 \;, \\ \nonumber
 \delta A(\xxz;\xxz,\gaz)     =
 -\sqdk \delta U(\zzz)               
 +\sqsc \delta U(\alz)               
 \\ \nonumber        
 -\sqdf \delta J(\xxz,\alz)  
 -\sqbs \delta U(\xxz,\yyz)          
 +\sqlq \delta U(\xxz,\zzz)          
 \\ \nonumber         
 -\sqsd \delta U(\agz,\alz) 
 -\sqfv \delta U(\xxz,\alz)          
 -\sqse \delta U(\agz,\gaz)          
 \\ \nonumber        
 +\sqdt \delta U(\zzz,\alz)  
 +\sqsf \delta J(\agz,\xxz)          
 -\sqbv \delta J(\agz,\alz)          
 \\ \nonumber        
 +\sqdc \delta J(\agz,\zzz)          
 \;, \\ \nonumber
 \delta S(\agz,\xxz;\zzz,\bez)=
 -\sqpc \delta U(\zzz)               
 +\sqpd \delta U(\alz)           
 \\ \nonumber             
 +\sqcm \delta J(\xxz,\alz) 
 +\sqpe \delta U(\xxz,\yyz)          
 -\sqpf \delta U(\agz,\alz)          
 \\ \nonumber         
 -\sqgj \delta U(\xxz,\alz) 
 +\sqdq \delta U(\agz,\bez)          
 +\sqcw \delta U(\agz,\gaz)          
 \\ \nonumber         
 -\sqaq \delta U(\xxz,\bez) 
 +\sqfa \delta U(\zzz,\alz)          
 -\sqcn \delta U(\alz,\bez)         
 \\ \nonumber         
 +\sqpg \delta J(\agz,\xxz)  
 -\sqph \delta J(\agz,\alz)          
 +\sqpi \delta J(\agz,\zzz)          
 \;, \\ \nonumber
 \delta S(\agz,\xxz;\bez,\zzz)=
 -\sqcu \delta U(\zzz)               
 +\sqpj \delta U(\alz)           
 \\ \nonumber             
 -\sqpk \delta J(\xxz,\alz) 
 -\sqcu \delta U(\xxz,\yyz)          
 +\sqbz \delta U(\xxz,\zzz)          
 \\ \nonumber         
 -\sqpl \delta U(\agz,\alz) 
 +\sqpm \delta U(\xxz,\alz)          
 +\sqfa \delta U(\agz,\bez)          
 \\ \nonumber         
 +\sqpn \delta U(\agz,\gaz) 
 -\sqaq \delta U(\xxz,\bez)          
 +\sqpo \delta U(\zzz,\alz)          
 \\ \nonumber         
 +\sqcn \delta U(\alz,\bez) 
 -\sqpp \delta J(\agz,\xxz)          
 -\sqpq \delta J(\agz,\alz)          
 \\ \nonumber         
 +\sqaf \delta J(\agz,\zzz)          
 \;, \\ \nonumber
 \delta S(\agz,\xxz;\gaz,\yyz)=
 -\sqca \delta U(\zzz)               
 +\sqpr \delta U(\alz)            
 \\ \nonumber            
 -\sqdq \delta J(\xxz,\alz) 
 -\sqps \delta U(\agz,\alz)          
 -\sqpt \delta U(\xxz,\alz)           
 \\ \nonumber         
 +\sqpu \delta U(\zzz,\alz)
 +\sqpv \delta J(\agz,\xxz)          
 -\sqpw \delta J(\agz,\alz)           
 \\ \nonumber         
 +\sqab \delta J(\agz,\zzz)          
 \;, \\ \nonumber
 \delta S(\agz,\yyz;\zzz,\alz)=
 \sqpx \delta U(\zzz)               
 -\sqpy \delta U(\alz)               
 +\sqas \delta J(\xxz,\alz)          
 \\ \nonumber
 +\sqcw \delta U(\agz,\alz)          
 +\sqcu \delta U(\xxz,\alz)          
 +\sqbd \delta J(\agz,\xxz)          
 \\ \nonumber
 +\sqcv \delta J(\agz,\alz)          
 -\sqjb \delta J(\agz,\zzz)          
 \;, \\ \nonumber
 \delta S(\agz,\yyz;\alz,\zzz)=
 \sqas \delta U(\zzz)               
 -\sqpz \delta U(\alz)           
 \\ \nonumber            
 -\sqpn \delta J(\xxz,\alz)  
 -\sqaa \delta U(\xxz,\yyz)          
 -\sqbz \delta U(\xxz,\zzz)           
 \\ \nonumber        
 +\sqqa \delta U(\agz,\alz) 
 +\sqqb \delta U(\xxz,\alz)          
 -\sqbz \delta U(\agz,\gaz)          
 \\ \nonumber         
 -\sqqc \delta U(\zzz,\alz) 
 -\sqqd \delta J(\agz,\xxz)          
 +\sqqe \delta J(\agz,\alz)          
 \\ \nonumber         
 -\sqag \delta J(\agz,\zzz)          
 \;, \\ \nonumber
 \delta S(\agz,\zzz;\alz,\yyz)=
 -\sqew \delta U(\zzz)               
 +\sqqf \delta U(\alz)              
 \\ \nonumber         
 +\sqqg \delta J(\xxz,\alz)  
 +\sqaa \delta U(\xxz,\yyz)          
 +\sqbz \delta U(\xxz,\zzz)          
 \\ \nonumber         
 -\sqqh \delta U(\agz,\alz) 
 -\sqqi \delta U(\xxz,\alz)          
 +\sqbz \delta U(\agz,\gaz)         
 \\ \nonumber          
 +\sqqc \delta U(\zzz,\alz) 
 +\sqqj \delta J(\agz,\xxz)          
 -\sqqk \delta J(\agz,\alz)         
 \\ \nonumber          
 +\sqaq \delta J(\agz,\zzz)          
 \;, \\ \nonumber
 \delta S(\agz,\zzz;\bez,\xxz)=
 -\sqce \delta U(\zzz)               
 +\sqql \delta U(\alz)             
 \\ \nonumber           
 +\sqca \delta J(\xxz,\alz) 
 +\sqbe \delta U(\xxz,\yyz)          
 -\sqbz \delta U(\xxz,\zzz)         
 \\ \nonumber          
 -\sqca \delta U(\agz,\alz) 
 -\sqfd \delta U(\xxz,\alz)          
 +\sqgj \delta U(\agz,\bez)         
 \\ \nonumber         
 -\sqpo \delta U(\agz,\gaz) 
 -\sqaq \delta U(\xxz,\bez)          
 +\sqca \delta U(\zzz,\alz)           
 \\ \nonumber        
 -\sqdo \delta U(\alz,\bez) 
 +\sqcj \delta J(\agz,\xxz)          
 -\sqqm \delta J(\agz,\alz)          
 \\ \nonumber         
 +\sqap \delta J(\agz,\zzz)          
 \;, \\ \nonumber
 \delta S(\agz,\alz;\gaz,\bez)=
 -\sqjb \delta U(\zzz)               
 +\sqqn \delta U(\alz)              
 \\ \nonumber          
 -\sqas \delta J(\xxz,\alz) 
 +\sqhs \delta U(\xxz,\yyz)          
 -\sqqo \delta U(\agz,\alz)          
 \\ \nonumber        
 -\sqpe \delta U(\xxz,\alz)  
 -\sqaq \delta U(\agz,\bez)          
 +\sqag \delta U(\agz,\gaz)          
 \\ \nonumber        
 +\sqaa \delta U(\zzz,\alz)  
 +\sqaq \delta U(\alz,\bez)          
 +\sqqp \delta J(\agz,\xxz)          
 \\ \nonumber        
 -\sqqq \delta J(\agz,\alz)          
 \\ \nonumber
 +\sqat \delta J(\agz,\zzz)          
 \;, \\ \nonumber
 \delta S(\xxz,\yyz;\bez,\alz)=
 -\sqcu \delta U(\zzz)               
 +\sqqr \delta U(\alz)             
 \\ \nonumber          
 -\sqcv \delta J(\xxz,\alz)  
 -\sqaf \delta U(\xxz,\yyz)          
 -\sqqs \delta U(\agz,\alz)          
 \\ \nonumber         
 -\sqdp \delta U(\xxz,\alz) 
 -\sqcm \delta U(\agz,\bez)          
 -\sqdp \delta U(\agz,\gaz)        
 \\ \nonumber          
 +\sqfa \delta U(\zzz,\alz)  
 +\sqcm \delta U(\alz,\bez)          
 +\sqqt \delta J(\agz,\xxz)          
 \\ \nonumber         
 -\sqqu \delta J(\agz,\alz) 
 +\sqaf \delta J(\agz,\zzz)          
 \;.
 \end{eqnarray}


\subsection{$f$-orbitals, group $D_{4h}$}
Independent parameter variations:
\begin{eqnarray}
 &&
 \delta U(\zzz)                , 
 \delta U(\alz)                , 
 \delta U(\gaz)                , 
 \delta U(\xxz,\yyz)           , 
 \delta U(\xxz,\zzz)           , 
 \\\nonumber 
 &&
 \delta U(\agz,\alz)           , 
 \delta U(\xxz,\alz)           , 
 \delta U(\agz,\gaz)           , 
 \delta U(\zzz,\alz)           , 
 \delta U(\xxz,\gaz)           , 
 \\\nonumber 
 &&
 \delta U(\zzz,\gaz)           , 
 \delta U(\alz,\bez)           , 
 \delta J(\xxz,\gaz)           , 
 \delta J(\xxz,\alz)           , 
 \delta J(\agz,\xxz)           , 
 \\\nonumber 
 &&
 \delta J(\agz,\zzz)           \; . 
 \end{eqnarray}
Dependent parameter variations:
 \begin{eqnarray}
 \delta U(\xxz,\bez)          =
 \sqai \delta U(\xxz,\alz)          
 \;, \\ \nonumber
 \delta U(\alz,\gaz)          =
 \sqai \delta U(\agz,\alz)          
 +\sqco \delta U(\xxz,\gaz)          
 +\sqgj \delta U(\zzz,\gaz)          
 \;, \\ \nonumber
 \delta J(\agz,\alz)          =
 -\sqgk \delta U(\zzz)               
 +\sqgl \delta U(\alz)               
 +\sqgm \delta U(\xxz,\yyz)          
 \\  \nonumber
 -\sqee \delta U(\agz,\alz)          
 -\sqgn \delta U(\xxz,\alz)          
 +\sqgo \delta U(\agz,\gaz)          
 \\  \nonumber
 +\sqgp \delta U(\zzz,\alz)          
 -\sqgq \delta U(\xxz,\gaz)          
 +\sqee \delta U(\zzz,\gaz)          
 \\  \nonumber
 -\sqgr \delta U(\alz,\bez)          
 +\sqgs \delta J(\xxz,\alz)          
 +\sqgt \delta J(\agz,\xxz)          
 \\  \nonumber
 +\sqgu \delta J(\agz,\zzz)          
 \;, \\ \nonumber
 \delta J(\agz,\gaz)          =
 \sqaq \delta U(\gaz)               
 -\sqaa \delta U(\agz,\gaz)          
 \;, \\ \nonumber
 \delta J(\xxz,\yyz)          =
 -\sqaa \delta U(\xxz,\yyz)          
 \;, \\ \nonumber
 \delta J(\xxz,\zzz)          =
 \sqgv \delta U(\zzz)               
 +\sqgo \delta U(\alz)               
 -\sqgw \delta U(\xxz,\yyz)          
 \\  \nonumber
 -\sqaa \delta U(\xxz,\zzz)          
 -\sqgx \delta U(\agz,\alz)          
 +\sqgy \delta U(\xxz,\alz)          
 \\  \nonumber
 -\sqgz \delta U(\agz,\gaz)          
 -\sqha \delta U(\zzz,\alz)          
 +\sqhb \delta U(\xxz,\gaz)          
 \\  \nonumber
 -\sqem \delta U(\zzz,\gaz)          
 +\sqhc \delta U(\alz,\bez)          
 -\sqhd \delta J(\xxz,\alz)          
 \\  \nonumber
 +\sqhe \delta J(\agz,\xxz)          
 -\sqhf \delta J(\agz,\zzz)          
 \;, \\  \nonumber
 \delta J(\yyz,\alz)          =
 \sqai \delta J(\xxz,\alz)          
 \;, \\ \nonumber
 \delta J(\zzz,\alz)          =
 \sqhg \delta U(\zzz)               
 -\sqhh \delta U(\alz)               
 +\sqhi \delta U(\xxz,\yyz)          
 \\  \nonumber
 +\sqhj \delta U(\agz,\alz)          
 -\sqhk \delta U(\xxz,\alz)          
 +\sqhl \delta U(\agz,\gaz)          
 \\  \nonumber
 +\sqhm \delta U(\zzz,\alz)          
 -\sqhn \delta U(\xxz,\gaz)          
 +\sqho \delta U(\zzz,\gaz)          
 \\  \nonumber
 -\sqhp \delta U(\alz,\bez)          
 +\sqhq \delta J(\xxz,\alz)          
 -\sqhm \delta J(\agz,\xxz)          
 \\  \nonumber
 +\sqhr \delta J(\agz,\zzz)          
 \;, \\ \nonumber
 \delta J(\zzz,\gaz)          =
 -\sqhs \delta U(\gaz)               
 +\sqht \delta J(\xxz,\gaz)          
 +\sqaj \delta U(\xxz,\gaz)          
 \\  \nonumber
 -\sqaa \delta U(\zzz,\gaz)          
 -\sqht \delta J(\agz,\xxz)          
 +\sqai \delta J(\agz,\zzz)          
 \;, \\ \nonumber
 \delta J(\alz,\bez)          =
 \sqaa \delta U(\alz)               
 -\sqaa \delta U(\alz,\bez)          
 \;, \\ \nonumber
 \delta J(\alz,\gaz)          =
 -\sqgk \delta U(\zzz)               
 +\sqgl \delta U(\alz)               
 -\sqas \delta U(\gaz)               
 \\  \nonumber
 +\sqgm \delta U(\xxz,\yyz)          
 -\sqee \delta U(\agz,\alz)          
 -\sqgn \delta U(\xxz,\alz)          
 \\  \nonumber
 +\sqgo \delta U(\agz,\gaz)          
 +\sqbb \delta J(\xxz,\gaz)          
 +\sqgp \delta U(\zzz,\alz)          
 \\  \nonumber
 +\sqhu \delta U(\xxz,\gaz)          
 -\sqhv \delta U(\zzz,\gaz)          
 -\sqgr \delta U(\alz,\bez)          
 \\  \nonumber
 +\sqgs \delta J(\xxz,\alz)          
 -\sqhw \delta J(\agz,\xxz)          
 +\sqgu \delta J(\agz,\zzz)          
 \;, \\ \nonumber
 \delta T(\agz;\xxz,\alz)     =
 -\sqku \delta U(\zzz)               
 +\sqlt \delta U(\alz)            
 \\  \nonumber           
 -\sqns \delta U(\xxz,\yyz)  
 -\sqnt \delta U(\agz,\alz)        
 \\  \nonumber                 
 +\sqlu \delta U(\xxz,\alz)    
 +\sqkv \delta U(\agz,\gaz) 
 +\sqkw \delta U(\zzz,\alz)          
 \\  \nonumber                 
 -\sqlv \delta U(\xxz,\gaz)  
 +\sqkx \delta U(\zzz,\gaz)          
 \\  \nonumber         
 +\sqlw \delta U(\alz,\bez)          
 -\sqlx \delta J(\xxz,\alz)          
 \\  \nonumber         
 -\sqly \delta J(\agz,\xxz) 
 +\sqky \delta J(\agz,\zzz)          
 \;, \\ \nonumber
 \delta T(\xxz;\yyz,\bez)     =
 \sqkz \delta U(\zzz)               
 +\sqla \delta U(\alz)            
 \\  \nonumber            
 -\sqlz \delta U(\xxz,\yyz) 
 -\sqma \delta U(\agz,\alz)           
 \\  \nonumber                 
 +\sqmb \delta U(\xxz,\alz) 
 -\sqlb \delta U(\agz,\gaz)          
 \\  \nonumber         
 -\sqmc \delta U(\zzz,\alz)          
 +\sqmd \delta U(\xxz,\gaz)          
 \\  \nonumber         
 -\sqlc \delta U(\zzz,\gaz) 
 +\sqld \delta U(\alz,\bez)          
 \\  \nonumber                  
 -\sqle \delta J(\xxz,\alz) 
 +\sqnx \delta J(\agz,\xxz)          
 \\  \nonumber         
 -\sqme \delta J(\agz,\zzz)          
 \;, \\ \nonumber
 \delta T(\xxz;\zzz,\gaz)     =
 \sqmf \delta U(\zzz)               
 -\sqnk \delta U(\alz)               
 \\  \nonumber           
 +\sqdd \delta U(\gaz)    
 +\sqnu \delta U(\xxz,\yyz)          
 -\sqlf \delta U(\xxz,\zzz)         
 \\  \nonumber          
 +\sqmg \delta U(\agz,\alz) 
 -\sqmh \delta U(\xxz,\alz)           
 \\  \nonumber                 
 +\sqlg \delta U(\agz,\gaz) 
 -\sqlh \delta J(\xxz,\gaz)          
 \\  \nonumber         
 -\sqmi \delta U(\zzz,\alz)          
 -\sqnv \delta U(\xxz,\gaz)           
 \\  \nonumber        
 +\sqmj \delta U(\zzz,\gaz) 
 -\sqmk \delta U(\alz,\bez)          
 \\  \nonumber                  
 +\sqml \delta J(\xxz,\alz) 
 +\sqnw \delta J(\agz,\xxz)          
 \\  \nonumber         
 -\sqmm \delta J(\agz,\zzz)          
 \;, \\ \nonumber
 \delta T(\gaz;\xxz,\alz)     =
 \sqku \delta U(\zzz)               
 -\sqlt \delta U(\alz)               
 \\  \nonumber         
 +\sqns \delta U(\xxz,\yyz) 
 +\sqnt \delta U(\agz,\alz)           
 \\  \nonumber                  
 -\sqlu \delta U(\xxz,\alz)
 -\sqkv \delta U(\agz,\gaz)           
 \\  \nonumber         
 -\sqkw \delta U(\zzz,\alz)          
 -\sqmn \delta U(\xxz,\gaz)          
 \\  \nonumber         
 +\sqmo \delta U(\zzz,\gaz) 
 -\sqlw \delta U(\alz,\bez)           
 \\  \nonumber                 
 +\sqlx \delta J(\xxz,\alz)
 +\sqly \delta J(\agz,\xxz)          
 \\  \nonumber          
 -\sqky \delta J(\agz,\zzz)          
 \;, \\ \nonumber
 \delta A(\agz;\agz,\alz)     =
 \sqli \delta U(\zzz)               
 -\sqmp \delta U(\alz)             
 \\  \nonumber          
 +\sqlj \delta U(\xxz,\yyz)  
 +\sqmi \delta U(\agz,\alz)          
 \\  \nonumber                  
 +\sqlk \delta U(\xxz,\alz)
 -\sqmq \delta U(\agz,\gaz)          
 \\  \nonumber          
 -\sqll \delta U(\zzz,\alz)          
 +\sqmr \delta U(\xxz,\gaz)          
 \\  \nonumber         
 -\sqms \delta U(\zzz,\gaz) 
 -\sqlm \delta U(\alz,\bez)           
 \\  \nonumber                 
 +\sqmt \delta J(\xxz,\alz) 
 -\sqln \delta J(\agz,\xxz)           
 \\  \nonumber        
 -\sqlo \delta J(\agz,\zzz)          
 \;, \\ \nonumber
 \delta A(\xxz;\xxz,\alz)     =
 \sqkz \delta U(\zzz)               
 +\sqla \delta U(\alz)                
 \\  \nonumber        
 -\sqlz \delta U(\xxz,\yyz) 
 -\sqma \delta U(\agz,\alz)         
 \\  \nonumber                  
 +\sqmb \delta U(\xxz,\alz) 
 -\sqlb \delta U(\agz,\gaz)        
 \\  \nonumber           
 -\sqmc \delta U(\zzz,\alz)          
 +\sqmd \delta U(\xxz,\gaz)          
 \\  \nonumber         
 -\sqlc \delta U(\zzz,\gaz) 
 +\sqld \delta U(\alz,\bez)          
 \\  \nonumber                 
 -\sqle \delta J(\xxz,\alz) 
 +\sqnx \delta J(\agz,\xxz)          
 \\  \nonumber         
 -\sqme \delta J(\agz,\zzz)          
 \;, \\ \nonumber
 \delta A(\xxz;\xxz,\gaz)     =
 -\sqlp \delta U(\zzz)               
 +\sqnl \delta U(\alz)               
 \\  \nonumber            
 -\sqdv \delta U(\gaz)   
 -\sqny \delta U(\xxz,\yyz)          
 +\sqlq \delta U(\xxz,\zzz)           
 \\  \nonumber         
 -\sqmu \delta U(\agz,\alz)
 +\sqnm \delta U(\xxz,\alz)            
 \\  \nonumber                 
 -\sqle \delta U(\agz,\gaz)
 +\sqdd \delta J(\xxz,\gaz)            
 \\  \nonumber        
 +\sqlr \delta U(\zzz,\alz)          
 +\sqmv \delta U(\xxz,\gaz)         
 \\  \nonumber         
 -\sqmw \delta U(\zzz,\gaz)  
 +\sqmx \delta U(\alz,\bez)           
 \\  \nonumber                 
 -\sqmy \delta J(\xxz,\alz) 
 -\sqmz \delta J(\agz,\xxz)          
 \\  \nonumber         
 +\sqna \delta J(\agz,\zzz)          
 \;, \\ \nonumber
 \delta A(\yyz;\xxz,\alz)     =
 -\sqkz \delta U(\zzz)               
 -\sqla \delta U(\alz)              
 \\  \nonumber          
 +\sqlz \delta U(\xxz,\yyz) 
 +\sqma \delta U(\agz,\alz)          
 \\  \nonumber                  
 -\sqmb \delta U(\xxz,\alz) 
 +\sqlb \delta U(\agz,\gaz)          
 \\  \nonumber         
 +\sqmc \delta U(\zzz,\alz)          
 -\sqmd \delta U(\xxz,\gaz)         
 \\  \nonumber          
 +\sqlc \delta U(\zzz,\gaz) 
 -\sqld \delta U(\alz,\bez)          
 \\  \nonumber                 
 +\sqle \delta J(\xxz,\alz) 
 -\sqnx \delta J(\agz,\xxz)         
 \\  \nonumber          
 +\sqme \delta J(\agz,\zzz)          
 \;, \\ \nonumber
 \delta A(\gaz;\xxz,\alz)     =
 -\sqli \delta U(\zzz)               
 +\sqmp \delta U(\alz)                
 \\  \nonumber           
 -\sqlf \delta U(\gaz)   
 -\sqlj \delta U(\xxz,\yyz)          
 -\sqmi \delta U(\agz,\alz)         
 \\  \nonumber         
 -\sqlk \delta U(\xxz,\alz)  
 +\sqmq \delta U(\agz,\gaz)          
 +\sqbu \delta J(\xxz,\gaz)          
 \\  \nonumber         
 +\sqll \delta U(\zzz,\alz)
 +\sqnz \delta U(\xxz,\gaz)            
 \\  \nonumber                 
 -\sqnb \delta U(\zzz,\gaz)
 +\sqlm \delta U(\alz,\bez)            
 \\  \nonumber        
 -\sqmt \delta J(\xxz,\alz)          
 -\sqnc \delta J(\agz,\xxz)           
 \\  \nonumber        
 +\sqlo \delta J(\agz,\zzz)          
 \;, \\ \nonumber
 \delta S(\agz,\xxz;\yyz,\zzz)=
 \sqmf \delta U(\zzz)               
 -\sqnk \delta U(\alz)             
 \\  \nonumber          
 +\sqnu \delta U(\xxz,\yyz)  
 -\sqlf \delta U(\xxz,\zzz)          
 +\sqmg \delta U(\agz,\alz)          
 \\  \nonumber         
 -\sqmh \delta U(\xxz,\alz) 
 +\sqlg \delta U(\agz,\gaz)           
 \\  \nonumber                 
 -\sqmi \delta U(\zzz,\alz) 
 -\sqnd \delta U(\xxz,\gaz)           
 \\  \nonumber        
 +\sqne \delta U(\zzz,\gaz)          
 -\sqmk \delta U(\alz,\bez)          
 \\  \nonumber         
 +\sqml \delta J(\xxz,\alz) 
 +\sqoa \delta J(\agz,\xxz)            
 \\  \nonumber                
 -\sqmm \delta J(\agz,\zzz)          
 \;, \\ \nonumber
 \delta S(\agz,\xxz;\zzz,\yyz)=
 -\sqlp \delta U(\zzz)               
 +\sqnl \delta U(\alz)           
 \\  \nonumber             
 -\sqny \delta U(\xxz,\yyz) 
 +\sqlq \delta U(\xxz,\zzz)           
 \\  \nonumber                 
 -\sqmu \delta U(\agz,\alz) 
 +\sqnm \delta U(\xxz,\alz)           
 \\  \nonumber        
 -\sqle \delta U(\agz,\gaz)          
 +\sqlr \delta U(\zzz,\alz)         
 \\  \nonumber         
 +\sqnf \delta U(\xxz,\gaz)  
 -\sqls \delta U(\zzz,\gaz)          
 \\  \nonumber                  
 +\sqmx \delta U(\alz,\bez) 
 -\sqmy \delta J(\xxz,\alz)          
 \\  \nonumber         
 +\sqnn \delta J(\agz,\xxz)          
 +\sqna \delta J(\agz,\zzz)          
 \;, \\ \nonumber
 \delta S(\agz,\xxz;\zzz,\bez)=
 -\sqjq \delta U(\zzz)               
 +\sqjr \delta U(\alz)             
 \\  \nonumber           
 -\sqjs \delta U(\xxz,\yyz) 
 -\sqhx \delta U(\agz,\alz)          
 -\sqjt \delta U(\xxz,\alz)          
 \\  \nonumber         
 -\sqhy \delta U(\agz,\gaz) 
 -\sqep \delta U(\zzz,\alz)          
 +\sqju \delta U(\xxz,\gaz)        
 \\  \nonumber         
 -\sqhz \delta U(\zzz,\gaz) 
 +\sqia \delta U(\alz,\bez)          
 -\sqib \delta J(\xxz,\alz)          
 \\  \nonumber         
 -\sqic \delta J(\agz,\xxz) 
 +\sqjv \delta J(\agz,\zzz)          
 \;, \\ \nonumber
 \delta S(\agz,\xxz;\gaz,\yyz)=
 -\sqid \delta U(\zzz)               
 +\sqie \delta U(\alz)               
 -\sqbz \delta U(\gaz)               
 \\  \nonumber
 -\sqif \delta U(\xxz,\yyz)          
 -\sqig \delta U(\agz,\alz)          
 +\sqih \delta U(\xxz,\alz)          
 \\  \nonumber
 -\sqho \delta U(\agz,\gaz)          
 +\sqaa \delta J(\xxz,\gaz)          
 +\sqii \delta U(\zzz,\alz)          
 \\  \nonumber
 +\sqno \delta U(\xxz,\gaz)          
 -\sqij \delta U(\zzz,\gaz)          
 +\sqjw \delta U(\alz,\bez)          
 \\  \nonumber
 -\sqik \delta J(\xxz,\alz)          
 -\sqil \delta J(\agz,\xxz)          
 +\sqim \delta J(\agz,\zzz)          
 \;, \\ \nonumber
 \delta S(\agz,\xxz;\gaz,\bez)=
 -\sqli \delta U(\zzz)               
 +\sqmp \delta U(\alz)                
 \\  \nonumber             
 -\sqlq \delta U(\gaz) 
 -\sqlj \delta U(\xxz,\yyz)          
 -\sqmi \delta U(\agz,\alz)          
 \;, \\ \nonumber         
 -\sqlk \delta U(\xxz,\alz) 
 +\sqmq \delta U(\agz,\gaz)          
 +\sqbt \delta J(\xxz,\gaz)           
 \\  \nonumber         
 +\sqll \delta U(\zzz,\alz)
 +\sqng \delta U(\xxz,\gaz)           
 \\  \nonumber                  
 -\sqnh \delta U(\zzz,\gaz) 
 +\sqlm \delta U(\alz,\bez)          
 \\  \nonumber         
 -\sqmt \delta J(\xxz,\alz)          
 -\sqll \delta J(\agz,\xxz) 
 +\sqlo \delta J(\agz,\zzz)          
 \;, \\ \nonumber
 \delta S(\agz,\yyz;\alz,\zzz)=
 \sqdj \delta U(\zzz)               
 -\sqjx \delta U(\alz)               
 \\  \nonumber         
 +\sqjy \delta U(\xxz,\yyz) 
 -\sqbz \delta U(\xxz,\zzz)          
 +\sqin \delta U(\agz,\alz)          
 \\  \nonumber         
 +\sqjz \delta U(\xxz,\alz) 
 -\sqka \delta U(\agz,\gaz)          
 -\sqio \delta U(\zzz,\alz)          
 \\  \nonumber         
 +\sqnp \delta U(\xxz,\gaz) 
 -\sqkb \delta U(\zzz,\gaz)          
 -\sqkc \delta U(\alz,\bez)          
 \\  \nonumber         
 +\sqkd \delta J(\xxz,\alz) 
 +\sqip \delta J(\agz,\xxz)          
 -\sqke \delta J(\agz,\zzz)          
 \;, \\ \nonumber
 \delta S(\agz,\yyz;\alz,\gaz)=
 \sqku \delta U(\zzz)               
 -\sqlt \delta U(\alz)              
 \\  \nonumber          
 +\sqns \delta U(\xxz,\yyz) 
 +\sqnt \delta U(\agz,\alz)           
 \\  \nonumber                 
 -\sqlu \delta U(\xxz,\alz) 
 -\sqkv \delta U(\agz,\gaz) 
 -\sqkw \delta U(\zzz,\alz)           
 \\  \nonumber                 
 -\sqni \delta U(\xxz,\gaz) 
 +\sqnj \delta U(\zzz,\gaz)            
 \\  \nonumber               
 -\sqlw \delta U(\alz,\bez)   
 +\sqlx \delta J(\xxz,\alz)           
 \\  \nonumber               
 +\sqly \delta J(\agz,\xxz) 
 -\sqky \delta J(\agz,\zzz)          
 \;, \\ \nonumber
 \delta S(\agz,\zzz;\alz,\yyz)=
 \sqee \delta U(\zzz)               
 -\sqiq \delta U(\alz)            
 \\  \nonumber            
 -\sqir \delta U(\xxz,\yyz) 
 +\sqbz \delta U(\xxz,\zzz)          
 -\sqis \delta U(\agz,\alz)          
 \\  \nonumber         
 +\sqit \delta U(\xxz,\alz) 
 +\sqiu \delta U(\agz,\gaz)          
 +\sqiv \delta U(\zzz,\alz)          
 \\  \nonumber         
 -\sqiw \delta U(\xxz,\gaz)
 +\sqkf \delta U(\zzz,\gaz)          
 +\sqkg \delta U(\alz,\bez)          
 \\  \nonumber         
 +\sqix \delta J(\xxz,\alz)
 +\sqiy \delta J(\agz,\xxz)          
 -\sqiz \delta J(\agz,\zzz)          
 \;, \\ \nonumber
 \delta S(\agz,\alz;\gaz,\bez)=
 -\sqja \delta U(\zzz)               
 +\sqkh \delta U(\alz)               
 \\  \nonumber            
 -\sqjb \delta U(\gaz)   
 -\sqki \delta U(\xxz,\yyz)          
 -\sqjc \delta U(\agz,\alz)          
 \\  \nonumber         
 +\sqjd \delta U(\xxz,\alz) 
 +\sqkj \delta U(\agz,\gaz)          
 +\sqat \delta J(\xxz,\gaz)          
 \\  \nonumber         
 +\sqgs \delta U(\zzz,\alz) 
 +\sqkk \delta U(\xxz,\gaz)          
 -\sqje \delta U(\zzz,\gaz)          
 \\  \nonumber         
 +\sqjf \delta U(\alz,\bez) 
 -\sqjg \delta J(\xxz,\alz)          
 -\sqkl \delta J(\agz,\xxz)           
 \\  \nonumber        
 +\sqhe \delta J(\agz,\zzz)          
 \;, \\ \nonumber
 \delta S(\agz,\bez;\gaz,\xxz)=
 -\sqli \delta U(\zzz)               
 +\sqmp \delta U(\alz)              
 \\  \nonumber              
 -\sqlq \delta U(\gaz)  
 -\sqlj \delta U(\xxz,\yyz)          
 -\sqmi \delta U(\agz,\alz)         
 \\  \nonumber         
 -\sqlk \delta U(\xxz,\alz)  
 +\sqmq \delta U(\agz,\gaz)          
 +\sqbt \delta J(\xxz,\gaz)           
 \\  \nonumber        
 +\sqll \delta U(\zzz,\alz) 
 +\sqng \delta U(\xxz,\gaz)          
 \\  \nonumber                 
 -\sqnh \delta U(\zzz,\gaz) 
 +\sqlm \delta U(\alz,\bez)          
 \\  \nonumber         
 -\sqmt \delta J(\xxz,\alz)          
 -\sqll \delta J(\agz,\xxz) 
 +\sqlo \delta J(\agz,\zzz)         
 \;, \\ \nonumber         
 \delta S(\xxz,\yyz;\alz,\bez)=
 \sqjh \delta U(\zzz)               
 -\sqkm \delta U(\alz)           
 \\  \nonumber             
 +\sqkn \delta U(\xxz,\yyz) 
 +\sqji \delta U(\agz,\alz)          
 -\sqja \delta U(\xxz,\alz)          
 \\  \nonumber         
 +\sqko \delta U(\agz,\gaz) 
 -\sqjj \delta U(\zzz,\alz)          
 -\sqnq \delta U(\xxz,\gaz)          
 \\  \nonumber         
 +\sqkp \delta U(\zzz,\gaz) 
 -\sqkq \delta U(\alz,\bez)          
 +\sqjk \delta J(\xxz,\alz)            
 \\  \nonumber        
 +\sqjl \delta J(\agz,\xxz)
 -\sqjm \delta J(\agz,\zzz)          
 \;, \\ \nonumber
 \delta S(\xxz,\yyz;\bez,\alz)=
 -\sqjh \delta U(\zzz)               
 +\sqkm \delta U(\alz)          
 \\  \nonumber             
 -\sqkn \delta U(\xxz,\yyz)  
 -\sqji \delta U(\agz,\alz)          
 +\sqja \delta U(\xxz,\alz)           
 \\  \nonumber        
 -\sqko \delta U(\agz,\gaz) 
 +\sqjj \delta U(\zzz,\alz)          
 +\sqnq \delta U(\xxz,\gaz)          
 \\  \nonumber         
 -\sqkp \delta U(\zzz,\gaz) 
 +\sqkq \delta U(\alz,\bez)          
 -\sqjk \delta J(\xxz,\alz)          
 \\  \nonumber         
 -\sqjl \delta J(\agz,\xxz) 
 +\sqjm \delta J(\agz,\zzz)          
 \;, \\ \nonumber
 \delta S(\xxz,\zzz;\alz,\gaz)=
 -\sqee \delta U(\zzz)               
 +\sqiq \delta U(\alz)                 
 \\  \nonumber        
 +\sqir \delta U(\xxz,\yyz)
 -\sqbz \delta U(\xxz,\zzz)          
 +\sqis \delta U(\agz,\alz)         
 \\  \nonumber         
 -\sqit \delta U(\xxz,\alz)  
 -\sqiu \delta U(\agz,\gaz)          
 -\sqiv \delta U(\zzz,\alz)         
 \\  \nonumber         
 -\sqkr \delta U(\xxz,\gaz)  
 +\sqjn \delta U(\zzz,\gaz)          
 -\sqkg \delta U(\alz,\bez)         
 \\  \nonumber         
 -\sqix \delta J(\xxz,\alz) 
 -\sqiy \delta J(\agz,\xxz)          
 +\sqiz \delta J(\agz,\zzz)          
 \;, \\ \nonumber
 \delta S(\xxz,\zzz;\gaz,\alz)=
 -\sqdj \delta U(\zzz)               
 +\sqjx \delta U(\alz)                
 \\  \nonumber            
 -\sqas \delta U(\gaz)  
 -\sqjy \delta U(\xxz,\yyz)          
 +\sqbz \delta U(\xxz,\zzz)           
 \\  \nonumber        
 -\sqin \delta U(\agz,\alz) 
 -\sqjz \delta U(\xxz,\alz)          
 +\sqka \delta U(\agz,\gaz)         
 \\  \nonumber         
 +\sqag \delta J(\xxz,\gaz)  
 +\sqio \delta U(\zzz,\alz)          
 +\sqnr \delta U(\xxz,\gaz)          
 \\  \nonumber         
 -\sqks \delta U(\zzz,\gaz) 
 +\sqkc \delta U(\alz,\bez)          
 -\sqkd \delta J(\xxz,\alz)           
 \\  \nonumber         
 -\sqjo \delta J(\agz,\xxz)
 +\sqke \delta J(\agz,\zzz)          
 \;, \\ \nonumber
 \delta S(\xxz,\alz;\gaz,\zzz)=
 -\sqjq \delta U(\zzz)               
 +\sqjr \delta U(\alz)              
 \\  \nonumber             
 -\sqcu \delta U(\gaz)   
 -\sqjs \delta U(\xxz,\yyz)          
 -\sqhx \delta U(\agz,\alz)           
 \\  \nonumber         
 -\sqjt \delta U(\xxz,\alz)
 -\sqhy \delta U(\agz,\gaz)          
 +\sqaf \delta J(\xxz,\gaz)          
 \\  \nonumber         
 -\sqep \delta U(\zzz,\alz) 
 +\sqkt \delta U(\xxz,\gaz)          
 -\sqhz \delta U(\zzz,\gaz)          
 \\  \nonumber         
 +\sqia \delta U(\alz,\bez) 
 -\sqib \delta J(\xxz,\alz)          
 -\sqjp \delta J(\agz,\xxz)           
 \\  \nonumber                  
 +\sqjv \delta J(\agz,\zzz)          \;.
 \\  \nonumber         
 \end{eqnarray}

\bibliographystyle{unsrt}
\bibliography{bib5}

\end{document}